\documentclass[a4paper,11pt]{article}
\pdfoutput=1 

\usepackage{jheppub} 

\usepackage[T1]{fontenc} 

\usepackage{color}
\usepackage{latexsym}
\usepackage{mathrsfs}
\usepackage{amssymb}
\usepackage{amsmath}
\usepackage{amsfonts, epsf,epsfig,color}
\usepackage{graphicx}
\usepackage{tensor}
\usepackage{manfnt}
\usepackage{slashed}
\usepackage[all]{xy}
\include{macro}

\newcommand{\C}{\mathbb{C}}

\newcommand{\R}{\mathbb{R}}

\newcommand{\Z}{\mathbb{Z}}

\newcommand{\rd}{\, \mathrm{d}}

\newcommand{\be}{\begin{equation}\label}
\newcommand{\ee}{\end{equation}}
\newcommand{\bea}{\begin{eqnarray}\label}
\newcommand{\eea}{\end{eqnarray}}

\title{\textbf{ A Superstring Field Theory for Supergravity}}

\author{R. A. Reid-Edwards and D. A.  Riccombeni}

\affiliation{The Milne Centre for Astrophysics \\ \& School of Mathematics and Physical Sciences, \\University of Hull, Cottingham Road, Hull,\\ HU6 7RX, UK}

\emailAdd{r.reid-edwards@hull.ac.uk, d.riccombeni@hull.ac.uk}

\abstract{A covariant closed superstring field theory, equivalent to classical ten-dimensional Type II supergravity, is presented. The  defining conformal field theory is the ambitwistor string worldsheet theory of Mason and Skinner. This theory is known to reproduce the scattering amplitudes of Cachazo, He and Yuan in which the scattering equations play an important role and the string field theory naturally incorporates these results. We investigate the operator formalism description of the ambitwsitor string and propose an action for the string field theory of the bosonic and supersymmetric theories. The correct linearised gauge symmetries and spacetime actions are explicitly reproduced and evidence is given that the action is correct to all orders. The focus is on the Neveu-Schwarz sector and the explicit description of tree level perturbation theory about flat spacetime. Application of the string field theory to general supergravity backgrounds and the inclusion of the Ramond sector are briefly discussed.
}

\begin{document} 
\maketitle
\flushbottom

\section{Introduction}\label{introduction}

Well-understood backgrounds in string theory are few and far between and those that are understood often have a high degree of symmetry which enables the problem of finding the worldsheet theory to be tractable. As such, Supergravity has long been a useful indirect tool to gain insight into string theories in non-trivial backgrounds. Supergravity has also served as a source of inspiration for non-conventional, or stringy, backgrounds that are currently inaccessible to a full analysis at the worldsheet level\footnote{Examples include T-folds \cite{Hull:2006va,Hull:2005hk,Hull:2006tp}, Double Field Theory \cite{Hull:2009mi,Siegel:1993xq,Siegel:1993th}, U-duality \cite{Hull:1994ys}, M-theory \cite{Witten:1995ex}, flux compactifications and G-structures \cite{Gauntlett:2002sc,Grana:2005jc}, to name but a few.}.

Although there have been a number applications of the worldline formalism to certain problems \cite{Bastianelli:2006rx}, the overwhelming volume of work on supergravity has been from the perspective of the spacetime Einstein-Hilbert action or equations of motion. This is hardly surprising, given its conceptual elegance and historic achievements; however, the language in which the Einstein-Hilbert formulation is  written can make it difficult to generalise lessons from supergravity to the full string theory. For example, the conformal invariance that plays such an important role in the worldsheet theory, though implicit, is not easy to recognise in the target space formulation. And the worldline approach, though similar to the worldsheet theory in some respects, does not have many of the features central to our current understanding of string theory.

In this paper we take the first steps in developing an alternative approach to ten-dimensional supergravity based on the ambitwistor \emph{worldsheet} model of \cite{Mason:2013sva}. This ambitwistor string describes Type II supergravity in ten dimensions in terms of the chiral embedding of a worldsheet $\Sigma$ into ambitwistor space. The worldsheet action for the ambitwistor string is
\begin{equation}\label{action01}
S=\int_{\Sigma}P_{\mu}\bar{\partial}X^{\mu}+ \frac{1}{2}eP^2+...,
\end{equation}
where $(P_{\mu},X^{\mu})$ take values in the cotangent bundle of spacetime, $e$ is a Lagrange multiplier imposing the constraint $P^2=0$, and the ellipsis denotes fermion and ghost contributions. All fields are holomorphic on the worldsheet $\Sigma$.

The ambitwistor string theory (\ref{action01}) is thought, with good reason, to be equivalent to a perturbative description of ten-dimensional Type II supergravity. Though written as a chiral worldsheet theory, with superconformal invariance very similar to that found in the conventional superstring, the spectrum of the ambitwistor theory is massless, it has the correct S-matrix, and the supergravity equations of motion are reproduced as the condition that an anomaly vanishes \cite{Adamo:2014wea}. There are no higher derivative corrections. In this paper we begin the systematic study of the ambitwistor string as a covariant string field theory. Our ultimate hope is that this will provide a useful toy model that will eventually cast some light on some of the outstanding problems in conventional string theory. We also hope to better understand this interesting class of chiral string theories in their own right.

The origin of the ambitwistor string lies in recent progress on the study of scattering amplitudes. In \cite{Cachazo:2014nsa,Cachazo:2013iea,Cachazo:2013hca} Cachazo, He and Yuan (CHY) proposed remarkably compact expressions for tree-level scattering amplitudes of gravity and Yang-Mills, the key ingredient of which are the scattering equations for $n$ momentum eigenstates with null momenta $k_i$
\begin{equation}\label{SE}
\sum_{j\neq i}\frac{k_i\cdot k_j}{z_i-z_j}=0,
\end{equation}
first found by Fairlie and Roberts \cite{Fairlie:1972zz} and later, in a very different context, by Gross and Mende \cite{Gross:1987ar}. The solutions of (\ref{SE}) determine $n$ marked points $z_i$ on a sphere or, in more suggestive language, they determine a point on the moduli space ${\cal M}_{n,0}$ of a $n$-punctured, genus zero, Riemann surface. Given this connection between  ${\cal M}_{n,0}$ and tree-level scattering, it would be odd if there was not some way of understanding these results in terms of a worldsheet theory. Indeed one might ask whether there is a worldsheet formulation of the theory that generates these compact expressions directly and naturally. The answer, to the best of our knowledge today, is a qualified yes; the qualification being that it is only the ten-dimensional Type II supergravity amplitudes that have been understood as critical string theories thus far. However, some understanding of the origin of other CHY amplitudes has been found. The ambitwistor string of \cite{Mason:2013sva} that describes type II supergravity has been generalised to other situations, but these other theories \cite{Ohmori:2015sha,Casali:2015vta,Geyer:2014fka} do not have the same status as the original ambitwistor string as they either are not in the critical dimension or do not have a sensible critical dimension. These other constructions are useful in understanding the CHY amplitudes but will not be studied here.

 The aim of this paper is to take the first steps in constructing a string field theory for perturbative classical Type II supergravity on flat spacetime. The basic ingredient is Type II ambitwistor string theory (\ref{action01}). Following the basic structure of covariant closed bosonic string theory \cite{Kugo:1989aa,Kugo:1989tk,Zwiebach:1992ie} and the proposed supersymmetric extension \cite{Sen:2015uaa}, we construct a superstring field theory for supergravity based on the ambitwistor string theory. The action will be of the form
\begin{eqnarray}\label{action0}
S[\Psi]=\langle \Psi |c_0Q|\Psi\rangle+\sum_{n>2} \frac{1}{n!} \{\Psi^n\},
\end{eqnarray}
where $Q$ is the BRST operator of the worldsheet theory, $c_0$ is a ghost zero mode, and $\{\Psi^n\}$ are $n$-point interaction terms for the string field $\Psi$.

A necessary step in the construction of the theory is to clarify the oscillator mode structure of the ambitwistor string theory and the constraints that must be imposed on the string fields. We shall see that the oscillator decomposition is subtly different from that of the conventional string. In particular, the $X^{\mu}$ and $P_{\mu}$ fields are independent in the gauge we work in and are composed of independent, conjugate oscillators. The supersymmetric theory is thought to be equivalent to Type II supergravity and so the superstring field theory is expected to be equivalent to perturbative Type II supergravity. In support of this we study the metric as a fluctuation $h_{\mu\nu}$ about a Minkowski background, we shall show that the quadratic term gives the correct linearised action for the Type II supergravity 
\begin{eqnarray}
\langle\Psi|c_0Q|\Psi\rangle&=&\int \rd^{10} x\left(
\frac{1}{4}h_{\mu\nu}\Box h^{\mu\nu}+\frac{1}{2}(\partial^{\nu}h_{\mu\nu})^2 +\frac{1}{2}h\partial^{\mu}\partial^{\nu}h_{\mu\nu}-\frac{1}{4}h\Box h\right.\nonumber\\
&&\left. -4\phi\Box \phi+2h\Box\phi-2\phi\partial^{\mu}\partial^{\nu}h_{\mu\nu}
-\frac{1}{12}H_{\mu\nu\lambda}H^{\mu\nu\lambda}
\right),\nonumber
\end{eqnarray}
We then argue that a proposed cubic interaction term is correct. Finally we consider the complete abstract string field theory to all orders. As we will show, the on-shell correlation functions implied by these interaction terms produce the correct on-shell scattering amplitudes and, once a gauge is fixed, the quadratic term produces a reasonable spacetime propagator. We comment on the application of this string field theory to curved backgrounds towards the end of the paper. The question of how the string field theory (\ref{action0}) might make contact with the Einstein-Hilbert action for a general spacetime, compactly written in terms of the Ricci scalar, will be discussed elsewhere.

During the course of this paper we shall see many ways in which the ambitwsitor string field theory mirrors the conventional string field theory superficially but differs in important, and often elegant, ways when studied in detail. It should be stressed from the outset that we are interested in a string field theory of \emph{classical} supergravity. As such, we do not consider loops. Though the theory is fully quantum mechanical on the worldsheet, it is classical in spacetime. That is not to say that the question of loops is not interesting \cite{Adamo:2013tsa,Adamo:2014wea,Geyer:2015jch}, just that it is not one we consider here and it would be interesting to see how the formalism presented here is extended to loops.

We have attempted to strike a balance between making the paper reasonably self-contained and keeping it to a reasonable length. As such, we have tried to sketch key ideas from ambitwistor string theory and string field theory that are necessary for our construction; however, we have omitted many of the technical details which may be followed up in the references given. In the next section we give a brief overview of classical ambitwistor string theory and present its quantisation in the operator formalism - the natural language of string field theory - paying particular care to those aspects that will be of importance for the construction of the string field theory. This section introduces most of the key ingredients that are needed to construct the bosonic ambitwsitor string field theory which is then presented in section \ref{Bosonic Ambitwistor String Field Theory}. Section \ref{The Supersymmetric theory} introduces the formalism and quadratic action for the supersymmetric ambitwistor theory and then, in section \ref{Interaction Terms}, we discuss the interaction terms in the supersymmetric theory. We present in this paper the first steps in a formalism that we feel has a rich structure and many potential directions of development. A number of directions for future work are discussed in section \ref{Discussion}.

\section{Ambitwistor String Theory}\label{Ambitwistor String Theory}

Ambitwistor space $\mathbb{A}$ is the space of null geodesics \cite{LeBrun,Isenberg:1978kk,Baston:1987av,LeBrun:1991jh}. This may be constructed simply as a sub-bundle of the cotangent bundle $T^*M$ of the spacetime $M$, which will be Minkowski spacetime for most of this paper. In most cases we shall be interested in the complexification of $M$ and $T^*M$ is the holomorphic tangent bundle. Natural coordinates on $T^*M$ are $x^{\mu}$ and $p_{\mu}$, where $x^{\mu}$ are coordinates on $M$. The null cotangent bundle $T_N^*M$ is then defined as
$$
T_N^*M=\{(x,p)\in T^*M|p^2=0\},
$$
where $p^2$ has been constructed using the metric on $M$. This is not quite the space of null lines since, given a point $x^{\mu}_0$ on a null line, the family of points $x^{\mu}_0+\alpha p^{\mu}$ all lie on the same null line for any constant $\alpha$. Shifts along the line are generated by the vector field
\begin{equation}\label{V}
\mathscr{V}=p^{\mu}\frac{\partial}{\partial x^{\mu}},
\end{equation}
and so ambitwistor space $\mathbb{A}$ is given by the quotient of $T_N^*M$ by the action of $\mathscr{V}$. The projective ambitwistor space $P\mathbb{A}$ is given by a further quotient of $\mathbb{A}$ by the action of the Euler vector field
$$
\Upsilon=p_{\mu}\frac{\partial}{\partial p_{\mu}}.
$$
This quotients out by the scale of $p_{\mu}$, giving $P\mathbb{A}$ as the space of scaled null geodesics in $M$. The ambitwistor string \cite{Mason:2013sva} is a sigma model describing the embedding of a worldsheet $\Sigma$ into ambitwistor space $\mathbb{A}$.  The map from $\Sigma$ to $T^*M$ is realised by elevating the coordinates $x^{\mu}$ and $p_{\mu}$ to (holomorphic) worldsheet fields $P_{\mu}(z)$ and $X^{\mu}(z)$. A simple Lagrangian on $T^*M$ is given by the $\beta\gamma$ system ${\cal L}=P_{\mu}\bar{\partial}X^{\mu}$, which is simply the chiral pull-back of the natural contact structure $\theta=p_{\mu} \rd x^{\mu}$ on $P\mathbb{A}$ to $\Sigma$. At the level of the worldsheet, the null constraint is imposed by introducing the Lagrange multiplier field $e(z)$, a Beltrami differential, giving the Lagrangian
$$
{\cal L}=P_{\mu}\bar{\partial}X^{\mu}+\frac{1}{2}eP^2.
$$
The symmetry associated to this constraint is equivalent, at the level of the worldsheet, to the quotient by the vector field $\mathscr{V}$.

The only outstanding issue at the classical level is that of the worldsheet metric or, equivalently, the worldsheet complex structure. This is not treated explicitly and is assumed fixed by the usual Faddeev-Popov technique, resulting in the introduction of a holomorphic $(b,c)$ ghosts system\footnote{One could argue that a half-twisting procedure along the lines discussed in \cite{Mason:2007zv,ReidEdwards:2012tq} allow for such a purely holomorphic construction to arise from a chiral topological twisting of a theory with a more conventional worldsheet gravity. Though interesting, this possibility will not be explored here and the action (\ref{action1}) will be taken as the definition of the theory.}. The bosonic  ambitwistor string action is taken to be
\begin{equation}\label{action1}
S=\int_{\Sigma}P_{\mu}\bar{\partial}X^{\mu}+ \frac{1}{2}eP^2+b\bar{\partial}c.
\end{equation}
Ideally, one would gauge fix $e(z)=0$ globally but, as discussed in \cite{Mason:2013sva,Adamo:2013tsa} and reviewed in section \ref{Symmetries}, this is not possible in general. The OPEs of the constituent fields are
\begin{equation}\label{OPE}
P_{\mu}(z)X^{\nu}(\omega)=\frac{\delta_{\mu}^{\nu}}{z-\omega}+...,	\qquad	b(z)c(\omega)=\frac{1}{z-\omega}+...,
\end{equation}
where the ellipsis denote terms that are non-singular in the $z\rightarrow \omega$ limit, with all other OPE's being trivial in the sense that they have no singular terms.

\subsection{Symmetries and quantization}\label{Symmetries and quantization}

Our ultimate goal is the construction of a covariant string field theory for the ambitwistor string. The main ingredients of the construction will be the BRST charge of the first quantised worldsheet theory and a translation of the first quantised theory into the language of the operator formalism. The main features of the operator formalism will be discussed in section \ref{The Operator Formalism for First Quantised Ambitwistor String Theory}. In this section we first review the BRST quantisation of \cite{Mason:2013sva,Adamo:2013tsa} and then lay the foundations for recasting the theory in the operator formalism of section \ref{The Operator Formalism for First Quantised Ambitwistor String Theory}.

\subsubsection{Symmetries}\label{Symmetries}

The ambitwistor string worldsheet fields transform under the (holomorphic) conformal transformations $z\rightarrow z+v(z)$ for which the fields transform as
\begin{equation}\label{conformal}
\delta(v) X^{\mu}=v\partial X^{\mu},	\qquad	\delta(v)P_{\mu}=\partial(v P_{\mu}),	\qquad	\delta(v) e=v\partial e-e\partial v.
\end{equation}
The conformal transformations are generated by the stress tensor $T(z)=P_{\mu}\partial X^{\mu}+T_{\text{gh}}$, where $T_{\text{gh}}$ are ghost contributions that will be described in more detail later. For a given vector field, $v(z)$ the transformation is generated by
$$
{\cal T}(v):=\oint\rd z\,v(z)T(z),
$$
so that the action on the field $\Phi(z)$ is $\delta(v)\Phi(z)=[{\cal T}(v,)\Phi(z)]$, where $\Phi(z)$ is a generic field of the worldsheet theory.

In addition to the conformal symmetry, a version of which exists for the conventional string, there is an additional gauge symmetry on the worldsheet that ensures the theory describes an embedding into ambitwistor space, rather than simply $T^*M$. The quotient by the vector field $\mathscr{V}$ in (\ref{V}) is achieved in the string theory by the gauge symmetry \cite{Mason:2013sva}
\begin{equation}\label{ambi}
\tilde{\delta}(v)X^{\mu}=v P^{\mu},	\qquad	\tilde{\delta}(v)P_{\mu}=0,	\qquad	\tilde{\delta}(v)e=\bar{\partial}v,
\end{equation}
where $v(z)$ is a $(1,0)$ worldsheet vector field. As commented upon in \cite{Mason:2013sva}, this symmetry has no counterpart in the conventional bosonic string and is a central feature of the ambitwistor string theory. This gauge symmetry is generated by ${\cal H}(v)$ where
$$
{\cal H}(v):=\oint\rd z\,v(z)H(z),	\qquad	H(z)=\frac{1}{2}P^2(z).
$$
As we shall see, $H(z)$ plays the role of a Hamiltonian in the ambitwistor theory. Indeed, the spacetime propagator to be discussed in section \ref{A brief sketch of perturbation theory} is effectively the inverse of the zero mode of $H(z)$.

Combining these transformations gives the classical algebra
$$
[{\cal T}(v_1),{\cal T}(v_2)]=-{\cal T}\big([v_1,v_2]\big),	\qquad	[{\cal T}(v_1),{\cal H}(v_2)]=-{\cal H}\big([v_1,v_2]\big),
$$
\begin{equation}\label{algebra}
	[{\cal H}(v_1),{\cal H}(v_2)]=0,
\end{equation}
where the commutator of the worldsheet vector fields takes the standard form $[v_1,v_2]=v_1\partial v_2-v_2\partial v_1$. We may think of this algebra acting naturally on the space ${\cal Y}\rightarrow \Sigma$, a bundle over the wordsheet. The abelian gauge symmetry generated by $H(z)$ acts on the fibres of ${\cal Y}$ and the conventional conformal symmetry generated by $T(z)$ acts on the base $\Sigma$.

The (holomorphic) worldsheet diffeomorphisms have been gauge-fixed in the usual way with the introduction of a $(b,c)$ ghost system and the additional gauge transformations (\ref{ambi}) are fixed by the usual Faddeev-Popov method, introducing ghosts $\tilde{b}$ and $\tilde{c}$. The constraints $T(z)=0$ and $H(z)=0$ are imposed in the standard way by introducing the BRST charge
\begin{equation}\label{Q}
Q=\oint\rd z\,j(z),
\end{equation}
with current
$$
j(z)=c(z)\left(T(z)+\widetilde{T}_{\text{gh}}(z)+\frac{1}{2}T_{\text{gh}}(z)\right)+\tilde{c}(z)H(z)
$$
where $T_{\text{gh}}$ and $\widetilde{T}_{\text{gh}}$ are stress tensors for the $(b,c)$ ghosts and  the $(\tilde{b},\tilde{c})$ ghosts respectively\footnote{For a standard $\beta\gamma$ system with $\beta$ of weight $\lambda$, the stress tensors take the conventional form $T_{\lambda}=(\partial\beta)\gamma-\lambda\partial(\beta\gamma)$. 
The stress tensor $T(z)$ for the action (\ref{S}) then has matter and ghost contributions $T(z)+T_{\text{gh}}(z)+\widetilde{T}_{\text{gh}}(z)$, where
$$
T(z)=P_{\mu}\partial X^{\mu},	\qquad	T_{\text{gh}}(z)=(\partial b)c-2\partial(bc),	\qquad	\widetilde{T}_{\text{gh}}(z)=(\partial \tilde{b})\tilde{c}-2\partial(\tilde{b}\tilde{c}).
$$
}. The origin of the ghost terms in the action will be of central importance later on so we pause here to repeat the arguments of \cite{Adamo:2013tsa}, which discuss the gauge-fixing of the action. The presentation closely follows that of \cite{Adamo:2013tsa} where further details may be found.

The BRST operator acts within a given Dolbeault cohomology class and we cannot set $e(z)=0$ globally. The best we can do is to set
$$
e(z)=\sum_as^a\mu^a(z),
$$
where $\{\mu_a\}$ is a basis of Beltrami differentials for $\Sigma$, where $a=1,2,...,n-3$. This is done by introducing the gauge-fixing fermion $F(e)$ and extending the action to
$$
\widehat{S}=\int_{\Sigma}P_{\mu}\bar{\partial}X^{\mu}+b\bar{\partial}c+Q\tilde{b}F(e).
$$
A useful choice is
$$
F(e)=e-\sum_{a=1}^{n-3}s_a\mu^a,
$$
where $\{\mu^a\}$ is a basis for $H^{0,1}(\Sigma,T_{\Sigma}(-z_1-...-z_n))$, the $z_i$ are points on $\Sigma$, and where the gauge transformation generated by $H(z)$ vanishes. The action of $Q$ on the fields is $Q \tilde{b}=\pi$,	 $Q e=\bar{\partial}\tilde{c}$, and $Q s^a=q^a$ and so
$$
Q\int_{\Sigma}\tilde{b}F(e)=\int_{\Sigma}\pi F(e)+\int_{\Sigma} \tilde{b}\bar{\partial}\tilde{c}-\sum_{a=1}^{n-3}q_a\int_{\Sigma}\tilde{b}\mu^a.
$$
Integrating out the Lagrange multiplier $\pi$ sets $F(e)=0$ and so the action is
$$
\widehat{S}=S -\frac{1}{2}\sum_{a=1}^{n-3}s_a\int_{\Sigma}\mu^aP^2 -\sum_{a=1}^{n-3}\int_{\Sigma}q_a\tilde{b}\mu^a,
$$
where
\begin{equation}\label{S}
S=\int_{\Sigma}\left(P_{\mu}\bar{\partial}X^{\mu}+b\bar{\partial}c+ \tilde{b}\bar{\partial}\tilde{c}\right).
\end{equation}
Integrating out the auxiliary fields $s_a$ and $q_a$ leads to an insertion of
\begin{equation}\label{ghost}
\prod_{a=1}^{n-3}\;\bar{\delta}\left(\int_{\Sigma}\mu^a(z)H(z)\right)\int_{\Sigma}\mu^a(z)\tilde{b}(z)\int_{\Sigma}\mu^a (z)b(z),
\end{equation}
into the path integral, where $a$ indicates the modulus associated with the deformation of the worldsheet moduli corresponding to a particular Beltrami differential. An alternative perspective on the origin of these delta-function insertions will be reviewed in section \ref{new}. It will turn out that this alternative viewpoint is more useful in studying the string field theory.

\subsubsection{Operator Quantization}
\label{Operator Quantization}

In this section we shall assume the gauge has been fixed as described above and $\Sigma$ is a genus zero Riemann surface with $n$ punctures.  Let us consider the case of a single puncture to begin with. In the conventional string, with equations of motion $\Box X^{\mu}=J^{\mu}$, where $J^{\mu}$ is some source, possibly due to a vertex operator inserted at a puncture, the natural oscillator expansion
includes the zero mode contributions $X^{\mu}=x^{\mu}+p^{\mu}\ln(t)+...$, where the ellipsis denote oscillator modes and we can think of $t$ as a local coordinate around the puncture. The puncture may be thought of as residing in the infinite past in worldsheet time and the relation between the operator inserted at $t=0$ and the state is given by the usual state-operator correspondence. The centre of mass momentum $p_{\mu}$ appears as one of two zero modes in the expansion for $X^{\mu}$. 

By contrast, in the ambitwistor string, we incorporate the momentum zero mode $p_{\mu}$ into a mode expansion for $P_{\mu}(z)$ and take the independent $X^{\mu}(z)$ mode expansion as
$$
X^{\mu}(z)=x^{\mu}-\sum_{n\neq 0}\frac{\tilde{\alpha}_n^{\mu}}{n}z^{-n},
$$
where we define the zero mode as $x^{\mu}\equiv \tilde{\alpha}^{\mu}_0$. In contrast to the conventional string, $X^{\mu}$ is a conformal field (of weight zero). As noted in \cite{Mason:2013sva}, this fact restricts the allowed vertex operators to a massless sector. Note that the absence of a logarithmic term means that
$$
\partial X^{\mu}(z)=\sum_{n\neq 0}\tilde{\alpha}_n^{\mu}z^{-n-1},
$$
does not have a zero mode\footnote{The zero mode $x^{\mu}$ does not appear in $\partial X^{\mu}(z)$.}. The conjugate field $P_{\mu}(z)$ is a conformal field of weight one and has the conventional expansion
$$
P_{\mu}(z)=\sum_n\alpha_{n\mu}z^{-n-1},
$$
where $p_{\mu}:= \alpha_{0\mu}$ is the momentum zero mode. We impose the commutation relations\footnote{One might like to think of these as `equal $\bar{z}$' commutation relations. Note also that the commutator does not depend on the spacetime metric.}
\begin{equation}\label{commutator}
[P_{\mu}(\sigma),X^{\nu}(\sigma')]=-i\delta^{\nu}_{\mu}\delta(\sigma-\sigma'),	\qquad		[P_{\mu}(\sigma),P_{\nu}(\sigma')]=0,	\qquad	[X^{\mu}(\sigma),X^{\nu}(\sigma')]=0,
\end{equation}
where $z=e^{i\sigma}$ and $z'=e^{i\sigma'}$. The commutation relations (\ref{commutator}) are satisfied if the mode operators satisfy the commutation relations
$$
[\alpha_{n\mu},\tilde{\alpha}_m^{\nu}]=-in\delta_{\mu}^{\nu}\delta_{n+m,0},	\qquad	[\alpha_{n\mu},\alpha_{m\nu}]=0,	\qquad	[\tilde{\alpha}_n^{\mu},\tilde{\alpha}_m^{\nu}]=0,
$$
for $n \neq 0$ and
$$
[\alpha_{0\mu},\tilde{\alpha}_0^{\nu}]=[p_{\mu},x^{\nu}]=-i\delta_{\mu}^{\nu},
$$
when $n=0$. We quantise on the vacuum $|0\rangle$ defined by\footnote{For a discussion of this and an alternative choice of vacuum see \cite{Casali:2016atr}.}
$$
\alpha_n|0\rangle=0,	\quad n\geq 0,\qquad \text{and}	\qquad
\tilde{\alpha}_n|0\rangle=0, \quad n>0.
$$
Notice that we do not require that the $X^{\mu}$ zero mode annihilates the vacuum. It is not hard to show that
$$
\left\langle X^{\mu}(z)P_{\nu}(w)\right\rangle=\frac{\delta^{\mu}_{\nu}}{z-w},
$$
as we would expect.

If we have more than one puncture it is natural to define local coordinates $t_i$ in a small disc about each puncture and employ the same oscillator expansions as above in terms of the local $t_i$ coordinate for each of the punctures. Conformal maps $h_i:t_i\rightarrow z$ may then be used to describe the  expressions in terms of a coordinate $z$ on the complex plane, such that the location of the punctures in the new coordinates is given by $z_i=h_i(0)$ - the origin of the local coordinate system. The oscillator expansions of the worldsheet fields will in general take on a more complicated form when written in the $z$ coordinates. For example, a simplistic (and somewhat naive \cite{DiVecchia:1986mb}) map would be $t_i=z-z_i$.

Let us consider this in more detail. We first consider the situation for the conventional string. Following \cite{DiVecchia:1986mb}, we can require the worldsheet punctures at $z_i$ to coincide with $n$ asymptotic states at points in spacetime $x_i$ by inserting $\prod_{i=1}^n\delta^D(X(z_i)-x_i)$ into the path integral. Taking the Fourier transform to momentum space results in the insertion of the distribution $J=\sum_{i=1}^nk_i\delta^2(z-z_i)$, familiar from calculations of Tachyon scattering amplitudes. This gives a source for the classical fields which obey the equation of motion
$$
\Box X_{\text{cl}}^{\mu}=J^{\mu}.
$$
Thus the field may be written as $X^{\mu}=X_{\text{cl}}^{\mu}+X_{\text{q}}^{\mu}$, where $X_{\text{q}}^{\mu}$ is a quantum fluctuation and $X_{\text{cl}}^{\mu}$ is the classical solution, at genus zero, given by
$$
X_{\text{cl}}^{\mu}(z)=\sum_{i=1}^nk^{\mu}_i\ln|z-z_i|^2.
$$
It is then natural to write $X^{\mu}$ as a sum $X^{\mu}=\sum_{i=1}^nX_i^{\mu}$, where $X^i$ is written in terms of the Hilbert space defined at the $i$'th puncture. For the ambitwistor string the punctures amount to inserting the current $J=\sum_{i=1}^nk_i\bar{\delta}(z-z_i)$ and the $P(z)$ equation of motion $\bar{\partial}P_{\text{cl}}=J$ has classical solution \cite{DiVecchia:1986mb}
\begin{equation}\label{P}
P_{\text{cl}}(z)=\sum_{i=1}^n\frac{k_i}{z-z_i}.
\end{equation}
It is then natural to expand $P(z)$ as $P=\sum_{i=1}^nP_i$, where each $P_i$ is written, by a conformal transformation, as an oscillator expansion using the $i$'th Hilbert space and local coordinates $t_i$ at the $i$'th puncture. We see that the naive choice $t_i=z-z_i$ gives rise to the expression (\ref{P}) for the zero modes contribution. This will be discussed at greater  length in section \ref{Scattering Amplitudes and the Scattering Equations} and more details of the general construction may be found in \cite{Reid-Edwards:2015stz} for the ambitwistor string and \cite{DiVecchia:1986mb,AlvarezGaume:1988bg,Vafa:1987es,LeClair:1988sp} for the conventional string. In the presence of more than one puncture the commutator relations generalise in the obvious way
$$
[\alpha^{(i)}_{n\mu},\tilde{\alpha}_m^{(j)\nu}]=-in\delta^{ij}\delta_{\mu}^{\nu}\delta_{n+m,0},	\qquad	[\alpha^{(i)}_{n\mu},\alpha^{(j)}_{m\nu}]=0,	\qquad	[\tilde{\alpha}_n^{(i)\mu},\tilde{\alpha}_m^{(j)\nu}]=0,
$$
for $m,n \neq 0$ and
$$
[\alpha^{(i)}_{0\mu},\tilde{\alpha}_0^{(j)\nu}]=[p^{(i)}_{\mu},x^{(j)\nu}]=-i\delta^{ij}\delta_{\mu}^{\nu},
$$
if $m=n=0$. We note that the commutation relations do not depend on the background spacetime metric.

The association of a Hilbert space with each puncture also provides a helpful way of writing the ghost insertions (\ref{ghost}). It is useful to define a disc $\mathscr{D}_i$ about each puncture  given, in terms of the local coordinates $t_i$, as the region $|t_i|<1$. The Beltrami differential encodes changes in the moduli of the Riemann surface $\Sigma$ which may be also understood in terms of deforming the worldsheet in the region of a puncture. In the region $|t_i|<1+\epsilon$ for some small $\epsilon$, let the coordinates be changed to $t_i'$. On the region $|t_i|>1-\epsilon$ there is a patch with coordinate $t_i$. On the overlap given by the annulus of width $2\epsilon$ which contains the boundary $\partial\mathscr{D}_i$, the coordinates are related by $t_i'=t_i+v_i(t)$. In this overlap the Beltrami differentials may be written as $\mu_i=\bar{\partial}v_i$, leading to an alternative description of the ghost insertions. For example, associating the $i$'th puncture with an excised disc $\mathscr{D}_i$ such that $\partial\Sigma=\cup_{i=1}^n\partial\mathscr{D}_i$, gives
$$
\int_{\Sigma}\mu^a (z)b(z)=\sum_{i=1}^n\oint_{\partial\mathscr{D}_i}\rd z_i\,v^a_i(z_i)b^{(i)}(z),	
$$
where the $b^{(i)}$ are the ghost modes associated with the Hilbert space at the $i$'th puncture. The ghost insertion term (\ref{ghost}) may then be written as
$$
\prod_{a=1}^{n-3}\bar{\delta}\Big({\cal H}(\vec{\nu}^a)\Big)\tilde{\mathbf{b}}(\vec{\nu}^a)\mathbf{b}(\vec{\nu}^a),
$$
where
$$
\mathbf{b}(\vec{\nu}^a)= \sum_{i=1}^n\oint_{\partial\mathscr{D}_i}\rd z\,v^a_i(z)b^{(i)}(z),	\qquad	\tilde{\mathbf{b}}(\vec{\nu}^a)=\sum_{i=1}^n \oint_{\partial\mathscr{D}_i}\rd z\,v^a_i(z)\tilde{b}^{(i)}(z),	
$$
\begin{equation}\label{H}
{\cal H}(\vec{\nu}^a)=\sum_{i=1}^n\oint_{\partial\mathscr{D}_i}\rd z\,v^a_i(z)H^{(i)}(z).
\end{equation}
The integral is taken over a contour $\partial\mathscr{D}_i$ surrounding the disc\footnote{The discs are chosen so that they do not overlap and each disc contains only one puncture.} $\mathscr{D}_i$ which has the point $z_i$ at its centre. $H^{(i)}(z)$ and $\tilde{b}^{(i)}(z)$, like $b^{(i)}(z)$, are defined in the Hilbert space at the $i$'th puncture. These are precisely the insertions we will see in the interaction terms of the string field action. The notation $\vec{\nu}^a$ indicates the $n$ vector fields, located at each of the punctures $\vec{\nu}^a=\big(v_1^a,v_2^a,...,v^a_n\big)$.

\subsubsection{The extended Virasoro algebra}\label{The extended Virasoro algebra}

The stress tensor $T(z)=P_{\mu}\partial X^{\mu}$ and the coefficients of its mode expansion are related by
$$
T(z)=\sum_n L_n z^{-n-2},	\qquad	L_n=\oint\rd z z^{n+1}T(z).
$$
Explicitly, the stress tensor components are
$$
L_0=\frac{1}{2}\sum_{m> 0}(\alpha_{-m}\cdot\tilde{\alpha}_m+\tilde{\alpha}_{-m}\cdot\alpha_m),\qquad	L_n=\sum_{m\neq n}\tilde{\alpha}_{n-m}\cdot\alpha_m.
$$
Note that the dot denotes a Lorentz index contraction $\alpha\cdot\tilde{\alpha}:=\alpha_{\mu}\tilde{\alpha}^{\mu}$ and so the generators are independent of the background spacetime metric. Note that $\tilde{\alpha}_0$ does not appear in the expressions for the $L_n$ as it does not appear in the mode expansion of $\partial X^{\mu}$. The additional gauge symmetry generated by $H(z)$ which we expand as
$$
H(z)=\sum_{n}\widetilde{L}_n z^{-n-2}.
$$
The $\widetilde{L}_n$ modes may be written in terms of the $\alpha_{\mu}$ modes as
$$
\widetilde{L}_n=\frac{1}{2}\eta^{\mu\nu}\sum_{m}\alpha_{m \mu}\alpha_{n-m \nu},
$$
where all values of $m$ are summed over and there is no normal ordering ambiguity since the $\alpha_n$ all commute with each other. This is the expansion of $H(z)$ for flat backgrounds. For curved backgrounds the appropriate metric must be used in place of $\eta^{\mu\nu}$. After straightforward computation we find
$$
[L_m,L_n]=(m-n)L_{m+n}+\delta_{m+n,0}\frac{D}{6}m(m^2-1),	\qquad	[L_m,\widetilde{L}_n]=(m-n)\widetilde{L}_{m+n},	\qquad	[\widetilde{L}_m,\widetilde{L}_n]=0,
$$ 
where $D$ is the dimension of the spacetime; $\mu=1,2,...,D$. This should be compared with the $\delta_{m+n,0}\frac{D}{12}m(m^2-1)$ anomaly in the conventional bosonic string. The central charge contribution to a free $\beta\gamma$ system is
$$
c=\mp3(2\lambda-1)^2\pm1,
$$
where the upper (lower) sign is taken for fermions (bosons) and $\lambda$ is the conformal weight of the highest weight field. The $(b,c)$ and $(\tilde{b},\tilde{c})$ ghost systems, each with $\lambda=2$, each contribute $-26$ to the central charge and so the critical dimension\footnote{As in the conventional string, $D$ arises in the computation of the algebra from the trace of the spacetime metric $\eta_{\mu\nu}$. As such it computes the number of independent $(P,X)$ systems that are introduced. In this paper we take the spacetime to be complexified, so $D$ counts the complex dimension and we only make use of the holomorphic coordinates (the $\bar{X}^{\mu}$ play no role).} is $D=26$. This is in accordance with the central charge bookkeeping, given that a single $(X,P)$ system, where the conformal weight of $P$ is $+1$ contributes $c=2$ to the central charge, thus the total central charge is $c=2D-26-26$ which vanishes in the critical dimension.

\subsubsection{BRST Quantisation}\label{BRST Quantisation}

The constraints $T(z)=0$ and $H(z)=0$ are imposed in the standard way by introducing the BRST charge (\ref{Q}). If we have $n$ punctures, each with an associated Hilbert space, it is helpful to consider a BRST charge $Q^{(i)}$ constructed using the fields of the $i$'th Hilbert space. The total BRST charge is then $Q=\sum_{i=1}^nQ^{(i)}$. For now, we shall consider a single Hilbert space. The ghosts appearing in $Q$ have the standard expansions
$$
c(z)=\sum_nc_nz^{-n+1},	\qquad	b(z)=\sum_nb_nz^{-n-2}
$$
and similarly for the $\tilde{b}$ and $\tilde{c}$ ghosts. In terms of these oscillator components, the BRST charge may be written as
$$
Q=\sum_nc_{-n}\left(L^{(m)}_n+L_n^{(g)}+\widetilde{L}^{(g)}_n\right)+\sum_n\tilde{c}_{-n}\widetilde{L}^{(m)}_n
$$
where, to avoid confusion, we have now denoted the matter contribution to the Virasoro algebra discussed in section \ref{The extended Virasoro algebra} above by $L_n^{(m)}$ to distinguish them from the ghost modes
$$
L_n^{(g)}=\sum_m(n-m):b_{n+m}c_{-m}:-\delta_{n,0},
$$
and similarly for $\widetilde{L}^{(g)}_n$. The condition that the physical states of the string are massless means that the higher oscillator modes do not play a direct role and we may concentrate on the lower order modes. To leading order, the BRST operator terms give
$$
\sum_nc_{-n}L_n^{(m)}=c_0(\alpha_{-1}\cdot\tilde{\alpha}_1+\tilde{\alpha}_{-1}\cdot\alpha_1)+\alpha_0\cdot(c_1\tilde{\alpha}_{-1}+c_{-1}\tilde{\alpha}_1)+...
$$
and
$$
\sum_n\tilde{c}_{-n}\widetilde{L}^{(m)}_n=\frac{1}{2}\tilde{c}_0\alpha^2_0+\frac{1}{2}\tilde{c}_0\alpha_{-1}\cdot\alpha_1+\alpha_0\cdot(\tilde{c}_{-1}\alpha_1+\tilde{c}_1\alpha_{-1})+...
$$
Notice that the expression involving $H(z)$ depends on the background metric, whereas that involving $T(z)$ does not. This observation will be important when we come to study the string field in section \ref{Bosonic Ambitwistor String Field Theory}; the $T(z)=0$ constraints are imposed as part of the definition of the string field. Such a definition cannot be subject to perturbative changes in the background; whereas, the $H(z)=0$ condition is imposed at the level of the equations of motion which \emph{must} be modified by the interaction terms which encode the effect of the background fields.

The ghost contributions to the BRST charge take the conventional form. It will be useful to isolate those terms in $Q$ which carry a factor of $c_0$ and to write $Q$ as
 \begin{eqnarray}\label{Q2}
 Q&=&c_0{\cal L}_0+\frac{1}{2}\tilde{c}_0\alpha_0^2+\frac{1}{2}\tilde{c}_0\alpha_{-1}\cdot\alpha_1+\alpha_0\cdot(c_1\tilde{\alpha}_{-1}+c_{-1}\tilde{\alpha}_1+\tilde{c}_{-1}\alpha_1+\tilde{c}_1\alpha_{-1})\nonumber\\
 &&-2b_0c_{-1}c_1+2\tilde{b}_0(c_1\tilde{c}_{-1}+\tilde{c}_1c_{-1})+\tilde{c}_0(c_{-1}\tilde{b}_1+c_1\tilde{b}_{-1})+...
 \end{eqnarray}
 where the terms multiplying $c_0$ are given by
 \begin{equation}\label{L}
 {\cal L}_0=(\alpha_{-1}\cdot\tilde{\alpha}_1+\tilde{\alpha}_{-1}\cdot{\alpha_1})+(b_{-1}c_1+c_{-1}b_1-1)+(\tilde{b}_{-1}\tilde{c}_1+\tilde{c}_{-1}\tilde{b}_1-1)+....
 \end{equation}
and the ellipsis denote higher mode terms that, as we shall see later, do not play a role. Note that, neglecting ghosts, this may be written as
$$
{\cal L}_0=L_0-2+...
$$
This isolation of the $c_0{\cal L}_0$ term in $Q$ plays a central role in the construction of the ambitwistor string field action. We shall require that the string field $|\Psi\rangle$ satisfy the spacetime metric-independent constraint ${\cal L}_0|\Psi\rangle=0$ as part of its definition. The remaining constraints encoded in $Q$ are given by the action for the string field through the target space equations of motion and gauge invariances. We shall discuss these issues in detail in section \ref{The action to quadratic order}. The BRST operator is the key ingredient in the quadratic string field action. In order to be able to construct a complete non-linear action for the string field we need an operator description of the ambitwistor string interactions. This is the subject of section \ref{The Operator Formalism for First Quantised Ambitwistor String Theory}.

\subsection{The Operator Formalism for the First Quantised Ambitwistor String Theory}\label{The Operator Formalism for First Quantised Ambitwistor String Theory}

The operator formalism for ambitwistor strings was investigated in \cite{Reid-Edwards:2015stz}. Here we review and extend those results. The central idea of the operator formalism \cite{AlvarezGaume:1988bg,Vafa:1987es} is to express the $n$-punctured genus $g$ worldsheet in terms of a state $\langle\Sigma|$, called the surface state, which may be thought of as a map from the n-fold product of Hilbert spaces $\otimes_{i=1}^n{\cal H}_i$ to $\C$ such that,  if we associate each of the ${\cal H}_i$ with a puncture on the surface and contract with an asymptotic state $|V_i\rangle\in{\cal H}_i$, the resulting function
\begin{equation}\label{form10}
\langle\Sigma|B(\vec{\nu})|V_i\rangle...|V_n\rangle,
\end{equation}
integrated over an appropriate space $\Gamma_n$ is the scattering amplitude for these states, where $B(\vec{\nu})$ denotes appropriate ghost insertions described below. In the case of the conventional string the space $\Gamma_n$ is the moduli space ${\cal M}_{n,g}$ of Riemann surfaces. An appropriate candidate for $\Gamma_n$ for the ambitwistor string was discussed in \cite{Ohmori:2015sha} and will be considered in the context of ambitwistor string field theory in section \ref{Ambitwistor interactions as forms}. For now, we shall formally take\footnote{In \cite{Ohmori:2015sha} it was argued that (\ref{form10}) should be thought of as a top holomorphic form on the $2(2n-6)$ dimensional space $T^*{\cal M}_n$. A Morse theory argument was used to select a $2n-6$ dimensional cycle $\Gamma_n$ over which to integrate (\ref{form10}). It was also shown that, via a localisation argument, this form could be simplified and the amplitude could be formally written in terms of an integral over the $2n-6$ dimensional space ${\cal M}_n$.} $\Gamma_n$ to be ${\cal M}_{n,0}$ in accordance with \cite{Mason:2013sva}. For the rest of this section we shall focus on the case of ambitwistor string theory and, since our discussion will be limited to classical supergravity, we shall restrict attention to the $g=0$ case. This means that, although the theory is quantised at the level of the worldsheet, we are only considering classical spacetime physics. For on-shell states, the connection with the vertex operators $V(t)$ is simply
\begin{equation}\label{limit}
|V\rangle=\lim_{t\rightarrow 0}V(t)|0\rangle,
\end{equation}
where $t$ is a local coordinate that vanishes at the puncture. For example, the massless symmetric state with vertex operator\footnote{We shall only consider unintegrated vertex operators in this paper as we want to be able to easily extend the formalism to include Ramond states, for which no meaningful notion of an integrated vertex operator exists.}
\begin{equation}\label{vertex}
V(z)= c(z)\tilde{c}(z)\varepsilon^{\mu\nu}P_{\mu}(z)P_{\nu}(z)e^{ik\cdot X(z)}
\end{equation}
corresponds to the state
\begin{equation}\label{state}
|V\rangle=c_1\tilde{c}_1\varepsilon_{\mu\nu}\alpha^{\mu}_{-1}\alpha^{\nu}_{-1} |k\rangle,
\end{equation}
where the zero mode momentum eigenstate is $|k\rangle=e^{ik\cdot x}|0\rangle$ and $\varepsilon_{\mu\nu}$ is a polarisation tensor. The $n$-point scattering amplitude is then
$$
M_n=\int_{{\cal M}_n}\langle\Sigma|B(\vec{\nu})|V_1\rangle...|V_n\rangle.
$$
Notice that all dependence on the location $t$ of the operator insertion has been lost in the limit in (\ref{limit}). In the operator formalism the location of the vertex insertion is no longer encoded in the states $|V_i\rangle$. That information is described by the surface state $\langle\Sigma|$, which we turn to next.

\subsubsection{The Surface State}\label{The Surface State}
 
The surface state $\langle\Sigma|$ is the crucial ingredient in the operator description of conventional string theory \cite{AlvarezGaume:1988bg,Vafa:1987es}. It encodes the information of the conformal field theory (CFT) on a genus $g$ Riemann surface with $n$ punctures. Recalling the discussion in section \ref{Operator Quantization}, we introduce local coordinates $t_i$ around the $i$'th puncture, with respect to which, the fields have the standard oscillator expansion (with the puncture located at $t_i=0$). It is useful to define the conformal map $h_i$ from a neighbourhood of the $i$'th puncture to the complex plane with coordinate $z$, given by $z=h_i(t_i)$. The location of the puncture is then $z_i=h_i(0)$. In addition, $\langle \Sigma|$ encodes a choice of the local coordinates $t_i$ around each of the punctures. The data associated with string states at those punctures is encoded in states $|\Psi_i\rangle$ in the $i$'th Hilbert space ${\cal H}_i$ which are then contracted with the surface state to give a CFT correlation function. As such, the surface state may be thought of as a map from $\otimes_i{\cal H}_i$ to $\C$. The extension to off-shell correlation functions is straightforward.

 The surface state $\langle\Sigma|$ for ambitwistor string theory was explored in \cite{Reid-Edwards:2015stz}. We summarise and extend the results here. The surface state may be written as
\begin{equation}\label{Surface}
\langle \Sigma|=\int \rd^np\;\langle \vec{p}_n|\;\delta\left(\sum p_{(i)}\right)\;e^W \;{\cal Z},
\end{equation}
where the integral is over all external momenta\footnote{$\rd^n p= \prod_{i=1}^n\rd p_{(i)}$} and the delta function imposes overall momentum conservation. The exponent is a sum of matter and ghost contributions $W=V_{X,P}+V_{\text{gh}}$, which may be written schematically as
$$
V_{X,P}(z_1,...z_n)=\sum_{i,j}\oint_0\rd t_i\oint_0\rd t_j\frac{X(t_i)\cdot P(t_j)}{h_i(t_i)-h_j({t_j})},
$$
\begin{equation}\label{V}
V_{\text{gh}}(z_1,...z_n)= \sum_{i,j}\oint_0\rd t_i\oint_0\rd t_j\frac{b(t_i)c(t_j)}{h_i(t_i)-h_j({t_j})}+ \sum_{i,j}\oint_0\rd t_i\oint_0\rd t_j\frac{\tilde{b}(t_i)\tilde{c}(t_j)}{h_i(t_i)-h_j({t_j})},
\end{equation}
and $\langle p_{(i)};3|$ is shorthand for the $SL(2;\C)$-invariant vacuum $\langle p_{(i)};3|\equiv \langle p_{(i)}|\otimes\langle 3_i|\otimes\langle\tilde{3}_i|$. The ghost vacua have the standard normalisation and are given by $\langle 3|=\langle 0|c_{-1}c_0c_1$, with a similar expression for $\langle \tilde{3}|$ involving the $\tilde{c}$ ghosts and are normalised as $\langle 3|0\rangle=1$ and similarly for $\langle \tilde{3}|$. In (\ref{Surface}) we have adopted the shorthand notation $\langle \vec{p}_n|:=\langle p_{(1)};3| ...\langle p_{(n)};3|\;$. The integrals are taken around the location of the puncture given in terms of local coordinates $t_i$ around the $i$'th puncture (located at $t_i=0$). The expression for the surface state $(\ref{Surface})$ also includes the object ${\cal Z}$, defined as
$$
{\cal Z}=\prod_{r=-1}^{+1}Z_r \prod_{r=-1}^{+1}\widetilde{Z}_r,
$$
where $Z_r$ is given by
\begin{equation}\label{H}
Z_r=\sum_{i=1}^n\sum_{m=-1}^{\infty}{\cal M}_{r m}(z_i)\,b_m^i.
\end{equation}
The coefficients ${\cal M}_{nm}(z_i)$ are
$$
{\cal M}_{nm}(z_i)=\oint_{t_i=0} \frac{\rd t_i}{2\pi i}\;t_i^{-m-2}\Big(h_i'(t)\Big)^{-1}\Big(h_i(t)\Big)^{n+1}.
$$
There are similar expressions defining $\widetilde{Z}$ with $b^{(i)}$ replaced by $\tilde{b}^{(i)}$. These expressions may be derived from the path integral for the ghost zero modes. The net effect of these contributions is to remove the $c_1\tilde{c}_1$ factors in three of the asymptotic string states, effectively dividing out by an $SL(2;\R)$ factor for each of the products in ${\cal Z}$. This is discussed in more detail in \cite{Reid-Edwards:2015stz}.

It is often useful to write these expressions in terms of the oscillator modes
\begin{equation}\label{V2}
V_{X,P}=\sum_{m,n\geq 0}\sum_{i,j} {\cal S}^{mn}(z_i,z_j)\tilde{\alpha}_m^{(i)}\cdot \alpha^{(j)}_n,
\end{equation}
and for the ghosts
$$
V_{\text{gh}}=\sum_{i,j}\sum_{\substack{n\geq 2\\ m\geq -1}}{\cal K}_{nm}(z_i,z_j)\,c_n^{(i)}\,b_m^{(j)}+ \sum_{i,j}\sum_{\substack{n\geq 2\\ m\geq -1}} {\cal K}_{nm}(z_i,z_j)\,\tilde{c}_n^{(i)}\,\tilde{b}_m^{(j)}. 
$$
The functions ${\cal S}$ is given by
\begin{equation}\label{S}
 {\cal S}_{mn}(z_i,z_j)=\oint\frac{\rd t_i}{2\pi i}\oint\frac{\rd t_j}{2\pi i}h'_i(t_i)t^{-m}_it_j^{-n-1}\frac{1}{h_i(t_i)-h_j(t_j)}.
 \end{equation}
The corresponding functions for the ghosts are given by
$$
{\cal K}_{nm}(z_i,z_j)= -\oint\frac{\rd t_i}{2\pi i}\oint\frac{\rd t_j}{2\pi i}\; t_i^{-n+1}t_j^{-m-2} \;\Big(h_i'(t_i)\Big)^2 \Big(h_j'(t_j)\Big)^{-1}\;\frac{1}{h_i(t_i)-h_j(t_j)}.
$$
Details of the derivations of these expressions may be found in \cite{Reid-Edwards:2015stz} and a brief overview of one approach, as applied to the more familiar case of the conventional string, is given in Appendix \ref{Conformal maps and vertex functions}

\subsubsection{Scattering Amplitudes and the Scattering Equations}\label{Scattering Amplitudes and the Scattering Equations}

Given the surface state $\langle\Sigma|$, and on-shell\footnote{i.e. BRST-invariant states.} states $|\Psi_i\rangle\in{\cal H}_i$ in the $i$'th Hilbert space, we can construct a top form on the holomorphic cotangent bundle of the moduli space $T^*{\cal M}_n$ as
\begin{equation}\label{form1}
\Omega_{|\vec{V}\rangle}(\vec{v})=\langle\Sigma|B_{n-3}(\vec{v})|\vec{V}\rangle,
\end{equation}
where
\begin{equation}\label{B}
B_{n-3}(\vec{v})= \prod_{a=1}^{n-3}\tilde{\mathbf{b}}(\vec{\nu}_a)
 \prod_{a=1}^{n-3}\mathbf{b}(\vec{\nu}_a)\prod_{a=1}^{n-3}\bar{\delta}\Big({\cal H}(\vec{\nu}_a)\Big),
\end{equation}
with $\mathbf{b}(\vec{\nu}_a)$, $\tilde{\mathbf{b}}(\vec{\nu}_a)$, and ${\cal H}(\vec{\nu}_a)$ as defined in (\ref{H}). $|\vec{V}\rangle$ is short hand for the tensor product of asymptotic states $|V_1\rangle\otimes...\otimes|V_n\rangle$. The forms (\ref{form1}) are motivated by similar constructions in \cite{AlvarezGaume:1988bg,Ohmori:2015sha}. We shall argue in section \ref{Ambitwistor interactions as forms} that we may actually formally evaluate this integral over moduli space ${\cal M}_n$. Taking the $|V\rangle$ to be on-shell states (\ref{state}) corresponding, via the state-operator correspondence, to vertex operators (\ref{vertex}). The on-shell scattering amplitude is given by integrating over the moduli space ${\cal M}_n$
\begin{equation}\label{amp}
\langle V(z_1),...,V(z_n)\rangle=\int_{{\cal M}_n} \Omega_{|\vec{V}\rangle}(\vec{v}).
\end{equation}
One point of concern might be that, since $\langle \Sigma|$ depends on a choice of local coordinates $z_i=h_i(0)$ centred on each puncture, it is not at all obvious that $ \Omega_{|\vec{V}\rangle}(\vec{v})$ is well-defined on ${\cal M}_n$. The natural framework to describe $ \Omega_{|\vec{V}\rangle}(\vec{v})$ is the bundle over ${\cal M}_n$ with fibres given by an independent choice of local coordinates about each puncture. However, provided that the external states $|\vec{V}\rangle$ are on-shell, in other words that they are BRST-invariant, the integrand is invariant under local reparametrisations and so the form $ \Omega_{|\vec{V}\rangle}(\vec{v})$ does descend to a well-defined form on ${\cal M}_n$. Put another way, it does not matter which section of the bundle we choose to integrate over.

Let us consider the explicit example of the scattering of $n$ on-shell states $|V_i\rangle$, each of the form (\ref{state}). It is straightforward to show (see Appendix \ref{Further details on the derivation of the scattering equations}) that
$$
\langle\Sigma|\alpha_{-1}^{(i)}=\int \rd^np\;\langle \vec{p}_n|\;\delta\left(\sum p_{(j)}\right)\;e^W \;\sum_{j\neq i}\sum_{n\geq 0}{\cal S}_{1n}(z_i,z_j)\alpha_n^{(j)}\;{\cal Z},
$$
where ${\cal S}_{mn}(z_i,z_j)$ is given by (\ref{S}). We also have that
$$
\sum_{j\neq i}\sum_{n\geq 0}{\cal S}_{1n}(z_i,z_j)\alpha_n^{(j)} |k_j\rangle=\sum_{j\neq i}\frac{k_j}{z_i-z_j}\; |k_j\rangle,
$$
where we have used the fact that $\alpha_n^{(i)}|k_i\rangle=\delta_{n,0}k_i|k_i\rangle$ for $n\geq 0$ and ${\cal S}_{10}(z_i,z_j)=(z_i-z_j)^{-1}$. The $\alpha^{(i)}_n$ for $n\geq 0$ commute with all operators to the right in the expression for the amplitude until they hit the $|k_i\rangle$ of the asymptotic states. Thus, in evaluating the scattering amplitude the net effect is to make the replacement
$$
\alpha^{(i)}_{-1}\rightarrow \sum_{j\neq i}\frac{k_j}{z_i-z_j}.
$$
This is the operator statement of the path integral result
\begin{equation}\label{P}
P_{\text{cl}}(z)=\sum_{j}\frac{k_j}{z-z_j},
\end{equation}
which arises when the $X^{\mu}(z)$ path integral is done (see \cite{Mason:2013sva} for details.).

Let us now focus on the ghost terms. The gauge-fixing of the worldsheet complex structure and the gauge symmetry of the Beltrami differential $e(z)$ give the ghost contribution
$$
\prod_{a=1}^{n-3}\bar{\delta}\Big({\cal H}(\vec{\nu}_a)\Big)\tilde{\mathbf{b}}(\vec{\nu}_a)\mathbf{b}(\vec{\nu}_a),
$$
which we recognise as the $B_{n-3}(\vec{v})$ insertion in $ \Omega_{|\vec{\Psi}\rangle}(\vec{v})$. The $\textbf{b}$ and $\tilde{\textbf{b}}$ insertions are of the standard type. The ${\cal H}(\vec{\nu}^a)$ contribution requires more discussion.

A similar calculation to that above (which also may be found in Appendix \ref{Further details on the derivation of the scattering equations}) yields
\begin{equation}\label{I2}
\langle\Sigma|\alpha^{(i)}_{-m}\alpha^{(i)}_{-n}=\int \rd^np\;\langle \vec{p}_n|\;\delta\left(\sum p_{(j)}\right)\;e^W \;A^{(i)}_{-m}A^{(i)}_{-n}\;{\cal Z},
\end{equation}
where
$$
A^{(i)}_{-m}\equiv \alpha^{(i)}_{-m}+\sum_{j\neq i}\sum_{n\geq 0}{\cal S}_{mn}(z_i,z_j)\alpha_{n}^{(j)}.
$$
The relationship of the worldsheet vector fields $v(z)$ and the types deformations of the moduli space may be described simply. Given a disc $\mathscr{D}_i$ containing the $i$'th puncture, and the vector field $v_i(z)$ on the boundary of the disc we can ask if one may smoothly extend $v_i(z)$ outside the disc. Those $v$ that cannot be extended outwards provide interesting deformations. If $v(z)$ vanishes at the puncture it describes coordinate changes that do not affect the location of the punctures. Since these do not have any effect on the moduli space ${\cal M}_n$, we ignore these in the on-shell theory; however, they do play a role in the off-shell theory. Those $v(z)$ that do not vanish at the puncture act to move the location of the puncture and so do have an interesting action on ${\cal M}_n$. At tree level, the locations of the punctures are the only moduli, so it is this class of vector fields that are of interest to us. For completeness we mention that those vector fields that cannot be extended to the full interior of the disc encode changes in the moduli of the underlying, unmarked, Riemann surface. This classification is nicely summarised in Table 1 of \cite{Zwiebach:1992ie}

Let us focus then on those $v_i$ that do not vanish at the point $z_i$ and can be extended into the interior of the disc $\mathscr{D}_i$ surrounding the point $z_i$. These correspond to deformations that can move the location of the punctures
$$
z_i\rightarrow z_i+v^a_i\delta \tau_a.
$$
where $\tau_a$ are coordinates on the moduli space. Let us choose a basis for the $v^a_i(z)$ such that three of the punctures are kept fixed while $n-3$ are shifted by an amount given directly by a particular modulus, so that $v_i^a(z_i)=\delta^a_i$ for $i=1,2,...,n-3$ and $v^a_i=0$ for $i=n-2,n-1,n$.

Using this $v(z)$, we have
$$
{\cal H}(\vec{\nu}^a)=\sum_{i=1}^n\oint\rd z\;H^{(i)}(z)\delta^a_i=\widetilde{L}^{(a)}_{-1}.
$$
Using the identity (\ref{I2}), we have
\begin{eqnarray}
\langle\Sigma|\widetilde{L}^{(a)}_{-1}&=&\sum_{n\geq 0}\langle\Sigma|\alpha^{(a)}_n\cdot\alpha^{(a)}_{-n-1}\nonumber\\
&=&\int \rd^np\;\langle \vec{p}_n|\;\delta\left(\sum p_{(j)}\right)\;\sum_{n\geq 0} \alpha_n^{(a)}\cdot\left(\sum_{j\neq a}\sum_{m\geq 0} {\cal S}_{1+n,m}(z_a,z_j)\alpha_m^{(j)}\right) \;e^W\;{\cal Z},\nonumber
\end{eqnarray}
It is then straightforward to show that
\begin{eqnarray}
\sum_{n\geq 0} \alpha_n^{(a)}\cdot\left(\sum_{j\neq a}\sum_{m\geq 0} {\cal S}_{1+n,m}(z_a,z_j)\alpha_m^{(j)}\right)|k_1\rangle...|k_n\rangle&=&\sum_{j\neq a}\frac{k_a\cdot k_j}{z_a-z_j}|k_1\rangle...|k_n\rangle\nonumber\\
&=&k_a\cdot P_{\text{cl}}(z_a)|k_1\rangle...|k_n\rangle.\nonumber
\end{eqnarray}
Putting this all together gives the result
$$
\int \rd^np\;\delta\left(\sum p_{j}\right)\;\langle p_1|...\langle p_n|\;e^{V_{X,P} }   \prod_{a=1}^{n-3}\bar{\delta}\Big(\widetilde{L}^{(a)}_{-1}\Big)|k_1\rangle...|k_n\rangle=\delta\Big(\sum k_i\Big)\prod_{a=1}^{n-3}\left(k_a\cdot P_{\text{cl}}(z_a)\right),
$$
where $V_{X,P}$ is given by (\ref{V2}). This is the required scattering equation and momentum conservation contributions to the amplitude. The final steps in the calculation of the scattering amplitude are straightforward and given in more detail in \cite{Reid-Edwards:2015stz}. We present a quick overview below. The on-shell asymptotic states are given by (\ref{state}). Substituting these into (\ref{amp}) gives the scattering amplitude
$$
\int_{{\cal M}_n} \Omega_{|\vec{\Psi}\rangle}(\vec{v}) = \int_{{\cal M}_n} \langle \Sigma|\,\prod_{i=1}^3c_1^{(i)}\tilde{c}_1^{(i)}\epsilon^{\mu\nu}_i\alpha^{(i)}_{-1\mu}\alpha^{(i)}_{-1\nu}\,\prod_{i=4}^n\bar{\delta}\left(k\cdot P_{\text{cl}}\right)\epsilon^{\mu\nu}_i\alpha^{(i)}_{-1\mu}\alpha_{-1\nu}^{(i)} |k_1\rangle...|k_n\rangle,
$$
 where $n-3$ factors of $c(z)\tilde{c}(z)$ have been absorbed by the $\mathbf{b}(\vec{\nu}^a)\tilde{\mathbf{b}}(\vec{\nu}^a)$ insertions\footnote{Taking $v^a(z_i)=\delta^a_i$ for i=1,...,n-3 and zero otherwise gives
 $$
 \mathbf{b}(\vec{\nu}^a)=\oint\rd  z b^{(a)}(z)=b^{(a)}_{-1},
 $$
which removes the $c_1^{(a)}$ insertion on $n-3$ of the external states.}. The remaining $c$ and $\tilde{c}$ ghosts are eliminated by the ${\cal Z}$ factor in $\langle\Sigma|$.  It was shown above that the $\alpha_{-1}^{(i)}$ factors in the external states are converted into $P_{\text{cl}}(z_i)$ factors when inserted into correlation functions involving $\langle\Sigma|$. The net result is the bosonic scattering amplitude \cite{Mason:2013sva}
\begin{equation}\label{amplitude}
M_N=\delta^D\left(\sum k_i\right)\int_{{\cal M}_N}\rd^{N-3}z_i \frac{1}{\rd\omega}\prod_{i=1}^N\epsilon_i^{\mu\nu}P_{\text{cl}\;\mu}P_{\text{cl}\;\nu}\prod_i'\bar{\delta}\left(k_i\cdot P_{\text{cl}}\,(z_i)\right),
\end{equation}
where\footnote{$\rd\omega= \frac{\rd z_i\rd z_j\rd z_k}{(z_i-z_j)(z_j-z_k)(z_k-z_i)}.$}
$$
\prod_{i}' \bar{\delta}(k_i\cdot P_{\text{cl}}(z_i))=\frac{1}{\rd\omega}\prod_{i=4}^N \bar{\delta}(k_i\cdot P_{\text{cl}}(z_i)).
$$
As commented upon in \cite{Mason:2013sva}, this is not the correct tree amplitude for Einstein gravity; however, Einstein \emph{super}gravity is recovered from the $N=2$ supersymmetric extension of (\ref{action1}). The arguments above, including the emergence of the scattering equations, apply also in the supersymmetric case.

As a final comment, we note that the fact that, the form $ \Omega_{|\vec{V}\rangle}(\vec{v})$ is invariant under diffeomorphisms on the worldsheet and so it is well-defined on the moduli space will become important when we consider correlation functions involving states which are not BRST-invariant. When we consider the string field theory in the next section, we shall want to generalise this discussion to off-shell quantities where the form $ \Omega_{|\vec{V}\rangle}(\vec{v})$ will not be invariant under general diffeomorphisms and consequently will not be well-defined on the moduli space ${\cal M}_n$. In the off-shell case, a generalisation of the bundle over moduli space will be the framework we will be forced to work with.

\subsubsection{An alternative perspective}\label{new}

In \cite{Ohmori:2015sha} an alternative and, arguably, more fundamental perspective on the scattering equations and how they arise in the ambitwistor string was proposed. This perspective differs significantly from that of conventional string theory and provides a useful framework in which to discuss the string field interactions. We summarise the main ideas here and refer the interested reader to \cite{Ohmori:2015sha}. In this approach the observation that the algebra (\ref{algebra}) relates to $T^*{\cal M}$, rather than simply ${\cal M}$, plays a central role. Following \cite{Witten:2012bh}, \cite{Ohmori:2015sha} generalises the BRST operator such that $\{Q,\mu\}=\delta\mu$ and $\{Q,e\}=\delta e$ where $\delta\mu$ and $\delta e$ are anti-commuting fields. Note that $\mu$ and $\delta\mu$ depend on the coordinates of the base of $T^*{\cal M}$ but are independent of the fibre directions, whereas $e$ and $\delta e$ vary as we move in the fibre and the base. The action is not invariant under this extended BRST transformation; however, an extension, which leads to an action invariant under this generalised BRST transformation may be found. The invariant action may be written in the form
$$
S_0+\{Q,{\cal W}\},
$$
where $S_0=\int P_{\mu}\bar{\partial}X^{\mu}+b\bar{\partial} c+\tilde{b}\bar{\partial}\tilde{c}$. In our language ${\cal W}=\mathbf{b}(u)+\tilde{\mathbf{b}}(u,\tilde{u})$, where the arguments reflect the dependence of $\mathbf{b}$ and $\tilde{\mathbf{b}}$ on the base and fibre directions of $T^*{\cal M}$. In \cite{Ohmori:2015sha}, it is $S_0$ that is used to compute correlation functions and $\{Q,{\cal W}\}$ is treated as an operator insertion. The scattering amplitudes are given by
\begin{equation}\label{Oamplitude}
M_n=\int_{\Gamma\subset T^*{\cal M}} \widetilde{\Omega}_{|\vec{V}\rangle}(u,\tilde{u}),
\end{equation}
where
$$
\widetilde{\Omega}_{|\vec{V}\rangle}(u,\tilde{u})=\left\langle e^{-\{Q,\mathbf{b}(u)+\tilde{\mathbf{b}}(u,\tilde{u})\}}\;V_1...V_n\right\rangle_{S_0},
$$
is a middle-dimensional form on $T^*{\cal M}$, $\Gamma$ is a suitably chosen middle-dimensional cycle in $T^*{\cal M}$ and the $V_i$ are vertex operators. The first hint that $T^*{\cal M}$, rather than ${\cal M}$ itself plays the key role is the suggestive semi-direct product form of the worldsheet theory gauge algebra (\ref{algebra}), a point that will be taken up again in section \ref{Ambitwistor interactions as forms}. The question of how to choose a suitable $\Gamma\subset T^*{\cal M}$ was discussed in \cite{Ohmori:2015sha} and we shall present a brief review below. The correlation function is evaluated using $S_0$ as the classical action in the path integral. The pair of vectors $u$ and $\tilde{u}$ formally generalise $v$ and denote deformations along the base and fibre directions of $\Gamma\cap T^*{\cal M}$ respectively.

The $e^{-\{Q,{\cal W}\}}$ factor in the correlation function $\widetilde{\Omega}_{|\vec{V}\rangle}$ indicates that standard localisation arguments may be applied to perform the integral. Indeed this is the case and furthermore, only the critical points\footnote{At tree level, each critical point is associated to the location of $n-3$ punctures which are solutions to the scattering equarions.} $\tau^*\in T^*{\cal M}$ of an appropriate Morse function\footnote{See \cite{Ohmori:2015sha} for further details.} are required, not the detailed form of $\Gamma$. The critical points satisfy two conditions: the first imposes the scattering equations, the second selects a point in each of the fibres of $T^*{\cal M}$. Evaluating $M_n$ on the critical points $\tau^*$ gives
$$
M_n=\sum_{\tau^*}\left\langle \prod_{a=1}^{n-3}\mathbf{b}(v_a) \prod_{a=1}^{n-3}\mathbf{\tilde{b}}(v_a)(\det\Phi)^{-1}\;V_1...V_n\right\rangle_{S_0},
$$
where the precise form of $\Phi$ is given in \cite{Ohmori:2015sha}. The key point is that formally, this expression for the amplitude may be written as an integral over a copy\footnote{The simplest identification of the copy of ${\cal M}$ as the base of $T^*{\cal M}$ does not do the job \cite{Ohmori:2015sha}.} of ${\cal M}$ in $T^*{\cal M}$ with coordinates $\tau$ where delta-functions are introduced to single out the required critical points $\tau^*$. These delta functions have support precisely on the solutions to the scattering equations and so the amplitude may be equivalently written in the form (\ref{amplitude}). The advantage of the expression (\ref{amplitude}) for the amplitude is that, written as an integral over ${\cal M}$, the similarities with conventional string theory are emphasised; however, we shall see that formulation (\ref{Oamplitude}) has advantages when we consider interactions of the associated string field theory. As such, it is useful to consider briefly how the discussion of \cite{Ohmori:2015sha} might proceed in the operator formalism.

The operator formalism posits a surface state $\langle\Sigma|$ such that
$$
M_n=\int_{\cal M}\langle\Sigma|B_{n-3}|V_1\rangle...|V_n\rangle.
$$
where $B_{n-3}$ is given by (\ref{B}). We can separate off the delta-functions in $B_{n-3}$ and use them to do the integral over ${\cal M}$. The delta-functions have support on the scattering equations which may be solved to give the collection of marked points on $\Sigma$, denoted by $\tau^*$. The expression of the amplitude becomes
$$
M_n=\sum_{\tau^*}\langle\Sigma^*|\prod_{a=1}^{n-3}\mathbf{b}(v^*_a) \prod_{a=1}^{n-3}\mathbf{\tilde{b}}(v^*_a)\;(\det\Phi)^{-1}|V_1\rangle...|V_n\rangle,
$$
where $\langle\Sigma^*|$ is the surface state evaluated $\tau^*\in{\cal M}$. $\Phi$ denotes the Jacobian matrix that arises from evaluating the integral on the delta-functions\footnote{The scattering equations are functions of the moduli, so there will be a Jacobian factor when evaluating the integral against the delta functions. It is natural to absorb this Jacobian into the definition of $\langle\Sigma^*|$.}. One could then propose a reverse-engineering of the amplitude to construct a surface state appropriate for the localisation procedure described in \cite{Ohmori:2015sha}. Such a surface state $\langle\widetilde{\Sigma}|$ would satisfy
$$
M_n=\int_{\Gamma\subset T^*{\cal M}}\langle\widetilde{\Sigma}||V_1\rangle...|V_n\rangle,
$$
where $\langle\widetilde{\Sigma}|$ takes the same form as $\langle\Sigma|$ but also incorporates the $e^{-\{Q,\mathbf{b}\}}e^{-\{Q,\tilde{b}\}}$ insertion as part of its definition. The operator formalism tells us how to construct the appropriate worldsheet correlation function and then a region over which it must be integrated must be chosen. This integral is then performed as outlined in \cite{Ohmori:2015sha}. Given that localisation is such a powerful tool, evaluating the operator expression $\langle\widetilde{\Sigma}||V_1\rangle...|V_n\rangle$ directly is not particularly efficient. It is far more useful to work in terms of the worldsheet correlation function $ \left\langle e^{\{Q,{\cal W}\}}V_1...V_n\right\rangle_{S_0}$. In any event, one could always choose to express the interaction terms in terms of worldsheet correlation functions.

\section{Bosonic Ambitwistor String Field Theory}\label{Bosonic Ambitwistor String Field Theory}

In this section we construct a string field theory for the bosonic ambitwistor string. Although our main interest is in the supersymmetric ambitwistor theory, the basic ingredients of the bosonic and supersymmetric constructions are similar and it is helpful to first study the slightly simpler bosonic theory first. We shall find that much of the analysis presented for the bosonic string generalises straightforwardly to the superstring\footnote{With the exception of the issue of how to deal with picture changing operators, this is largely true for the Neveu-Schwarz (NS) sector in the small Hilbert space approach. The Ramond sector however, introduces new complications. This is briefly discussed in section \ref{Discussion}.} but there are some important differences. We begin with a brief overview of the standard approach to conventional covariant string field theory, before outlining the formal structure of the ambitwistor string field theory in section \ref{Second Quantization of closed string theories}. Section \ref{Second Quantization of closed string theories} is rather formal and so, starting in section \ref{The Ambitwistor string field} and continuing in sections \ref{The action to quadratic order} and \ref{Interactions} we provide explicit constructions of the string field and the action.

\subsection{Second Quantization of closed string theories}\label{Second Quantization of closed string theories}

The condition of BRST-invariance imposes the spacetime equations of motion and thus far we have considered only ambitwistor fields which satisfy the on-shell condition $Q|\Psi\rangle=0$. Starting in this section, we shall follow the well-trodden path to second quantisation given by string field theory. We shall begin with more general, off-shell, string fields that are not required to satisfy $Q|\Psi\rangle=0$. The condition of BRST invariance will only emerge as a consequence of the classical equations of motion of the string field action $S[\Psi]$.

To set the scene, we first review the covariant approach to the conventional string field theory. We shall begin by considering the slightly simpler case of the conventional covariant open string field \cite{Witten:1985cc}. In a conventional string theory a suitable background is associated with a worldsheet CFT, which gives a BRST operator $Q$. The on-shell states are solutions to $Q|\Psi\rangle=0$ which may be derived from the action $S_2[\Psi]=\frac{1}{2}\langle \Psi|Q|\Psi\rangle$. The story does not end here. The string theory is not a free theory and its interactions describe perturbations of the original background and so such interactions must be included in the string field action. For the classical open string a cubic term, which is constructed using the three-punctured sphere surface state $\langle\Sigma_3|$ and is denoted by $\{\Psi^3\}=\langle\Sigma_3||\Psi\rangle|\Psi\rangle|\Psi\rangle$, is sufficient to reconstruct the string perturbation theory at tree level\footnote{Of course, to go to loop level, a closed sector must also be included.}. The equations of motion now include a non-linear term which effectively corrects the original, background-dependent, statement of BRST-invariance for the new, perturbed, background. The string field action is
\begin{equation}\label{oaction}
S[\Psi]=\langle \Psi|Q|\Psi\rangle+\frac{g}{3!}\langle\Sigma_3||\Psi\rangle|\Psi\rangle|\Psi\rangle
\end{equation}
Perturbation theory proceeds by fixing the gauge, the Siegel gauge $b_0|\Psi\rangle=0$ being a convenient choice. The quadratic term reduces to $\langle \Psi|Q|\Psi\rangle\rightarrow\langle \Psi|c_0L_0|\Psi\rangle$ and the propagator may be written schematically as
\begin{equation}\label{oprop}
\frac{b_0}{L_0}=b_0\int_{0}^{\infty}\rd\tau \;e^{-\tau L_0},
\end{equation}
where $\tau$ becomes a real modulus that contributes to the moduli spaces of higher point Riemann surfaces with boundary. One might have hoped that the chiral nature of the ambitwistor string means that the corresponding string field theory is constructed along the lines of the conventional open string field. This does not appear to be the case, rather the conventional closed string field theory seems to be the more natural cousin of the ambitwistor string field theory.

The story for the conventional closed string has a number of additional subtleties. Consideration of the ghost number means that an additional ghost must be inserted into the quadratic part of the action which becomes $S_2[\Psi]=\langle \Psi|c^-_0Q|\Psi\rangle$, where $c_0^-=c_0-\bar{c}_0$. The problem then is that the equations of motion for $S_2[\Psi]$ require all but the $c_0^-$ parts of the BRST operator to annihilate $|\Psi\rangle$. Those parts of the BRST charge that include $c_0^-$ dependence must be required to vanish as an additional, off-shell, constraint on the string field $|\Psi\rangle$. Fortunately, the constraint is the reasonable condition\footnote{The condition is $(L_0-\overline{L}_0)|\Psi\rangle=0$, supplemented by $(b_0-\bar{b}_0)|\Psi\rangle=0$.} of level matching that one would expect the string field to satisfy. To complete the action, non-linear interactions terms must be added. This is well-understood for the bosonic string and we shall not discuss it further here as our main interest is in the ambitwistor string. Further details on conventional string field theory may be found in \cite{Kugo:1989aa,Kugo:1989tk,Zwiebach:1992ie} and also \cite{LeClair:1988sp,LeClair:1988sj,Saadi:1989tb,Sonoda:1989wa}. Much progress has been made recently on the conventional superstring field theory, details of which may be found in \cite{Saroja:1992vw,Sen:2015uaa}.

\subsubsection{The action and spacetime gauge symmetries}\label{The action and spacetime gauge symmetries}

We now turn to our main interest - the ambitwistor string field theory. We shall consider the (problematic) bosonic theory in this section and discuss the better founded supersymmetric theory in sections \ref{The Supersymmetric theory} and \ref{Interaction Terms}. Though the theory has some similarities with open string theory, the perturbative structure seems to rely on closed string moduli space which suggests that the covariant theory must be non-polynomial as we discuss below.

We begin the discussion at the linearised level. We look for an action for which $\delta |\Psi\rangle=Q|\Lambda\rangle$ is a target space gauge symmetry and the net effect of the equation of motion and other physical constraints imposes $Q|\Psi\rangle=0$, as in the conventional string field theories. The strategy will be to impose $Q|\Psi\rangle=0$ on a string field in two stages; The first will be to require the string field $|\Psi\rangle$ to obey the condition ${\cal L}_0|\Psi\rangle=0$, where ${\cal L}_0$ is that part of $Q$ which multiplies $c_0$ (\ref{L}). The remaining conditions from $Q|\Psi\rangle=0$ are imposed by the linearised equations of motion. We shall see that ${\cal L}_0|\Psi\rangle=0$ is a background independent constraint, much like the level-matching condition in conventional string field theory. The background-dependent parts of the BRST condition are then imposed by the equations of motion. This seems reasonable, as the background-dependent parts are expected to receive corrections from the non-linear interaction terms in the action, whereas the background-independent constraints do not depend on any interaction terms we may subsequently add\footnote{It is a much more primitive condition that ensures things like the field having the correct level.}. The effect of including non-linear interaction terms will be to alter the equations of motion $Q|\Psi\rangle+...=0$ and gauge transformations $\delta |\Psi\rangle=Q|\Lambda\rangle+...$ but not the constraint ${\cal L}_0|\Psi\rangle=0$.

We take the string action to have the same general form as the closed bosonic string field theory
\begin{equation}\label{SFTaction}
S[\Psi]=\langle \Psi |c_0Q|\Psi\rangle+\sum_{n>2} \frac{1}{n!} \{\Psi^n\},
\end{equation}
where the interaction terms $\{\Psi^n\}$ will be discussed further below. The quadratic part includes the $c_0Q=c_0(c_0{\cal L}_0+\tilde{c}_0\widetilde{L}_0+...)$ insertion. Since $c_0$ is grassmann, the ${\cal L}_0$ term drops out of the quadratic term. As stated above, we impose the condition ${\cal L}_0|\Psi\rangle=0$ as part of the definition of an ambitwistor string field. We shall also impose the condition $b_0|\Psi\rangle$, which arises naturally given that $\{Q,b_0\}={\cal L}_0$.

Although it will not play a significant role in this paper we recall the string product $|[\Psi_1,...,\Psi_n]\rangle$. Following \cite{Zwiebach:1992ie} it can be useful to write the interaction terms $\{\Psi^n\}$ in terms of the string product $|[\Psi_1,...,\Psi_n]\rangle$ as $\{\Psi^n\}=\langle\Psi_n|c_0|[\Psi_1,...,\Psi_{n-1}]\rangle$. The relationship may be inverted by introducing a complete basis of states $\{|\phi_r\rangle\}$ so that
$$
|[\Psi_1,...,\Psi_n]\rangle=\sum_rb^{(r)}_0|\phi_r\rangle\{\phi_r,\Psi_1,...,\Psi_n\}.
$$
In this way, the operation $|[\Psi_1,...,\Psi_n]\rangle$ may be thought of as a map from $\otimes^n{\cal H}$ to ${\cal H}$. The product is useful in writing the equations of motion of the action (\ref{SFTaction})
$$
Q|\Psi\rangle+\sum_{n>2} \frac{1}{n!}|[\Psi_1,...,\Psi_n]\rangle=0.
$$
In the conventional bosonic \cite{Kugo:1989aa,Kugo:1989tk,Zwiebach:1992ie} and supersymmetric \cite{Sen:2015uaa} string field theories, the objects $[\cdot,...,\cdot]$ satisfy an important identity called the \emph{main identity}, which may be written at tree level as
\begin{eqnarray}\label{main}
Q[\Psi_1,...,\Psi_n]&&+\sum_{i=1}^n(-1)^{|\Psi_1|+...+|\Psi_{i-1}|}[\Psi_1,...,Q\Psi_i,...,\Psi_n]\nonumber\\
&&+\sum_{\{i_{\ell},j_k\},l,k}\sigma(i_{\ell},j_k)[\Psi_{i_1},...,\Psi_{i_{\ell}},[\Psi_{j_1},..,\Psi_{j_k}]]=0,
\end{eqnarray}
where the set of vertices has been partitioned into sets $\{i_1,...,i_{\ell}\}$ and $\{j_1,...,j_k\}$ and $\sigma(i_{\ell},j_k)$ denotes an appropriate sign generated by moving the BRST operator onto $\Psi_i$. The expression relates the failure of the product $[\cdot,...,\cdot]$ to satisfy a Jacobi identity with the failure of the the BRST operator to act as a derivation on the product. At loop level, an additional term enters and the notion of the interaction term $\{\Psi^n\}$ must be generalised for higher genus. A detailed discussion and explicit proof of the main identity may be found in \cite{Zwiebach:1992ie}. Whilst we have not checked carefully, we believe there is good evidence that a similar identity holds for the ambitwistor superstring field theory and we hope to present a proof of this elsewhere.

\subsubsection{Target space gauge symmetries}\label{Target space gauge symmetries}

Though we have not proven the main identity for the ambitwistor string field, there is evidence that the action has (target space) gauge invariance 
$$
\delta|\Psi\rangle=Q|\Lambda \rangle+\sum_n \frac{1}{n!}|[\Lambda,\Psi_1,...,\Psi_n]\rangle,
$$
where $|\Lambda\rangle$ is a gauge parameter field. The standard way to deal with this spacetime gauge invariance is to employ the BV procedure \cite{Zwiebach:1992ie} and we shall not consider this further here, we shall study the linearised symmetries at length in section \ref{The Ambitwistor string field}.

We do not wish to quantise the target space theory as it describes classical supergravity in ten-dimensions and so no sensible quantum theory is expected to exist; however, gauge-fixing can be employed to simplify the theory and is a prerequisite for tree-level perturbation theory. An analogue of Siegel gauge for the ambitwistor string field is
$$
\tilde{b}_0|\Psi\rangle=0.
$$
In this gauge the kinetic term for the string field theory becomes
$$
\langle\Psi|c_0Q|\Psi\rangle=\langle\Psi|c_0\tilde{c}_0\widetilde{L}_0|\Psi\rangle,
$$
where we have used the fact that $\{Q,\tilde{b}_0\}=\widetilde{L}_0+...$, where the ellipsis denote ghost terms that vanish on $|\Psi\rangle$. The additional constraints on the string field are simply ${\cal L}_0|\Psi\rangle=0$, where ${\cal L}_0$ is that part of the BRST current that multiplies $c_0$, and $b_0|\Psi\rangle=0$. In this gauge, the propagator looks like
\begin{equation}\label{prop}
\frac{\delta({\cal L}_0)}{\widetilde{L}_0}\tilde{b}_0b_0|{\cal R}_{LR}\rangle,
\end{equation}
where $|{\cal R}_{LR}\rangle$ is the reflector state that relates the string fields in two Hilbert spaces and their conjugates\footnote{As will be discussed below, the notion of conjugate here is not the usual BPZ conjugate in the bosonic case. However, in the supersymmetric the conventional BPZ conjugate may be used.} as $\langle\Psi_L|{\cal R}_{LR}\rangle=|\Psi_R\rangle$. The reflector state appears naturally in the propagator as we may write the kinetic term as $\langle\Psi_L|\langle\Psi_R|c_0Q|{\cal R}_{LR}\rangle$.
The form of the propagator is what we might expect since $\widetilde{L}_0=\frac{1}{2}k^2+...$, which is the conventional kinetic term. The delta function simply imposes the required ${\cal L}_0|\Psi\rangle=0$ constraint. The way in which this propagator may be written as an integral, reminiscent of (\ref{oprop}), is discussed in section \ref{A brief sketch of perturbation theory}.

\subsubsection{Feynman rules and fundamental vertices}\label{Feynman rules and fundamental vertices}

In this section we briefly review the arguments for why the action (\ref{SFTaction}) is required to be non-polynomial. Readers familiar with these arguments may safely skip this section. A superior discussion may be found in section 5 of \cite{Zwiebach:1992ie}. 

As with the conventional bosonic string, finding the Feynman rules for the ambitwistor string amounts to finding a minimal set of vertices that, when supplemented by the propagator, may be used to construct a single cover of the moduli space of punctured Riemann surfaces ${\cal M}_n$. For example, we could start with the three-punctured sphere as a basic building block. The moduli space ${\cal M}_3$ is a point. One might then try to construct all possible four-punctured Riemann surfaces, the space ${\cal M}_4$ (the Riemann sphere), from sewing two three-point surfaces with a propagator, the two moduli coming from the propagator. As demonstrated in \cite{Giddings:1986bp,Peskin,Kaku:1988zw,Kaku:1988zv,Sonoda:1989sj}, this fails. There is a fundamental ``missing" region ${\cal D}_4\subset{\cal M}_4$ that cannot be constructed in this way\footnote{More correctly, one cannot obtain a single cover of ${\cal M}_4$ in this way.} and so must be added in as a fundamental 4-point interaction \cite{Saadi:1989tb}. Thus, it is not sufficient for the action of the closed theory to have just a quadratic term and a cubic interaction $\{\Psi^3\}$ as in (\ref{oaction}), we must also add in a quartic interaction $\{\Psi^4\}$ which encodes the `missing' region ${\cal D}_4$.

This result generalises to more punctures; a single cover of ${\cal M}_n$ cannot be constructed from fundamental regions ${\cal D}_m$ with $m<n$ and propagators alone, and so a new fundamental vertex must be introduced at each $n$ to fill in the missing region of moduli space. Thus, we are required to introduce interaction terms $\{\Psi^n\}$ (each encoding the missing region ${\cal D}_n$) for \emph{all} $n\geq 3$ and we see that the covariant action is non-polynomial. The converse of this is that the boundary of each fundamental region is described by Riemann surfaces that are built from vertices ${\cal D}_m$ with $m<n$ glued together with up to $n-3$ propagators.

The string vertex ${\cal V}_n$ is defined to be the set of Riemann surfaces in ${\cal D}_n$ with a choice of coordinate, up to a phase, around each puncture \cite{Zwiebach:1992ie}. As will be discussed in section \ref{Interactions} below, when dealing with off-shell objects, the natural object that appears is a generalisation of the moduli space ${\cal M}_n$; the bundle $\widehat{\cal P}_n$, whose base is ${\cal M}_n$ and whose (infinite-dimensional) fibres describe the choice of local coordinates around each puncture. Thus ${\cal V}_n\subset\widehat{\cal P}_n$. The ambitwistor string theory naturally defines forms $\Omega_{|\vec{\Psi}\rangle}(\vec{\nu})$ on $\widehat{\cal P}_n$ which are introduced below. The interactions are then written as
$$
\{\Psi^n\}=\int_{{\cal V}_n}\Omega_{|\vec{\Psi}\rangle}(\vec{\nu}).
$$
For the first quantised ambitwistor string, the forms $\Omega_{|\vec{\Psi}\rangle}(\vec{\nu})$ are those given in (\ref{form1}). In general, the string states $|\Psi\rangle$ will be the off-shell string fields discussed in the following section.

The fact that there are an infinite number of ${\cal V}_n$ that must be introduced gives rise to the non-polynomial structure of closed string field theory\footnote{There are a number of ways of providing concrete constructions of these regions. Originally this was done using restricted polyhedra \cite{Saadi:1989tb} which work well-enough at tree level. Later minimal area metrics and the introduction of `stubs' were used which provide a realisation that extended to loop level. Since we shall only be interested in tree level, either realisation could be used here. It would be interesting if the analysis of \cite{Ohmori:2015sha} could be applied in some way to provide an implicit construction using Morse theory. Such a construction might be more natural from the perspective of the ambitwsitor string but, as yet, we have no concrete way to realise this.}. The forms $\Omega$ have been constructed in \cite{AlvarezGaume:1988bg,Zwiebach:1992ie} for the bosonic string and progress has been made in the superstring in \cite{Saroja:1992vw}. We discuss the $\Omega$ forms for the ambitwistor string in section \ref{Ambitwistor interactions as forms}.

The boundary of any ${\cal V}_n$ is given by sets of ${\cal V}_{m<n}$ joined by propagators. This may be expressed as a recursion relation (see the Figure 5.1 in \cite{Zwiebach:1992ie}). This factorisation property, coupled with the fact that the BRST operator acts as a total derivative on the forms $\Omega$ are the key ingredients in deriving the main identity (\ref{main}). As mentioned previously, we shall discuss the main identity in the context of the ambitwistor superstring field theory elsewhere.

It should be stressed that the infinite number of fundamental vertices is characteristic of the theory built on \emph{closed} Riemann surfaces. The detail of the precise CFT under consideration does not change the basic story. As such we shall appropriate this construction to the ambitwistor string field theory.

\subsection{The Ambitwistor string field}\label{The Ambitwistor string field}

Thus far our discussion of the ambitwistor string field theory has been somewhat abstract, drawing heavily of the existing lore of string field theory. Beginning in this section, we provide a more explicit construction of the theory. In doing so, we hope to highlight the many similarities and differences between the conventional and ambitwistor string theories. It has been observed that, whilst the ambitwistor string shares many superficial similarities with the conventional string, there are a large number of important differences that arise as the detail of the theory has become better understood. We focus on the case of flat, empty spacetime and look for a string field that describes small, perturbative, fluctuations on that spacetime. The question of more general backgrounds and going beyond perturbation theory will be discussed briefly in section \ref{Discussion}.

The state operator correspondence gives the perturbative, momentum eigenstate, `graviton'\footnote{This theory does not describe conventional Einstein gravity, indeed there is evidence that it is not spacetime diffeomorphism invariant, and so we hesitate to call this state a graviton. The supersymmetric theory exhibits no such problems.} with polarisation $\varepsilon_{\mu\nu}$ as
$$
|\Psi\rangle=\varepsilon_{\mu\nu}\alpha_{-1}^{\mu}\alpha_{-1}^{\nu}c_1\tilde{c}_1|k\rangle,
$$
where $|k\rangle=e^{ik\cdot x}|0\rangle$. A more general state is given by a linear superposition of such states, weighted with a function $h_{\mu\nu}(k)$, thus we start with the minimal proposal for the string field
\begin{equation}\label{ansatz}
|\Psi\rangle=\int\rd k\left(-\frac{1}{2}h_{\mu\nu}(k)\; \alpha_{-1}^{\mu}\alpha_{-1}^{\nu}c_1\tilde{c}_1+...\right)|k\rangle,
\end{equation}
where $h_{\mu\nu}$ is a function of the momentum $k$ and $+...$ denote terms to be determined by the symmetries of the theory. This clearly generalises the graviton vertex operator. The linearised gauge transformation is $\delta|\Psi\rangle=Q|\Lambda\rangle$ for some parameter field $|\Lambda\rangle$. In order to have the linearised spacetime diffeomorphisms $\delta h_{\mu\nu}=\partial_{\mu}\lambda_{\nu}+\partial_{\nu}\lambda_{\mu}$ as a symmetry, we require at least the minimal gauge field
$$
|\Lambda\rangle=i\int\rd k \;\lambda_{\mu}(k)\;\alpha_{-1}^{\mu}c_1|k\rangle.
$$
Under the linearised gauge transformation $\delta|\Psi\rangle=Q|\Lambda\rangle$ we have, using the BRST charge (\ref{Q2}),
$$
Q|\Lambda\rangle=\int\rd k\left(\frac{i}{2}\, \tilde{c}_0c_1\,\alpha_0^2\,\lambda_{\mu}(k)\,\alpha_{-1}^{\mu}+i\tilde{c}_1c_1\,\alpha_{0\mu}\lambda_{\nu}(k)\alpha_{-1}^{\mu}\alpha_{-1}^{\nu}+ic_{-1}c_1\,\alpha^{\mu}_0\lambda_{\mu}(k)\right)|k\rangle.
$$
This gives the correct (momentum space) variation for $h_{\mu\nu}(k)$, which may be read off from the $\alpha_{-1}^{\mu}\alpha_{-1}^{\nu}c_1\tilde{c}_1$ coefficient. There are also terms proportional to $ \tilde{c}_0c_1$ and $c_{-1}c_1$, which have no origin in the first terms of (\ref{ansatz}) and so must correspond to the variation of terms denoted by $+...$ in (\ref{ansatz}). We introduce fields $f_{\mu}(k)$ and $e(k)$ to provide origins for these terms\footnote{The momentum space field $e(k)$ has no relation to the worldsheet Beltrami differential $e(z)$. Since these fields arise in quite different contexts, we hope that no confusion will arise.}. The simplest ansatz for the string field is then
\begin{equation}\label{ansatz2}
|\Psi\rangle=\int\rd k\left(-\frac{1}{2}h_{\mu\nu}(k)\,\alpha_{-1}^{\mu}\alpha_{-1}^{\nu}c_1\tilde{c}_1+\frac{1}{2}e(k)\,c_{-1}c_1+if_{\mu}(k)\,\alpha_{-1}^{\mu}\tilde{c}_0c_1\right)|k\rangle.
\end{equation}
We shall see that, in order to describe the graviton in the bosonic theory, no additional terms are required. Identifying $\alpha_{0\mu}|k\rangle=k_{\mu}|k\rangle$ and Fourier transforming to configuration space, the linearised gauge transformations may be read off as
\begin{equation}\label{g}
\delta h_{\mu\nu}(x)=\partial_{\mu}\lambda_{\nu}(x)+\partial_{\nu}\lambda_{\mu}(x),	\qquad	\delta f_{\mu}(x)=-\frac{1}{2}\Box\lambda_{\mu}(x),	\qquad	\delta e(x)=2\partial^{\mu}\lambda_{\mu}(x).
\end{equation}
Note that $tr(\delta h_{\mu\nu})=\delta e$ so we shall identify
\begin{equation}\label{e}
e(x)=\eta^{\mu\nu}h_{\mu\nu}(x).
\end{equation}
We shall see that is the correct identification when we compute the quadratic action.

In terms of worldsheet fields, the string field may be written as an off-shell CFT field
\begin{equation}\label{a}
\Psi(z)=\int\rd k\left(-\frac{1}{2}h_{\mu\nu}(k)\;P^{\mu}P^{\nu}c\tilde{c}+\frac{1}{2}e(k)\;\partial^2c c+if_{\mu}(k)\;P^{\mu}\partial \tilde{c}c\right)e^{ik\cdot X},
\end{equation}
where (\ref{ansatz2}) and (\ref{a}) are related by
$$
|\Psi\rangle=\lim_{z\rightarrow 0}\Psi(z)|0\rangle.
$$
This is a minimal string field, in that it contains all that is needed for a study of the quadratic action, which will be the topic of the next section. It is certainly plausible that other terms play a role in the interacting theory. For the considerations  we limit ourselves to here, this string field will be adequate.

For the quadratic action, especially in the bosonic theory, the oscillator decomposition (\ref{ansatz2}) is not too unwieldy; however, when we come to consider picture changing in the supersymmetric theory, the oscillator description can be a little involved and worldsheet field descriptions of the form (\ref{a}) are more useful. The gauge transformation of the worldsheet field is given by
$$
\delta\Psi(z)=\oint_{\cal C}\rd\omega j(\omega)\Psi(z).
$$
where the contour ${\cal C}$ surrounds the point $\omega=z$. The result is best computed using OPEs (\ref{OPE}), where the $\tilde{b}$ and $\tilde{c}$ ghosts satisfy the same OPE as the $b$ and $c$ ghosts, and reproduces the result (\ref{g}) found above.

\subsection{The action to quadratic order}\label{The action to quadratic order}

 A classic problem from conventional closed bosonic string theory is how to construct a quadratic term with the correct ghost number: The naive choice $\langle \Psi|Q|\Psi\rangle$ does not have the correct ghost number; however, the quadratic term  $\langle \Psi|c_0^-Q|\Psi\rangle$, where $c_0^{\pm}=c_0\pm\bar{c}_0$, does have the correct ghost number. The condition $L^-_0|\Psi\rangle=0$, wich does not arise from the equations of motion, must be imposed as an additional constraint on the string field and is supplemented by the condition $b^-_0|\Psi\rangle=0$. The $L_0^-=0$ condition is simply level matching. In the Siegel gauge $b_0^+|\Psi\rangle=0$, the quadratic action is $\langle \Psi|c_0^-c_0^+L_0^+|\Psi\rangle$ and the linearised equation of motion gives $L_0^+|\Psi\rangle=0$.

As discussed in section \ref{The action and spacetime gauge symmetries} a similar story holds for the ambitwistor string; however, the idea is modified in an important way. In the ambitwistor string field theory the role of $c_0^-$ and $L^-_0$ in the conventional closed string field theory are played by $c_0$ and ${\cal L}_0$ respectively, where ${\cal L}_0$ is given by (\ref{L}). In the ambitwistor string field theory the kinetic term thus takes the form
\begin{equation}\label{quad}
S_2[\Psi]=\langle \Psi|c_0Q|\Psi\rangle=\langle R_{LR}|c_0Q|\Psi_L\rangle|\Psi_R\rangle.
\end{equation}
and we require
$$
{\cal L}_0|\Psi\rangle=0,	\qquad	b_0|\Psi\rangle=0,
$$
as part of the definition of the string field $|\Psi\rangle$. The equation of motion depends on the spacetime metric through the $H(z)$ dependence in $Q$ and receives perturbative corrections through non-linear interaction terms as one might expect.

\subsubsection{The quadratic action}\label{The quadratic action}

 We shall see that the bosonic theory gives the standard Fierz-Pauli action of linearised gravity at quadratic order. At higher order we do not expect the bosonic ambitwistor string field theory to reproduce Einstein gravity as the on-shell scattering amplitudes, from which the surface states $\langle\Sigma|$ are constructed, are known not to be those of Einstein gravity \cite{Mason:2013sva}.

We take the quadratic action to be (\ref{quad}). An important point is that, for the bosonic theory, we shall not take $\langle \Psi|$ to be the usual BPZ conjugate of $|\Psi\rangle$. The standard BPZ conjugate is given by
$$
\int\rd k\langle \text{-}k|\left(-\frac{1}{2}h_{\mu\nu}(k)\;\alpha_{1}^{\mu}\alpha_{1}^{\nu}c_{-1}\tilde{c}_{-1}+\frac{1}{2}e(k)\;c_{1}c_{-1}-if_{\mu}(k)\;\alpha_{1}^{\mu}\tilde{c}_0c_{-1}\right).
$$
Instead, we define $\langle\Psi|$ as
$$
\langle\Psi|=\int\rd k\langle \text{-}k|\left(-\frac{1}{2}h_{\mu\nu}(k)\;\tilde{\alpha}_{1}^{\mu}\tilde{\alpha}_{1}^{\nu}\tilde{c}_{-1}c_{-1}+\frac{1}{2}e(k)\;\tilde{c}_{1}\tilde{c}_{-1}-if_{\mu}(k)\;\tilde{\alpha}_{1}^{\mu}\tilde{c}_0\tilde{c}_{-1}\right).
$$
The motivation for introducing such an operation is that the standard BPZ conjugate does not give a non-trivial quadratic action. Notice that this is the only place in which $\langle\Psi|$ appears in the action. An explicit expression for the reflector state is easily deduced\footnote{One could relate these two conjugates by introducing an operator ${\cal O}$ which maps oscillator operators as ${\cal O}:(\alpha_{\pm 1},c_{\pm 1},\tilde{c}_{\pm 1})\rightarrow (\tilde{\alpha}_{\pm 1},\tilde{c}_{\pm 1},c_{\pm 1})$ and has no effect on the $(\alpha_0,c_0,\tilde{c}_0)$.}. It must be stressed that this non-standard inner product is a feature of the bosonic theory only. The supersymmetric theory discussed later utilises the standard BPZ conjugate.

The $c_0$ term in the action ensures that the action has the correct ghost number for a closed string field theory, but it also projects out the $c_0{\cal L}_0|\Psi\rangle=0$ part of the BRST constraint, hence the imposition of ${\cal L}_0|\Psi\rangle=0$ is imposed as a separate condition.

\subsubsection{Recovering the Fierz-Pauli action}\label{Recovering the Fierz-Pauli action}

Substituting (\ref{ansatz2}) into (\ref{quad}), using the commutation relations and imposing the normalisation
$$
\langle k'|\tilde{c}_{-1}\tilde{c}_0\tilde{c}_1c_{-1}c_0c_1|k\rangle=\delta(k-k'),
$$
we find that
\begin{eqnarray}
S_2[\Psi]&=&\int\rd k\left(-\frac{1}{4}h_{\mu\nu}(-k)k^2h^{\mu\nu}(k)+2ih_{\mu\nu}(-k)k^{\mu}f^{\nu}(k)+\frac{1}{8}e(-k)k^2e(k)\right.\nonumber\\
&&-ie(-k)k^{\mu}f_{\mu}(k)-2f_{\mu}(-k)f^{\mu}(k)\Big).
\end{eqnarray}
The $f_{\mu}(k)$ have no kinetic term and so are auxiliary fields to be integrated out. In configuration space the linearised action is
\begin{eqnarray}
S_2[h,e]=\int\rd x\left(\frac{1}{4}h_{\mu\nu}\Box h^{\mu\nu}+2h_{\mu\nu}\partial^{\mu}f^{\nu}-\frac{1}{8}e\Box e-e\partial^{\mu}f_{\mu}-2f_{\mu}f^{\mu}\right),
\end{eqnarray}
where all fields are functions of $x$. The $f_{\mu}$ equation of motion is
\begin{equation}\label{fm}
f_{\mu}=-\frac{1}{2}\left(\partial^{\nu}h_{\mu\nu}-\frac{1}{2}\partial_{\mu}e\right).
\end{equation}
Such a relationship could be inferred from the gauge transformation of the components of the string field (\ref{g}). In other words, given the gauge transformation of $h_{\mu\nu}$ and $e$, the correct transformation for $f_{\mu}$ could be inferred from this equation of motion. Substituting (\ref{fm}) back in for $f_{\mu}$ in the action and integrating by parts where required gives
\begin{equation}\label{FP}
S_2[h]=\int\rd x\left(\frac{1}{4}h_{\mu\nu}\Box h^{\mu\nu}+\frac{1}{2}(\partial^{\nu}h_{\mu\nu})(\partial_{\lambda}h^{\mu\lambda})+\frac{1}{2}h\partial_{\mu}\partial_{\nu}h^{\mu\nu}-\frac{1}{4}h\Box h\right),
\end{equation}
 where we have imposed the identification (\ref{e}) of $e(x)$ with the trace of the metric fluctuation $h:=\eta^{\mu\nu}h_{\mu\nu}$. The action (\ref{FP}) is precisely the Fierz-Pauli action \cite{Fierz:1939ix} for linearised gravity. Note that it is the background Minkowski metric $\eta_{\mu\nu}$ and its inverse which is being used to lower and raise indices. The naive imposition of a Siegel type gauge $\tilde{b}_0|\Psi\rangle=0$, imposes the condition $f_{\mu}(k)=0$ which, noting (\ref{fm}), is precisely the harmonic (or de Donder) gauge for linearised gravity\footnote{\cite{Ortin:2015hya,Feynman:1996kb} contain nice reviews of linearised gravity.}. In this gauge the $h_{\mu\nu}$ equation on motion is simply $\Box h_{\mu\nu}=0$, which is consistent with the proposed propagator discussed above.

    \subsection{Interactions}\label{Interactions}

In this section we illustrate how correlation functions, interpreted as forms $\Omega_{|\vec{\Psi}\rangle}(\vec{\nu})$ in a bundle over moduli space, give the basic ingredient in the interaction terms $\{\Psi^n\}$. Such forms (\ref{form1}), constructed using on-shell asymptotic string states played a central role in the study of the on-shell scattering amplitudes discussed in section \ref{Scattering Amplitudes and the Scattering Equations}. In this section we are interested in generalising such objects to off-shell correlation functions as these are one of the central ingredients in the constructing the $\{\Psi^n\}$ terms. We start by giving the briefest of overviews of how this works in the conventional bosonic string field theory \cite{Zwiebach:1992ie} before describing how this story must be modified for the bosonic ambitwistor string.

\subsubsection{Interactions in conventional bosonic string field theory}\label{Interactions in conventional bosonic string field theory}

Tangent vectors  to the moduli space $V^a$, where $a=1,2,...,n-3$, provide a convenient way to think geometrically about deformations of the worldsheet theory at the level of the moduli space. Such deformations may also be considered at the level of the worldsheet by the effect of the worldsheet vector fields $v^a_i(z)$, based around the $i$'th puncture on $\Sigma$, which also change the moduli. As such we can think of the $\vec{\nu}^a=(v^a_1,...,v^a_n)$ as functions of the $V^a$.  Given a set of deformations $\vec{\nu}^a$ corresponding to tangent vectors $V^a$ of the moduli space ${\cal M}_n$, we can define a correlation function $\Omega_{|\vec{\Psi}\rangle}$ by 
\begin{eqnarray}\label{form}
\Omega_{|\vec{\Psi}\rangle}(\vec{\nu})=\langle\Sigma|\mathbf{b}(\vec{\nu}_1)...
\mathbf{b}(\vec{\nu}_{2n-6})|\vec{\Psi}\rangle,
\end{eqnarray}
where $|\vec{\Psi}\rangle$ is shorthand for a product of $n$ states $|\Psi_i\rangle$ and the ghost insertions are
$$
\textbf{b}(\vec{\nu}^a)=\sum_{i=1}^n\left(\oint\rd z\,b^{(i)}(z)v^a_i(z)+\oint\rd \bar{z}\,\bar{b}^{(i)}(\bar{z})\bar{v}^a_i(\bar{z})\right).
$$
From the perspective of the moduli space, $\Omega_{|\vec{\Psi}\rangle}(\vec{\nu})$ is a multilinear function of $2n-6$ tangent vectors\footnote{In addition to the $v_i^a$ there are also the complex conjugate fields $\bar{v}^a_i$. This is in contrast to the ambitwistor string in which the $\bar{v}^a_i$ do not appear.} to ${\cal M}_n$. It is therefore tempting to think of $\Omega_{|\vec{\Psi}\rangle}(\vec{\nu})$ as a top form on the moduli space. If the states in $|\vec{\Psi}\rangle$ are on-shell, then the form $\Omega_{|\vec{\Psi}\rangle}(\vec{\nu})$ in (\ref{form}) is indeed a well-defined top form on the moduli space ${\cal M}_n$ and the integral of $\Omega_{|\vec{\Psi}\rangle}(\vec{\nu})$ over ${\cal M}_n$ is a well-defined object. This is the case in first quantised string theory and is a key ingredient of the operator formalism for the conventional bosonic string \cite{AlvarezGaume:1988bg}. A detailed discussion of how the form (\ref{form1}) amounts to a measure on moduli space, in the case where the states are on-shell, may also be found in \cite{AlvarezGaume:1988bg}. This case closely parallels the discussion in section \ref{Scattering Amplitudes and the Scattering Equations} for the on-shell ambitwistor string.
 
More pertinant is the situation when the states in $|\vec{\Psi}\rangle$ are not on-shell. In this case $\Omega_{|\vec{\Psi}\rangle}(\vec{\nu})$ depends on the local coordinates $t_i$ defined about the punctures where the states $|\Psi_i\rangle$ are inserted. $\Omega_{|\vec{\Psi}\rangle}$ is then not well-defined on the moduli space; however,  $\Omega_{|\vec{\Psi}\rangle}$ is well defined on ${\cal P}_n$, the bundle over moduli space with base ${\cal M}_n$ and (infinite-dimensional) fibres given by the choice of local coordinate at each puncture. In fact,  $\Omega_{|\vec{\Psi}\rangle}$ descends to a well-defined form on the bundle $\widehat{\cal P}_n$ over ${\cal M}_n$ with (infinite-dimensional) fibres $\mathscr{T}$ given by a choice of local coordinate about each puncture up to a puncture-dependent phase $t_i\sim e^{i\theta_i}t_i$.
 $$
\begin{array}{ccc}
\mathscr{T} & \hookrightarrow & \widehat{\cal P}_n \\
& & \downarrow\\
& & {\cal M}_n
\end{array}
$$
Details may be found in \cite{Zwiebach:1992ie}. Since $\widehat{\cal P}_n$ is infinite-dimensional, $\Omega_{|\vec{\Psi}\rangle}(\vec{\nu})$ is no longer a top form but it can be integrated over $2n-6$ dimensional regions of $\widehat{\cal P}_n$.

\subsubsection{Ambitwistor interactions as forms}\label{Ambitwistor interactions as forms}

We now address how this story changes in the ambitwistor case. The relevant starting point is a form akin to (\ref{form}) with the exception that we do not want to include an anti-holomorphic sector as was the case for the conventional bosonic string. Instead it is clear that, in order to recover the correct scattering amplitudes (\ref{amplitude}), we must include a string of $n-3$ $\tilde{\mathbf{b}}(\vec{\nu}^a)$ ghost insertions. We also need to include the same number of $\bar{\delta}({\cal H}(\vec{\nu}^a))$ insertions. This suggests the generalisation of (\ref{form}) to
\begin{equation}\label{form2}
\Omega_{|\vec{\Psi}\rangle}(\vec{\nu})=\langle\Sigma |B_{n-3}(\vec{\nu})|\vec{\Psi}\rangle
\end{equation}
where $B_{n-3}(\vec{\nu})$ is given by (\ref{B}). How should we think about this correlation function? For on-shell momentum eigenstates $|\Psi\rangle$, this is simply the integrand of the on-shell scattering amplitude (\ref{amplitude}), a form on $T^*{\cal M}_n$. It is useful to think of the additional $\tilde{\mathbf{b}}$ insertions as describing the moduli associated with the Beltrami differential $e(z)$ in the worldsheet theory. These additional directions would then describe a space ${\cal N}_n\subset T^*{\cal M}_n$ which is the bundle over ${\cal M}_n$
 $$
{\cal N}_n\xrightarrow{\pi}{\cal M}_n,
$$
where the fibres of ${\cal N}_n$ are $n-3$ dimensional and describe the moduli of $e(z)$. One can see hints of this bundle structure in the algebra (\ref{algebra}) and a related construction has previously been noted, from a different perspective, in \cite{Ohmori:2015sha}. For on-shell $|\Psi\rangle$, the form (\ref{form2}) is well-defined on $T^*{\cal M}_n$. To determine the on-shell amplitude, we pick a section of ${\cal N}_n$ and formally integrate over the base ${\cal M}_n$
\begin{equation}\label{form3}
\int_{{\cal M}_n}\Omega_{|\vec{\Psi}\rangle}(\vec{\nu}).
\end{equation}
We argue that the choice of section does not matter and so the integral above is well-defined.  We anticipate that a general infinitesimal displacement in $T^*{\cal M}_n$, parametrised by the worldsheet vector $v(z)$, alters the surface state as
$$
\delta_{\vec{\nu}}\langle\Sigma|=\langle\Sigma|{\cal T}(\vec{\nu})+\langle\Sigma|{\cal H}(\vec{\nu}),
$$
which generalises the conventional bosonic string result \cite{AlvarezGaume:1988bg}. ${\cal T}(\vec{\nu})$ generates a displacement in the base ${\cal M}_n$, whilst  ${\cal H}(\vec{\nu})$ generates a displacement in the fibres of $T^*{\cal M}_n$. The $\delta\big({\cal H}(\vec{\nu})\big)$ insertions in $\Omega_{|\vec{\Psi}\rangle}(\vec{\nu})$ kill the ${\cal H}(\vec{\nu})$ component in $\delta_{\vec{\nu}}\langle\Sigma|$, giving
$$
\delta_{\vec{\nu}}\langle\Sigma|\delta\big({\cal H}(\vec{\nu})\big)=\langle\Sigma|{\cal T}(\vec{\nu})\delta\big({\cal H}(\vec{\nu})\big),
$$
so that, for a general displacement in $T^*{\cal M}_n$, only the change in the base coordinate gives rise to a change in $\Omega_{|\vec{\Psi}\rangle}(\vec{\nu})$. It appears that deformations in the fibre directions preserve $\Omega_{|\vec{\Psi}\rangle}(\vec{\nu})$ and so we may formally integrate over the base\footnote{Of course there may be global issues that we have not considered here.} and (\ref{form3}) is well-defined\footnote{It would be interesting to see how this relates to the Morse theory and localisation results in \cite{Ohmori:2015sha} which show how the expression for the on-shell scattering amplitude as integral of a form over a half-dimensional cycle $\Gamma_n\subset T^*{\cal M}_n$ formally reduces to an integral over the moduli space.}.

The generalisation to the off-shell case is now straightforward. We define an infinite dimensional bundle ${\cal A}_n$ with base $T^*{\cal M}_n$ and fibres given by a choice of local coordinate about each puncture. As in the conventional bosonic string, imposing the identification $z_i\sim e^{i\theta_i}z_i$ reduces us from ${\cal A}_n$ to a bundle which we shall refer to as $\widehat{\cal A}_n$.
 $$
\begin{array}{ccc}
\mathscr{T} & \hookrightarrow & \widehat{\cal A}_n \\
& & \downarrow\\
& & T^*{\cal M}_n
\end{array}
$$
For on-shell and off-shell states the form (\ref{form2}) is well-defined on both ${\cal A}_n$ and, more importantly, $\widehat{\cal A}_n$. In practice, we are interested in a $2n-6$ dimensional cycle in $T^*{\cal M}_n$ which we can formally identify as a copy of ${\cal M}_n$ in our expressions. As such, we might formally use $\widehat{\cal P}_n$ in place of $\widehat{\cal A}_n$. The vertices would then be formally defined in a way analogous to the conventional string field in terms of the vertices ${\cal V}_n\subset\widehat{\cal P}_n$ as
\begin{equation}\label{form4}
\{\Psi^n\}=\int_{{\cal V}_n}\Omega_{|\vec{\Psi}\rangle}(\vec{\nu}),
\end{equation}
where ${\cal V}_n\subset\widehat{\cal P}_n$ were discussed in \ref{Feynman rules and fundamental vertices} and the integrand is given by (\ref{form2}).

Whilst (\ref{form4}) provides a suitable formal generalisation of the vertex of the conventional closed string to the ambitwistor case and avoids the complications of dealing with a middle-dimensional cycle in $T^*{\cal M}$ directly, it may not be the most convenient description of the vertex for the calculation of perturbative amplitudes. 

In section \ref{new}, we saw that the holomorphic delta-functions appearing in the integrands of scattering amplitudes can be viewed as a formal device to capture the result of the Morse Theory prescription of \cite{Ohmori:2015sha}. For the first quantised operator formalism, the approach of \cite{Ohmori:2015sha} is not necessary and one can work entirely in terms of the moduli space ${\cal M}$, provided one is willing to accept the origin of these delta-functions in the surface states from gauge-fixing the ghosts. By contrast, in the second quantised theory the gluing of surface states via a propagator is important and additional holomorphic delta-functions must appear associated with the moduli of this propagator. It is unclear how to incorporate holomorphic delta functions directly into the propagator (\ref{prop}) that appears in the ambitwsitor string and so it seems necessary to work at the level of $T^*{\cal M}$ as proposed by \cite{Ohmori:2015sha}. The delta-functions then emerge, as in the first quantised case, as a formal device to compactly write the result of a Morse Theory evaluation of the amplitude. As such it is more useful to work with the generalised vertex where we integrate not over ${\cal V}_n\subset\widehat{\cal P}_n$, but rather over a middle-dimensional cycle $\Gamma_n\subset \widehat{\cal A}_n$, fixed by arguments similar to those of \cite{Ohmori:2015sha}. What is really required then is
\begin{equation}
\{\Psi^n\}=\int_{\Gamma_{{\cal V}_n}}\widetilde{\Omega}_{|\vec{\Psi}\rangle}(\vec{\nu}),
\end{equation}
where $\widetilde{\Omega}$ involves the surface state $\langle\widetilde{\Sigma}|$ as discussed in section \ref{new} and the holomorphic delta-function insertions are omitted. To complete the description of the vertex we would need a prescription to construct the $\Gamma_{{\cal V}_n}\subset T^*{\cal V}_n$ or, at the least, a way to determine the contribution of this term to the fixed points $\tau^*$ of any observable to which this vertex contributes. The naive choice of using the Morse function associated with $\widetilde{\Omega}_{|\vec{\Psi}\rangle}(\vec{\nu})$ and considering only those critical points contained in $T^*{\cal V}_n$ is appealing but requires further investigation.

\subsubsection{A brief sketch of perturbation theory}\label{A brief sketch of perturbation theory}

In this section we briefly outline the approach to perturbation theory in this formalism. The key ingredients, already discussed, are the propagator and the interaction vertices. The story closely follows that of the conventional bosonic string but, as many other authors have found \cite{Adamo:2014wea,Adamo:2013tsa,Geyer:2015jch,Geyer:2015bja,Reid-Edwards:2015stz}, the ambitwistor string differs in many important respects from the conventional string. We shall only give a sketch of the perturbation theory, giving a more complete treatment elsewhere. Although we focus on the bosonic fields, the general discussion also applies to the bosonic sector of the supersymmetric theory. In fact there is evidence that this construction will only work in the context of the supersymmetric theory. As such, this section should be read with a view to later application to the superstring field theory constructed in the following sections.

An important role is played by gluing lower point surfaces together. The picture we have is of an $n$-punctured Riemann surface $\Sigma_n$ constructed from a propagator connecting two Riemann surfaces which we denote by $\Sigma_L$ and $\Sigma_R$. These Riemann surfaces have $n_L$ and $n_R$ punctures respectively, where $n-2=n_L+n_R$.  The real dimensions of the moduli spaces of these Riemann surfaces are $dim({\cal M}_{L})=2n_L-6$ and $dim({\cal M}_{R})=2n_R-6$ and so for the moduli space of the $n$-punctured Riemann surface to be correct, the propagator must carry one complex modulus. This modulus is denoted by $q$ and appears in the gluing of local coordinates $z_L$ and $z_R$ in the regions of the propagator on $\Sigma_L$ and $\Sigma_R$ respectively as $z_Lz_R=q$. As we have seen in the ambitwistor string, each modulus comes with a holomorphic delta-function insertion. This raises the question of how the appropriate holomorphic delta-function associated with the modulus carried by the propagator arises from the propagator expression (\ref{prop}). A complete understanding of the ambitwistor string propagator is still lacking, although some recent progress has been made \cite{Ohmori:2015sha,Li:2017emw,Casali:2017zkz}. Here, we shall see that the perspective of \cite{Ohmori:2015sha} provides a more natural framework in which to understand how the holomorphic delta-function associated with the propagator modulus arises in perturbation theory.

The first step in any perturbation theory is to fix the spacetime gauge symmetries. As discussed in sections \ref{Target space gauge symmetries} and \ref{Recovering the Fierz-Pauli action}, there is a simple analogue of the Siegel gauge appropriate for the ambitwistor string; $\tilde{b}_0|\Psi\rangle=0$. The kinetic term then becomes
$$
S_2[\Psi]=\langle {\cal R}_{LR}|c_0\tilde{c}_0\widetilde{L}_0|\Psi_L\rangle|\Psi_R\rangle,
$$
where $\langle {\cal R}_{LR}|$ and $| {\cal R}_{LR}\rangle$ are appropriate reflection states. In the calculation of the quadratic action we used string fields that satisfied the constraint ${\cal L}_0=0$ or, focussing on the matter sector for simplicity, $L_0=2$. This suggests the propagator\footnote{An analysis of degenerating Riemann surfaces in the ambitwistor string in \cite{Ohmori:2015sha} gave a result which also included contributions for sectors of other conformal weights which gave rise to spurious singularities in the bososnic theory. In the treatment presented here, the constraint ${\cal L}_0|\Psi\rangle=0$ projects out such sectors and so such terms do not appear in our Siegel gauge propagator.}
$$
\tilde{b}_0b_0\frac{\delta(L_0-2)}{\widetilde{L}_0}| {\cal R}_{L,R}\rangle,
$$
where $\delta(L_0-2)$ projects onto states for which $L_0=2$ and the subscripts $L,R$ denote the Hilbert spaces to the left and right of the propagator. Given that we worked with string fields $|\Psi\rangle$ such that $L_0=2$, it is possible that a more general function of $L_0$ could also appear in the propagator, to which the analysis presented in section \ref{The Ambitwistor string field} would only record the appropriate factors of $2$. As mentioned briefly in section \ref{new}, we may appeal to a possibly more elegant way to understand the scattering amplitudes of the ambitwsitor string is in terms of localisation and Morse theory \cite{Ohmori:2015sha}. A study of factorisation limits in \cite{Ohmori:2015sha} suggests a propagator of the form
\begin{equation}\label{Oprop}
\int\rd s\rd\tilde{s}\;b_0\tilde{b}_0\;e^{-\{Q,sb_0+\tilde{s}\tilde{b}_0\}}| {\cal R}_{L,R}\rangle,
\end{equation}
which may be written as
$$
\frac{\tilde{b}_0b_0}{\widetilde{L}_0L_0}| {\cal R}_{L,R}\rangle.
$$
When considering the quadratic action we imposed $L_0=2$ as a condition on the string field and so, on the support of the projection $\delta(L_0-2)$ this propagator is, up to an overall factor, equivalent to
$$
\frac{\tilde{b}_0b_0}{\widetilde{L}_0}| {\cal R}_{L,R}\rangle.
$$
Thus, the quadratic action explored in section \ref{The action to quadratic order} is consistent with the propagator (\ref{Oprop}). We shall find that it is the form of the propagator given by (\ref{Oprop}) that is most useful in understanding perturbation theory in the context of the perspective of the scattering amplitudes derived using localisation in \cite{Ohmori:2015sha}.

It is tempting to interpret (\ref{Oprop}), supplemented with a projection onto $L_0=2$ states, as a closed string propagator for the bundle ${\cal Y}$. The projection onto the base $\Sigma$ is of the form (\ref{oprop}) in which all of the anti-holomorphic dependence has been suppressed and $q=e^{-s}$. The procedure outlined in \cite{Ohmori:2015sha} requires a choice of cycle $\Gamma\subset T^*{\cal M}$ which excludes the anti-holomorphic contributions from consideration. The additional $\tilde{s}\{Q,\tilde{b}_0\}$ contribution deals with propagation of the fibres of ${\cal Y}$. As already mentioned, the question of how to correctly understand the ambitwistor string propagator is an open one we shall not offer a significantly new perspective to the discussion here.

We now outline the contributions to the scattering amplitude. There will be contributions from integrating over the fundamental `missing' regions ${\cal D}_n$ of moduli space and those contributions coming from regions of the moduli space constructed from $m<n$ point vertices joined by propagators. Alternatively, we can consider these contributions as coming from a middle-dimensional region $\Gamma_{{\cal D}_n}\subset T^*{\cal D}_n$. The contributions from the fundamental regions are simply
\begin{eqnarray}\label{D}
M_n^{\Gamma_{{\cal D}_n}}&=&\int_{{\cal D}_n}\prod_{a=1}^{n-3}\rd\tau_a\langle\Sigma_n|\prod_{a=1}^{n-3}\tilde{\mathbf{b}}(\vec{\nu}^a)\mathbf{b}(\vec{\nu}^a)\bar{\delta}\Big({\cal H}(\vec{\nu}^a)\Big)|\vec{\Psi}_n\rangle\nonumber\\
&=&\int_{\Gamma_{{\cal D}_n}\subset T^*{\cal D}_n}\left\langle e^{-\{Q,\mathbf{b}\}}e^{-\{Q,\tilde{\mathbf{b}}\}}\vec{\Psi}_n \right\rangle_{S_0},
\end{eqnarray}
which gives the standard scattering amplitude integrand but integrated only over the region ${\cal D}_n\subset{\cal M}_n$ rather than the full moduli space, where the $\tau^a$ are (holomorphic) coordinates on ${\cal M}_n$. The other contributions to the scattering amplitude come from terms constructed using lower interaction terms glued together by propagators.

We consider next the contribution given by gluing pairs of lower point vertices by a single propagator. Working on $T^*{\cal M}$, we shall write the propagator in the form (\ref{Oprop}). Such terms take the form
$$
M_n^{{\cal R}_1}=\sum_{\sigma, \{n_L,n_R\}}\int_{\Gamma_L\in T^*{\cal M}_L}\int_{\Gamma_R\in T^*{\cal M}_R} \int \rd s\rd \tilde{s}\langle\widetilde{\Sigma}_L|\langle\widetilde{\Sigma}_R|\;e^{-\{Q,sb_0+\tilde{s}\tilde{b}_0\}}|{\cal R}_{L,R}\rangle |\vec{\Psi}_L\rangle |\vec{\Psi}_R\rangle,
$$
where the sum denotes a double sum over all $\{n_L,n_R\}$ such that $n_L+n_R-2=n$ and $n_L,n_R\geq 3$, and $\sigma$ denotes a sum over all permutations of external states. $|\vec{\Psi}_L\rangle=|\Psi_1\rangle|\Psi_2\rangle...|\Psi_{n_L-1}\rangle$ is a product of asymptotic states located at each of the punctures of $\Sigma_L$ not connected to the propagator and similarly for  $|\vec{\Psi}_R\rangle$. Inserting a complete set of states $\Phi_{L,R}$ and using $\langle\widetilde{\Sigma}||\Psi_1\rangle...|\Psi_n\rangle=\langle e^{-\{Q,{\cal W}\}}\Psi_1...\Psi_n\rangle_{S_0}$, we may write this in terms of correlation functions on the component Riemann surfaces
\begin{eqnarray}
M_n^{{\cal R}_1}&=&\sum_{\sigma, \{n_L,n_R\}}\int_{\Gamma_{{\cal R}_1}}\sum_{\Phi_L,\Phi_R}\left\langle e^{-\{Q,\mathbf{b}(u_L)\}}e^{-\{Q,\tilde{\mathbf{b}}(u_L,\tilde{u}_L)\}} \vec{\Psi}_L \Phi_L\right\rangle_{S_0}\nonumber\\
&&\times\langle\Phi_L| \;e^{-\{Q,sb_0+\tilde{s}\tilde{b}_0\}}|\Phi_R\rangle\left\langle e^{-\{Q,\mathbf{b}(u_R)\}}e^{-\{Q,\tilde{\mathbf{b}}(u_R,\tilde{u}_R)\}}\Phi_R\vec{\Psi}_R\right\rangle_{S_0}\nonumber,
\end{eqnarray}
where the union of the integration regions has been written as $\Gamma_{{\cal R}_1}$ for simplicity. Following \cite{Ohmori:2015sha}, this may be written as the correlation function on the $n=n_L+n_R-2$ punctured Riemann surface $\Sigma_n$ as
\begin{equation}\label{R1}
M_n^{{\cal R}_1}=\sum_{\sigma, \{n_L,n_R\}}\int_{\Gamma_{{\cal R}_1}} \left\langle e^{-\{Q,\mathbf{b}(u)\}}e^{-\{Q,\tilde{\mathbf{b}}(u,\tilde{u})\}}\vec{\Psi},_n\right\rangle_{S_0}
\end{equation}
where $(u_L,\tilde{u}_L)$, $(u_R,\tilde{u}_R)$, and $(s,\tilde{s})$ have been combined into $(u,\tilde{u})$ for simplicity of notation. The $\Gamma_{{\cal R}_1}$ is a half-dimensional cycle in $T^*{\cal M}_L\times T^*{\cal M}_R\times\C$, fixed by Morse theory. This result seems to imply that the surface states obey an analogue of the generalised glueing and re-smoothing theorem $\langle\widetilde{\Sigma}_n|=\langle\widetilde{\Sigma}_{n_L}|\langle\widetilde{\Sigma}_{n_R}|e^{-\{Q,sb_0+\tilde{s}\tilde{b}_0\}}|{\cal R}_{LR}\rangle$, adapted to glue the fibres of the bundle ${\cal Y}$ as well as the base $\Sigma$\footnote{There is a question here of what the correct form of the reflector state $|{\cal R}_{LR}\rangle$ is. The observation that the conventional BPZ conjugation does not lead to a non-trivial quadratic action in the bosonic theory suggests a non-conventional reflector state, involving $\tilde{\alpha}$ but not $\alpha$ modes, must be used. It is not clear that such a reflector state will give the identity suggested; however, in the supersymmetric case which is discussed in the following sections, prospects are much better. Indeed, the supersymmetric theory does require the conventional BPZ conjugate be used, giving rise to a reflector state of the conventional form (involving $\alpha$ and $\tilde{\alpha}$ modes). We hope to report on these issues elsewhere.} \cite{LeClair:1988sp,LeClair:1988sj}. This relationship seems reasonable, at least in the supersymmetric theory, but has not yet been proven directly in the operator formalism. The integrands in expressions (\ref{D}) and (\ref{R1}) have the same form and we take the final answer to be this integrand, integrated over the union of the regions of moduli space described by the separate components. In general there will be contributions from terms $A_n^{{\cal R}_m}$ built from a number of Riemann surface sub-units glued together using a number of propagators, with associated integration regions $\Gamma_{{\cal R}_m}$, leading to 
$$
M_n^{\Gamma_{{\cal D}_n}}+M_n^{{\cal R}_1}+M_n^{{\cal R}_2}+...= \int_{\Gamma_n}\left\langle e^{-\{Q,\mathbf{b}(u)\}}e^{-\{Q,\tilde{\mathbf{b}}(u,\tilde{u})\}}\vec{\Psi}_n\right\rangle_{S_0}
$$
where $\Gamma_n=\Gamma_{{\cal D}_n}\cup\Gamma_{{\cal R}_1}\cup\Gamma_{{\cal R}_2}+...\subset T^*{\cal M}$. The idea is that this final integral is then done using Morse Theory as outlined in \cite{Ohmori:2015sha} and may formally be written as a integration over the full moduli space ${\cal M}_n$, giving the amplitude in the more familiar form in terms of the scattering equations. The decomposition into a `missing' region and terms constructed using propagators and lower point vertices is most simply seen in the $n=4$ case, where the only contributions comes from the fundamental vertex $\{\Psi^4\}$ and two cubic terms $\{\Psi^3\}$ glued together by a propagator.
 
\section{The Supersymmetric Theory}\label{The Supersymmetric theory}

It was argued in \cite{Mason:2013sva} that the bosonic ambitwistor string does not to describe conventional Einstein gravity; however, a supersymmetric extension does describe Einstein \emph{super}gravity. Many extensions and generalisations of the ambitwistor string have been explored \cite{Casali:2015vta,Geyer:2014fka}, here we consider the simplest generalisation of extending the bosonic theory to an ${\cal N}=2$ (chiral) supersymmetric theory\footnote{The critical dimension on the ambitwistor string extended in this way is a positive integer only for ${\cal N}=2$ and ${\cal N}=4$, where the critical dimension is 10 and 2 respectively. As far as we know there has not, as yet, been a systematic study of the ${\cal N}=4$ case. It is also possible that the dimension counting for the ${\cal N}=4$ is not straightforward (cf. the conventional ${\cal N}=2$ string mentioned in section \ref{The N=2 String}).}. The theory has the symmetry (\ref{ambi}) under which the fermions transform trivially and also a natural extension of the bosonic conformal symmetry (\ref{conformal}) to a superconformal symmetry. After gauge fixing the worldsheet complex structure and the Beltrami differential $e(z)$, the ${\cal N}=2$ ambitwistor superstring has action
$$
S=\int_{\Sigma}P_{\mu}\bar{\partial}X^{\mu}+b\bar{\partial}c+\tilde{b}\bar{\partial}\tilde{c}+\eta_{\mu\nu}\psi^{\mu}\bar{\partial}\psi^{\nu}+\eta_{\mu\nu}\widetilde{\psi}^{\mu}\bar{\partial}\widetilde{\psi}^{\nu}+\chi P_{\mu}\psi^{\mu}+\widetilde{\chi} P_{\mu}\widetilde{\psi}^{\mu},
$$
where $\chi$ and $\widetilde{\chi}$ are the worldsheet gravitini and $\psi^{\mu}$ and $\widetilde{\psi}^{\mu}$ are holomorphic worldsheet spinors. We shall restrict to the case where the worldsheet spinors are Neveu-Schwarz (NS). The Ramond case is discussed briefly in section \ref{Discussion}.

As in the conventional string, the two gravitini $\chi$ and $\widetilde{\chi}$ may be gauge-fixed to vanish everywhere except at $n-2$ points, where we insert picture changing operators (PCOs). The usual Faddeev-Popov procedure results in the introduction of $(\beta,\gamma)$ and $(\widetilde{\beta},\widetilde{\gamma})$ superghost systems to gauge fix the $\chi$ and $\widetilde{\chi}$ respectively. The gauge-fixed action is then
\begin{equation}\label{sS}
S=\int_{\Sigma}P_{\mu}\bar{\partial}X^{\mu}+\eta_{\mu\nu}\psi^{\mu}\bar{\partial}\psi^{\nu}+\eta_{\mu\nu}\widetilde{\psi}^{\mu}\bar{\partial}\widetilde{\psi}^{\nu}+b\bar{\partial}c+\tilde{b}\bar{\partial}\tilde{c}+\beta\bar{\partial}\gamma+\widetilde{\beta}\bar{\partial}\widetilde{\gamma}.
\end{equation}
The non-trivial OPEs for the new fields are
$$
\psi^{\mu}(z)\psi^{\nu}(\omega)=\frac{\eta^{\mu\nu}}{z-\omega}+...,	\qquad	\beta(z)\gamma(\omega)=\frac{1}{z-\omega}+...,
$$
and similarly for the $(\tilde{\beta},\tilde{\gamma})$ and $\widetilde{\psi}^{\mu}$ fields. The OPEs for the fields that were present already in the bosonic theory are unchanged.

\subsection{Symmetries}

The gravitini act as Lagrange multipliers which impose the vanishing of the fermionic currents $G=P_{\mu}\psi^{\mu}$ and $\widetilde{G}=P_{\mu}\widetilde{\psi}^{\mu}$, which in turn generate the two worldsheet supersymmetries. As in the bosonic case, the stress tensor $T(z)$ generates the conformal transformations with the additional transformations of the worldsheet fermions
$$
\delta_vX^{\mu}=v\partial X^{\mu},	\qquad	\delta_vP_{\mu}=\partial(v P_{\mu}),	\qquad	\delta_v\psi^{\mu}=\frac{1}{2}(\partial v)\psi^{\mu}+v\partial\psi^{\mu}	\qquad	\delta_v\widetilde{\psi}^{\mu}=\frac{1}{2}(\partial v)\widetilde{\psi}^{\mu}+v\partial\widetilde{\psi}^{\mu}.
$$
The worldsheet spinors are invariant under the worldsheet gauge transformations generated by $H(z)$, which enforce the null condition $P^2(z)=0$ which acts on $X^{\mu}$ as $\tilde{\delta}_vX^{\mu}=v P_{\mu}$. The new ingredient is the ${\cal N}=2$ worldsheet supersymmetry. The $G(z)$ supercurrent generates the transformations
$$
\delta_{\epsilon}X^{\mu}=\epsilon\psi^{\mu},	\qquad	\delta_{\epsilon}\psi^{\mu}=\epsilon P^{\mu},	\qquad	\delta_{\epsilon}\widetilde{\psi}^{\mu}=0,	\qquad	\delta_{\epsilon}P_{\mu}=0,
$$
and the $\widetilde{G}(z)$ supercurrent generates the transformations
$$
\widetilde{\delta}_{\epsilon}  X^{\mu}=\epsilon\widetilde{\psi}^{\mu},	\qquad	\widetilde{\delta}_{\epsilon} \psi^{\mu}=0,	\qquad	\widetilde{\delta}_{\epsilon}  \widetilde{\psi}^{\mu}=\epsilon P^{\mu},	\qquad	\widetilde{\delta}_{\epsilon} P_{\mu}=0.
$$
Following on from ${\cal T}(\nu)$ and ${\cal H}(\nu)$ in the bosonic theory, it is useful to introduce the generators
$$
{\cal G}(\varepsilon)=\oint\rd z\, \varepsilon(z)\,G(z),
$$
and similarly for $\widetilde{\cal G}(\varepsilon)$, where $\varepsilon$ is a spin-valued worldsheet vector. The superalgebra is then easily deduced
$$
[{\cal T}(v_1),{\cal T}(v_2)]=-{\cal T}([v_1,v_2]),	\qquad	[{\cal T}(v_1),{\cal H}(v_2)]=-{\cal H}([v_1,v_2]),
$$
$$
[{\cal T}(v),{\cal G}(\varepsilon)]=-{\cal G}([v,\varepsilon]),	\qquad	[{\cal T}(v),\widetilde{\cal G}(\varepsilon)]=-\widetilde{\cal G}([v,\varepsilon]),
$$
$$
[{\cal G}(\varepsilon_1),{\cal G}(\varepsilon_2)]=-{\cal H}([\varepsilon_1,\varepsilon_2]),	\qquad	[\widetilde{\cal G}(\varepsilon_1),\widetilde{\cal G}(\varepsilon_2)]=-{\cal H}([\varepsilon_1,\varepsilon_2]),
$$
with all other commutators vanishing. Note here that, in the $[{\cal G},{\cal G}]$ commutator, ${\cal H}$ is playing the role of a worldsheet Hamiltonian.

\subsubsection{The SuperVirasoro Algebra}

With the exception of a brief discussion in section \ref{Discussion}, we shall restrict our attention to the Neveu-Schwarz sector. In terms of modes, the fermionic fields are written as
$$
\psi^{\mu}(z)=\sum_{r\in\Z+\frac{1}{2}}\psi^{\mu}_rz^{-r-\frac{1}{2}},	\qquad
G(z)=\sum_{r\in\Z+\frac{1}{2}}G_rz^{-r-\frac{3}{2}},
$$
and similarly for $\widetilde{\psi}^{\mu}(z)$ and $\widetilde{G}(z)$. The modes of the fermionic currents are given by
$$
G_r=\sum_{n\in\Z}\alpha_{n\mu}\;\psi^{\mu}_{r-n},	\qquad	
\widetilde{G}_r=\sum_{n\in\Z}\alpha_{n\mu}\; \widetilde{\psi}^{\mu}_{r-n}.
$$
The superalgebra is given in terms of these modes by
$$
[L_m,L_n]=(m-n)L_{m+n}+\delta_{m+n,0}\frac{D}{6}m(m^2-1),	\qquad	[L_m,\widetilde{L}_n]=(m-n)\widetilde{L}_{m+n},	\qquad	[\widetilde{L}_m,\widetilde{L}_n]=0,
$$
$$
[L_m,G_r] = \frac{(m-2r)}{2}\,G_{m+r},  \qquad  [L_m,\widetilde{G}_r] = \frac{(m-2r)}{2}\,\widetilde{G}_{m+r},
$$
$$
\{G_r,G_s \}=2\,\widetilde{L}_{r+s},	\qquad\qquad	\{G_r,\widetilde{G}_s\}=0,	\qquad\qquad	\{\widetilde{G}_r,\widetilde{G}_s\}=2\widetilde{L}_{r+s}.
$$
The matter stress tensor $T(z)$ includes contributions from the fermions $\psi^{\mu}$ and $\widetilde{\psi}^{\mu}$ as well as the $(X^{\mu},P_{\mu})$ system; whereas, $H(z)$ is identical to that in the bosonic theory. It is a simple exercise in central charge bookkeeping \cite{Mason:2013sva} to show that the critical dimension of the supersymmetric theory is $10$.

\subsubsection{The ${\cal N}=2$ String}\label{The N=2 String}

We note in passing that a number of useful comparisons \cite{Mason:2013sva,Casali:2016atr} have been made between the ambitwistor string and the holomorphic sector of the conventional type II string but in many ways there are also similarities with the less frequently discussed ${\cal N}=2$ string \cite{Ademollo:1976wv,Ademollo:1975an,Marcus:1992wi}. Upon gauge-fixing, this theory has action
$$
S=-\frac{1}{2\pi}\int\rd^2z \partial_{\alpha} X^{\mu}\partial^{\alpha}\overline{X}_{\mu}-i\bar{\psi}\rho^{\alpha}\partial_{\alpha}\psi,
$$
where the target space is complexified $X^{\mu}=X_1^{\mu}+iX^{\mu}_2$ and the fermions naturally appear in complex pairs $\psi^{\mu}=\psi_1^{\mu}+i\psi^{\nu}_2$. The critical dimension is two complex (four real) dimensions\footnote{The history of the dimension counting of this theory is a little convoluted, as recounted in \cite{Marcus:1992wi}. The target space has (4,0) or (2,2) signature.}. The oscillator algebra obeys
$$
[\alpha_m^{\mu},\bar{\alpha}_n^{\nu}]=m\delta_{m+n}\eta^{\mu\nu},	\qquad	[\alpha_m^{\mu},\alpha_n^{\nu}]=0,	\qquad	[\bar{\alpha}_m^{\mu},\bar{\alpha}_n^{\nu}]=0,
$$
which can be compared with the algebra of the modes of the $X^{\mu}(z)$ and $P_{\mu}(z)$ fields in the ambitwistor string. Also the super-Virasoro algebra has many similarities to the ambitwistor string. However, the target space theory of the ${\cal N}=2$ string contains self-dual gravity rather than Einstein gravity and so a detailed comparison may not prove fruitful.

\subsection{BRST Operator}\label{BRST Operator}

As with the bosonic string field theory, the crucial ingredients in the construction of the supersymmetric ambitwistor string field theory are; the string field $|\Psi\rangle$, the surface state $\langle\Sigma|$ (which contains information about interactions), and the BRST charge $Q$ (which provides the propagator of the theory). We shall consider the surface state $\langle\Sigma|$ in later sections, here we focus on the BRST charge $Q$ and how it may be used to constrain the form of the string field. The BRST charge may be written in terms of the current $j(z)$ where
$$
Q=\oint\rd z\,j(z).
$$
For the ${\cal N}=2$ ambitwistor string under consideration, the BRST current is given by
$$
j(z)=c\left(T_m+T_{\beta\gamma}+\widetilde{T}_{\beta\gamma}\right)+\gamma G+\tilde{\gamma}\widetilde{G}+bc\partial c+\tilde{b}\tilde{c}\partial\tilde{c}+\frac{1}{2}\gamma^2\tilde{b}+\frac{1}{2}\tilde{\gamma}^2\tilde{b}+\tilde{c}H,
$$
where $T_{\beta\gamma}$ and $\widetilde{T}_{\beta\gamma}$ are superghost stress tensors and the matter stress tensor now includes contributions from the worldsheet fermions $T_m(z)=P_{\mu}\partial X^{\mu}+ \psi^{\mu}\partial\psi_{\mu}+ \widetilde{\psi}^{\mu}\partial\widetilde{\psi}_{\mu}$.
The currents $G(z)$ and $\widetilde{G}(z)$ where given in the previous section. In terms of oscillator modes the relevant terms in the BRST charge are
\begin{eqnarray}\label{sQ}
Q&=&c_0{\cal L}_0+\frac{1}{2}\tilde{c}_0\alpha_0^2+\frac{1}{2}\tilde{c}_0\alpha_{-1}\cdot\alpha_1+\alpha_0\cdot(c_1\tilde{\alpha}_{-1}+c_{-1}\tilde{\alpha}_1+\tilde{c}_{-1}\alpha_1+\tilde{c}_1\alpha_{-1}) \nonumber\\
&&-2b_0c_{-1}c_1+2\tilde{b}_0(c_1\tilde{c}_{-1}+\tilde{c}_{-1}c_{-1})+\tilde{c}_0(c_{-1}\tilde{b}_1+c_1\tilde{b}_{-1}) \nonumber\\
&&+ \gamma_{-\frac{1}{2}}\alpha_0\cdot\psi_{\frac{1}{2}}+ \gamma_{\frac{1}{2}}\alpha_0\cdot\psi_{-\frac{1}{2}} +\widetilde{\gamma}_{-\frac{1}{2}}\alpha_0\cdot\widetilde{\psi}_{\frac{1}{2}}+\widetilde{\gamma}_{\frac{1}{2}}\alpha_0\cdot\widetilde{\psi}_{-\frac{1}{2}}
\nonumber\\
&&-2\tilde{b}_0( \gamma_{-\frac{1}{2}}\gamma_{\frac{1}{2}}+ \tilde{\gamma}_{-\frac{1}{2}}\tilde{\gamma}_{\frac{1}{2}})+...
\end{eqnarray}
where the $+...$ denotes terms that depend on oscillator modes that commute with all oscillators that will appear in the string field. As in the bosonic case, the BRST charge appears in the quadratic part of the string field action multiplied by the ghost zero mode $c_0$. This means that all terms in $Q$ involving $c_0$ are projected out of the quadratic part of the action and we must therefore deal with these terms separately. In the above oscillator expansion of $Q$ those terms that multiply a $c_0$ factor have been isolated and written as\footnote{We used ${\cal L}_0$ to denote the corresponding object in the bosonic string field. From this point on ${\cal L}_0$ refers to (\ref{L2}).} ${\cal L}_0$. Since that part of the constraint given by ${\cal L}_0$ cannot be imposed on-shell by the string field equations of motion, since it is projected out of the quadratic action, this constraint must be imposed on the string field directly and may be seen, as in the bosonic case, as part of the definition of the string field $|\Psi\rangle$.  Thus we would like a superstring field such that
$$
{\cal L}_0|\Psi\rangle=0,	\qquad	b_0|\Psi\rangle=0.
$$
To show that these conditions are naturally satisfied by a reasonable $|\Psi\rangle$ we need an explicit expression for the superstring field. Finding such an explicit expression will be the task of the next section. The ${\cal L}_0$ operator which is required to annihilate the string field is given by
\begin{eqnarray}\label{L2}
{\cal L}_0&=&(\alpha_{-1}\cdot\tilde{\alpha}_1+\tilde{\alpha}_{-1}\cdot\alpha_1)+\frac{1}{2}( \psi_{-\frac{1}{2}} \cdot  \psi_{\frac{1}{2}} + \widetilde{\psi}_{-\frac{1}{2}} \cdot \widetilde{\psi}_{\frac{1}{2}} )+(b_{-1}c_1+c_{-1}b_1)+(\tilde{b}_{-1}\tilde{c}_1+\tilde{c}_{-1}\tilde{b}_1)\nonumber\\
&&-\frac{1}{2}( \gamma_{-\frac{1}{2}} \beta_{\frac{1}{2}}- \beta_{-\frac{1}{2}} \gamma_{\frac{1}{2}} )-\frac{1}{2}(\tilde{\gamma}_{-\frac{1}{2}} \tilde{\beta}_{\frac{1}{2}}-\tilde{\beta}_{-\frac{1}{2}} \tilde{\gamma}_{\frac{1}{2}})-1.
\end{eqnarray}

\subsection{Gauge Transformations and the Superstring Field}\label{Gauge Transformations and the Superstring Field}

In this section we shall derive the picture $(-1,-1)$ ambitwistor superstring field. When dealing with picture changing later on it will be simpler to work with the `bosonised' superghosts
$$
\beta=\partial\xi e^{-\phi},	\qquad	\gamma=\eta e^{\phi},	\qquad	\tilde{\beta}=\partial\tilde{\xi} e^{-\tilde{\phi}},	\qquad	\tilde{\gamma}=\tilde{\eta} e^{\tilde{\phi}}.
$$
In this form, the superghost stress tensor contribution is $T_{\beta\gamma}=\frac{1}{2}\partial\phi\partial\phi-\partial^2\phi-\eta\partial\xi$, and similarly for $T_{\tilde{\beta}\tilde{\gamma}}$. The two sets of superghosts are independent of each other and as such we label the vacuum with two independent picture numbers $(q,\tilde{q})$. Though the notation is similar, this should not be confused with the independent holomorphic and anti-holomorphic picture labels $(q,\bar{q})$ in the conventional string; the ambitwistor string is purely holomorphic and $(q,\tilde{q})$ labels a product of holomorphic superghost vacua. We shall be working in the small Hilbert space description of the theory, where the zero modes of the fields $\xi$ and $\tilde{\xi}$ are excluded\footnote{Note that only derivatives of $\xi$ and $\tilde{\xi}$ enter into the definition of the superghosts
.}. This will be realised by the additional constraint on the string field $\eta_0|\Psi\rangle=\tilde{\eta}_0|\Psi\rangle=0$ and so the list of constraints that the string field is required to satisfy is
$$
{\cal L}_0|\Psi\rangle=0,	\qquad	b_0|\Psi\rangle=0,	\qquad	\eta_0|\Psi\rangle=0,	\qquad	\tilde{\eta}_0|\Psi\rangle=0.
$$
We take these constraints as part of the definition of $|\Psi\rangle$. We follow the same procedure used to find the bosonic string field in section \ref{The Ambitwistor string field}. That is, we propose the linearised transformation $\delta|\Psi\rangle=Q|\Lambda\rangle$ corresponding to linearised gauge transformations in spacetime. Taking inspiration from the the vertex operators\footnote{We know that the on-shell correlation functions involving the string fields must reduce to the integrand of the on-shell scattering amplitude.} a natural ansatz for the picture $(-1,-1)$ string field is
\begin{equation}\label{guess}
\Psi(z)=\int\rd k\left(E_{\mu\nu}(k)\;\psi^{\mu}\,\widetilde{\psi}^{\nu}\;e^{-\phi-\tilde{\phi}} c\tilde{c}+...\right)\;e^{ik\cdot X},
\end{equation}
where $E_{\mu\nu}(k)$ is a momentum space field which is a sum of parts symmetric and antisymmetric in the $\mu$ and $\nu$ indices. Knowing in advance that we want to recover, at the very least, the linearised target space diffeomorphisms from $Q|\Lambda\rangle$, we shall take the gauge parameter field to be
$$
\Lambda(z)=-\int\rd k\left( i\lambda_{\mu}(k)\;\psi^{\mu}\,\partial\tilde{\xi}\;e^{-2\tilde{\phi}-\phi}-  i\widetilde{\lambda}_{\mu}(k)\;\widetilde{\psi}^{\mu}\,\partial\xi\;e^{-2\phi-\tilde{\phi}} +\Omega(k)\;\partial\tilde{c}\,\partial\xi\,\partial\tilde{\xi}\;e^{-2\phi-2\tilde{\phi}} \right)c\tilde{c}\;e^{ik\cdot X},
$$
where $\lambda$, $\widetilde{\lambda}$ and $\Omega$ are momentum-dependent parameters. The gauge transformation of the string field $\Psi(z)$ to linear order is given by
$$
\delta\Psi(z)=\oint_z\rd\omega\; j(\omega)\,\Lambda(z),
$$
where $j(\omega)$ is the BRST current. Using the OPEs given in (\ref{OPE}) and
\begin{equation}\label{OPE2}
\xi(z)\eta(\omega)=\frac{1}{z-\omega}+...,
\qquad
e^{\ell_1\phi(z)}e^{\ell_2\phi(\omega)}=(z-\omega)^{-\ell_1\ell_2}e^{(\ell_1+\ell_2)\phi(\omega)}+...,
\end{equation}
with similar expressions for the fields $(\tilde{\phi},\tilde{\eta},\tilde{\xi})$ the transformation given by $\oint_z\rd\omega\; j(\omega)\,\Lambda(z)$ may be computed directly. The result $Q\Lambda(z)$ contains terms that cannot be interpreted as target space transformations of $E_{\mu\nu}(k)$ in the limited ansatz (\ref{guess}) above. It is therefore necessary to generalise the ansatz (\ref{guess}) to include additional terms. The procedure is analogous to that described for the bosonic string field so we shall not present the details. The resulting minimal ansatz for the superstring field is
\begin{eqnarray}\label{field}
\Psi(z)&=&\int\rd k\left(E_{\mu\nu}(k)\;\psi^{\mu}\,\widetilde{\psi}^{\nu}\;e^{-\phi-\tilde{\phi}} + 2e(k)\;\eta\,\partial\tilde{\xi}\,e^{-2\tilde{\phi}} + 2\tilde{e}(k)\;\tilde{\eta}\,\partial\xi\,e^{-2\phi}\right.\nonumber\\
&&\left.\qquad\qquad+ i f_{\mu}(k)\;\psi^{\mu}\, \partial\tilde{\xi}\;e^{-2\tilde{\phi}-\phi}\,\partial\tilde{c} + i \tilde{f}_{\mu}(k)\;\widetilde{\psi}^{\mu} \,\partial\xi\;e^{-2\phi-\tilde{\phi}}\,\partial\tilde{c}   \right)c\tilde{c}\;e^{ik\cdot X}.
\end{eqnarray}
The argument may also be understood from the perspective of the mode decomposition and is presented in Appendix \ref{Alternative derivation of the superstring field} where further details may be found. The linearised transformations of the momentum space component fields are then
$$
\delta E_{\mu\nu}(k)=ik_{\mu}\widetilde{\lambda}_{\nu}(k)+ik_{\nu}\lambda_{\mu}(k)	\qquad	\delta e(k)=-\frac{i}{2}k^{\mu}\lambda_{\mu}(k)+\Omega(k),	\qquad	\delta \tilde{e}(k)=\frac{i}{2}k^{\mu}\widetilde{\lambda}_{\mu}(k)+\Omega(k)
$$
$$
\delta f_{\mu}(k)=\frac{1}{2}k^2\lambda_{\mu}(k)+ik_{\mu}\Omega(k),	\qquad	\delta \tilde{f}_{\mu}(k)=-\frac{1}{2}k^2\widetilde{\lambda}_{\mu}(k)+ik_{\mu}\Omega(k),
$$
where $k^2=\eta^{\mu\nu}k_{\mu}k_{\nu}$. Fourier transforming, the linearised transformations become in configuration space\footnote{We have kept the arguments of the fields explicit in the hope that a momentum space field $E_{\mu\nu}(k)$ will not be confused with the corresponding, Fourier-transformed, configuration space field $E_{\mu\nu}(x)$.}
$$
\delta E_{\mu\nu}(x)=\partial_{\mu}\widetilde{\lambda}_{\nu}(x)+\partial_{\nu}\lambda_{\mu}(x)	\qquad	\delta e(x)=-\frac{1}{2}\partial^{\mu}\lambda_{\mu}(x)+\Omega(x),	\qquad	\delta \tilde{e}(x)=\frac{1}{2}\partial^{\mu}\widetilde{\lambda}_{\mu}(x)+\Omega(x)
$$
$$
\delta f_{\mu}(x)=-\frac{1}{2}\Box\lambda_{\mu}(x)+\partial_{\mu}\Omega(x),	\qquad	\delta \tilde{f}_{\mu}(x)=\frac{1}{2}\Box\widetilde{\lambda}_{\mu}(x)+\partial_{\mu}\Omega(x),
$$
where $\Box=\eta^{\mu\nu}\partial_{\mu}\partial_{\nu}$. Note that
$$
\delta \tilde{f}_{\mu}(x)=\frac{1}{2}\partial^{\nu}\Big(\delta E_{\nu\mu}(x)\Big)+\partial_{\mu}\Big(\delta e(x)\Big),	\qquad	\delta f_{\mu}(x)=-\frac{1}{2}\partial^{\nu}\Big(\delta E_{\mu\nu}(x)\Big)+\partial_{\mu}\Big(\delta \tilde{e}(x)\Big),
$$
suggesting that the associated fields should be identified. We shall see in the next section that this is indeed the case and the $f_{\mu}(x)$ and $\tilde{f}_{\mu}(x)$ are auxiliary fields which may be written in terms of the other spacetime fields $E_{\mu\nu}(x)$, $e(x)$, and $\tilde{e}(x)$. As with the bosonic string field, this superstring field is complete with regards to the linearised theory. We cannot rule out other terms playing a role when we consider the interaction terms.

\subsection{The quadratic action}\label{The quadratic action2}

Using the picture $(-1,-1)$ string field constructed in the previous section, we may now give a concrete proposal for the quadratic ambitwistor superstring action. When we come to consider interaction terms and picture changing operators in section \ref{Interactions}, the string field (\ref{field}) given in terms of the bosonised superghosts will be most useful. For the quadratic action, which does not involve any picture changing operators if we use $(-1,-1)$ picture string fields, the corresponding state $|\Psi\rangle$, written in terms of the superghosts $(\beta,\gamma)$ is more conveneint. In terms of mode oscillators the picture $(-1,-1)$ superstring field is (see Appendix \ref{Alternative derivation of the superstring field} for the derivation of this form of the superstring field)
\begin{eqnarray}\label{state2}
|\Psi\rangle&=&\int\rd k \left(E_{\mu\nu}(k)\;\psi^{\mu}_{-\frac{1}{2}}\widetilde{\psi}^{\nu}_{-\frac{1}{2}} +2e(k)\;\gamma_{-\frac{1}{2}}\widetilde{\beta}_{-\frac{1}{2}}+2\tilde{e}(k)\;\widetilde{\gamma}_{-\frac{1}{2}}\beta_{-\frac{1}{2}}\right.\nonumber\\
&&\left.+if_{\mu}(k) \;\psi^{\mu}_{-\frac{1}{2}}\widetilde{\beta}_{-\frac{1}{2}}\tilde{c}_0 +i\tilde{f}_{\mu}(k) \;\widetilde{\psi}^{\mu}_{-\frac{1}{2}}\beta_{-\frac{1}{2}}\tilde{c}_0 \right)c_1\tilde{c}_1\;|\text{-1,-1},k\rangle.
\end{eqnarray}
We take the conjugate string field to be
\begin{eqnarray}\label{state3}
\langle\Psi|&=&\int\rd k\;\langle \text{-1,-1},\text{-}k|\;c_{-1}\tilde{c}_{-1} \left(E_{\mu\nu}(k)\;\psi^{\mu}_{\frac{1}{2}}\widetilde{\psi}_{\frac{1}{2}}^{\nu} +2e(k)\;\gamma_{\frac{1}{2}}\widetilde{\beta}_{\frac{1}{2}}+2\tilde{e}(k)\;\widetilde{\gamma}_{\frac{1}{2}}\beta_{-\frac{1}{2}}\right.\nonumber\\
&&\left.+if_{\mu}(k) \;\psi^{\mu}_{\frac{1}{2}}\widetilde{\beta}_{\frac{1}{2}}\tilde{c}_0 +i\tilde{f}_{\mu}(k)\; \widetilde{\psi}^{\mu}_{\frac{1}{2}}\beta_{\frac{1}{2}}\tilde{c}_0 \right).
\end{eqnarray}
Note that, in contrast with the bosonic case, the conjugation is the standard BPZ conjugation\footnote{The $\alpha_{n\mu}$ and $\tilde{\alpha}^{\mu}_n$ mode operators do not appear in (\ref{field})} and the reflector state that appears in the propagator will be closer in spirit to that which appears in the conventional superstring. The arguments leading to the construction of the quadratic action for the bosonic ambitwistor string field also apply to the supersymmetric case. The quadratic action is therefore
\begin{equation}\label{action3}
S_2[\Psi]=\frac{1}{2}\langle \Psi|c_0Q|\Psi\rangle.
\end{equation}
Substituting the string fields (\ref{state2}), (\ref{state3}) and BRST operator (\ref{sQ}) into the quadratic action (\ref{action3}) gives an expression of the form
$$
S_2[\Psi]=\int\rd k\rd k'\langle\text{-1,-1},\text{-}k'|\;c_{-1}\tilde{c}_{-1}c_0c_1\tilde{c}_1\;{\cal F}\;|\text{-1,-1},k\rangle,
$$
where ${\cal F}={\cal F}(\tilde{c}_0,\tilde{b}_0, \alpha_0, \psi_{\pm\frac{1}{2}}, \gamma_{\pm\frac{1}{2}}, \beta_{\pm\frac{1}{2}})$ is a function that is not annihilated by the $c_{\pm 1}$, $\tilde{c}_{\pm 1}$ or $c_0$ ghosts. The vacuum is normalised to
$$
\langle \text{-1,-1},\text{-}k'|c_{-1}\tilde{c}_{-1}c_0\tilde{c}_0c_1\tilde{c}_1|\text{-1,-1},k\rangle=\delta(k+k'),
$$
and so the only contributions that come from the ${\cal F}$ function are those proportional to $\tilde{c}_0$. After some straightforward algeba, we find
\begin{eqnarray}
S_2[\Psi]&=&\int \rd k\Bigg(
-\frac{1}{4}E_{\mu\nu}(\text{-}k)\,k^2\, E^{\mu\nu}(k)-2\tilde{e}(\text{-}k)\,p^2\,e(k)-if^{\mu}(\text{-}k)\,k^{\nu}\,E_{\mu\nu}(k)+i\tilde{f}^{\nu}(\text{-}k)\,k^{\mu}\,E_{\mu\nu}(k)\nonumber\\
&&+2if^{\mu}(\text{-}k)\,k_{\mu}\,\tilde{e}(k)+2i\tilde{f}^{\mu}(\text{-}k)\,k_{\mu}\,e(k)-f_{\mu}(\text{-}k)\,f^{\mu}(k)-\tilde{f}^{\mu}(\text{-}k)\,\tilde{f}_{\mu}(k)
\Bigg)\nonumber.
\end{eqnarray}
We Fourier transform to bring this action to a form written in terms of configuration space fields
\begin{eqnarray}
S_2[\Psi]&=&\int \rd x\Bigg(
\frac{1}{4}E_{\mu\nu}(x)\Box E^{\mu\nu}(x)+2\tilde{e}(x)\Box e(x)-f_{\mu}(x)f^{\mu}(x)-\tilde{f}^{\mu}(x)\tilde{f}_{\mu}(x)\nonumber\\
&&-f^{\mu}(x)\Big[\partial^{\nu}E_{\mu\nu}(x)-2\partial_{\mu}\tilde{e}(x)\Big]+\tilde{f}^{\nu}(x)\Big[\partial^{\mu}E_{\mu\nu}(x)+\partial_{\nu}e(x)\Big]\Bigg).
\end{eqnarray}
As anticipated, there are no kinetic terms for the $f(x)$ and $\tilde{f}(x)$ fields, so they are auxiliary fields which should be integrated out. The equations of motion for these auxiliary fields are
\begin{equation}\label{f}
f_{\mu}(x)=-\frac{1}{2}\Big(\partial^{\nu}E_{\mu\nu}(x)-2\partial_{\mu}\tilde{e}(x)\Big),	\qquad	\tilde{f}_{\mu}(x)=\frac{1}{2}\Big(\partial^{\nu}E_{\nu\mu}(x)+2\partial_{\mu}e(x)\Big).
\end{equation}
And these expressions for $f(x)$ and $\tilde{f}(x)$ are 
substituted back into the action to give\footnote{Alternatively, if one were attepting to quantise, they may be integrated out in the configuration space path integral.}
\begin{equation}\label{action}
S_2[E,e,\tilde{e}]=\int \rd x\left(
\frac{1}{4}E_{\mu\nu}(x)\Box E^{\mu\nu}(x)+2\tilde{e}(x)\Box e(x)+f_{\mu}(x)f^{\mu}(x)+\tilde{f}^{\mu}(x)\tilde{f}_{\mu}(x)\right),
\end{equation}
where the fields $f_{\mu}(x)$ and $\tilde{f}_{\mu}(x)$ are now understood as shorthand for the expressions in (\ref{f}). A crucial point to notice is that $E_{\mu\nu}(x)$ does not have definite symmetry so there is more than simply the graviton in the spectrum of the theory. This is what we expect from the massless NS sector of Type II supergravity, yet to see the connection with the linearisation of the standard Type II supergravity action we must do a little more work.

\subsubsection{Field Redefinitions}\label{Field Redefinitions}

The quadratic action (\ref{action}) does not yet bear an obvious relationship to Type II supergravity; however, a similar construction emerges from conventional \emph{bosonic} string field theory on toroidal backgrounds in the derivation of Double field theory \cite{Hull:2009mi}, which describes the physics of the massless NS sector of the bosonic string\footnote{The novelty in \cite{Hull:2009mi} of course is that the zero mode contributions from winding strings are also included so the argument goes beyond the usual supergravity approximation.}. In what follows we closely mirror the extraction of the familiar massless NS sector from an action of the form (\ref{action}) to show that the ambitwistor string field does give rise to the correct supergravity limit. Note that, since we shall only be discussing configuration space fields in this section, we shall not explicitly include the $x$-dependence. Following \cite{Hull:2009mi}, it is useful to define
$$
\vartheta^{\pm}=\frac{1}{2}\big(e\pm\tilde{e}\big).
$$
Notice that $\vartheta^+$ and $\vartheta^-$ transform as
$$
\delta\vartheta^+=\frac{1}{2}\partial^{\mu} \epsilon_{\mu}+\Omega,	\qquad	\delta \vartheta^-=-\frac{1}{2}\partial^{\mu}\zeta_{\mu},
$$
 where we have defined
 $$
\zeta_{\mu}=\frac{1}{2}(\lambda_{\mu}+\widetilde{\lambda}_{\mu}),	\qquad	\epsilon_{\mu}=-\frac{1}{2}(\lambda_{\mu}-\widetilde{\lambda}_{\mu}).
$$
We see that $\vartheta^+$ is a Stueckelberg field and we can fix the $\Omega$ transformation such that $\vartheta^+=0$. In this case $e=-\tilde{e}$ and the action becomes, after some integrations by parts
$$
S_2=\int \rd x\left(
\frac{1}{4}E_{\mu\nu}\Box E^{\mu\nu}-4\vartheta^-\Box \vartheta^-+\frac{1}{4}(\partial^{\nu}E_{\mu\nu})^2-2\vartheta^-(\partial^{\mu}\partial^{\nu}E_{\mu\nu})+\frac{1}{4}(\partial^{\nu}E_{\nu\mu})^2
\right).
$$
The $E_{\mu\nu}$ field may be split into symmetric and antisymmetric parts; $h_{\mu\nu}$ and $b_{\mu\nu}$ respectively to give
$$
S_2=\int \rd x\left(
\frac{1}{4}h_{\mu\nu}\Box h^{\mu\nu}+\frac{1}{2}(\partial^{\nu}h_{\mu\nu})^2 - 2\vartheta^-(\partial^{\mu}\partial^{\nu}h_{\mu\nu})-4\vartheta^-\Box \vartheta^-+\frac{1}{4}b_{\mu\nu}\Box b^{\mu\nu}+\frac{1}{2}(\partial^{\nu}b_{\mu\nu})^2
\right).
$$
This is the linearised action for a metric, $B$-field and scalar field $\vartheta^-$. A more natural field choice includes the dilaton $\phi$ defined by
$$
\phi=\vartheta^-+\frac{1}{4}h,
$$
where $h=\eta^{\mu\nu}h_{\mu\nu}=\eta^{\mu\nu}E_{\mu\nu}$ is the trace of the graviton. The motivation for the field redefinition is that this dilaton is invariant under the linearised gauge transformations. Integrating the $b_{\mu\nu}$ terms by parts, one can massage this action into the more familiar form
$$
\frac{1}{4}b_{\mu\nu}\Box b^{\mu\nu}+\frac{1}{2}(\partial^{\nu}b_{\mu\nu})^2
\approx
-\frac{1}{12}H_{\mu\nu\lambda}H^{\mu\nu\lambda},
$$
where $\approx$ denotes equality up to total derivatives and the Kalb-Ramond field strength takes its usual form $H_{\mu\nu\lambda}=\partial_{[\mu}b_{\nu\lambda]}$. The linearised action is then
\begin{eqnarray}\label{action2}
S_2[h,b,\phi]&=&\int \rd x\left(
\frac{1}{4}h_{\mu\nu}\Box h^{\mu\nu}+\frac{1}{2}(\partial^{\nu}h_{\mu\nu})^2 +\frac{1}{2}h\partial^{\mu}\partial^{\nu}h_{\mu\nu}-\frac{1}{4}h\Box h\right.\nonumber\\
&&\left. -4\phi\Box \phi+2h\Box\phi-2\phi\partial^{\mu}\partial^{\nu}h_{\mu\nu}
-\frac{1}{12}H_{\mu\nu\lambda}H^{\mu\nu\lambda}
\right).
\end{eqnarray}
If we set $b_{\mu\nu}=0$ and $\phi=0$, we recover the Fierz-Pauli action (\ref{FP}) for the graviton $h_{\mu\nu}$. Noting that the linearised Ricci scalar is $R=\partial^{\mu}\partial^{\nu}h_{\mu\nu}-\Box h$ giving the standard dilaton coupling, we see that the action (\ref{action2}) is simply the linearised action for the NS sector of the Type II supergravity, the full non-linear action for which, up to Weyl rescaling, is
$$
S=\int e^{-\phi}\left(R*1-\frac{1}{2}H_{(3)}\wedge *H_{(3)}+*\rd\phi\wedge\rd\phi\right).
$$
To complete the discussion at the linearised level, we note that the gauge transformations of the component fields are
$$
\delta h_{\mu\nu}=\partial_{\mu}\zeta_{\nu}+\partial_{\nu}\zeta_{\mu},	\qquad	\delta b_{\mu\nu}=\partial_{\mu}\epsilon_{\nu}-\partial_{\nu}\epsilon_{\mu},		\qquad	\delta\phi=0.
$$
These are the standard linearised gauge transformations of the graviton, Kalb-Ramond field and dilaton.

The proposed quadratic term (\ref{action3}) does indeed produce the correct linearised theory, transforming under the correct linearised gauge transformations (diffeomorphisms and antisymmetric tensor transformations). We could go on to gauge fix and compute the propagator for the theory. The Siegel gauge $\tilde{b}_0|\Psi\rangle=0$ would be a good candidate for a gauge choice, leading to the quadratic action
$$
S_2[\Psi]=\frac{1}{2}\langle\Psi|c_0\tilde{c}_0\widetilde{L}_0|\Psi\rangle,
$$
as in the bosonic case, except that the string fields are those of the supersymmetric theory and $\langle\Psi|$ is related to $|\Psi\rangle$ by BPZ conjugation. The propagator then has the same form as (\ref{prop}), the only differences being that the reflector state is that appropriate for the supersymmetric theory and we must also insert a GSO projector to ensure that only GSO projected states propagate. Note that we have only considered the NS sector. We shall briefly discuss the Ramond sector in section \ref{Discussion}.

\section{Interaction Terms}\label{Interaction Terms} 

The ambitwistor superstring field theory has a non-polynomial action of the form (\ref{SFTaction}) with $|\Psi\rangle$ now given by the $(-1,-1)$ picture NS superstring field (\ref{field}) and the BRST charge (\ref{sQ}). Having introduced the superstring field and discussed the quadratic action at length in sections \ref{Gauge Transformations and the Superstring Field} and \ref{The quadratic action2}, we turn now to the interaction terms $\{\Psi^n\}$. In this section the interaction terms of the superstring field theory will be sketched. In many ways the novel features are present already in the bosonic case and, with the bosonic case understood, the application of the current state of the art of conventional superstring field theory to the ambitwistor theory is expected to be straightforward.

The supersymmetric generalisation of the bosonic surface state discussed in section \ref{The Surface State} will be presented in section \ref{The surface superstate}.  The ghost insertions are the same as those in the bosonic case; however gauge fixing the gravitini leads to superghost insertions that are best understood in terms of picture changing operators \cite{Friedan:1985ge,Verlinde:1987sd} in this framework. The picture changing operators (or PCOs) are the main qualitative modification to structure of the bosonic interaction terms (\ref{form2}). We consider the PCOs in the next section and explicitly derive superstring fields in the $(0,0)$, $(-1,0)$ and $(0,-1)$ pictures.

Of course, the whole issue of PCOs may be avoided by working explicitly in terms of super-Riemann surfaces, where one hopes the story may ultimately prove simpler \cite{Witten:2012bh}. As we are only interested in tree level\footnote{The Type II supergravity does not exist as a spacetime quantum theory; however, there have been a  number of interesting developments in relating the form of one-loop ambitwistor string amplitudes to results obtained directly from supergravity \cite{Geyer:2015jch,Geyer:2015bja}.} we do not have to worry about obstructions to this approach at higher genus \cite{Donagi:2013dua}.

\subsection{Picture Changing}\label{Picture Changing}

In the first quantised ambitwitor string, one requires operators in different pictures to compute scattering amplitudes. The BRST current may be written as
$$
j(z)=c(z)T(z)+\gamma(z)S(z)+\tilde{\gamma}(z)\widetilde{S}(z)+\tilde{c}(z)H(z),
$$
where the supercurrents are $S(z)$ and $\widetilde{S}(z)$. The gravitini may be set to zero everywhere except at $n-2$ points. On a super Riemann surface of sufficiently low genus the integrating out of the odd moduli results in the insertion of picture changing operators ${\cal X}(z)$ and $\widetilde{\cal X}(z)$ at $n-2$ points, which may be written in terms of the supercurrents as ${\cal X}(z)=\delta(\beta(z))S(z)$ and $\widetilde{\cal X}(z)=\delta(\widetilde{\beta}(z))\widetilde{S}(z)$, respectively.

These basic issues carry over to the insertion of picture changing operators in the string field theory; however, various technical issues make the story here a little more subtle. Ideally one would choose a picture number in which to represent the string field and then incorporate PCO insertions at $n-2$ points into the definition of the forms $\Omega_{|\vec{\Psi}\rangle}$ \cite{Saroja:1992vw}. The canonical choice would be to use superstring fields in the $(-1,-1)$ picture as the basic ingredient. Any correlation function we compute must have total picture number $(-2,-2)$, so using string fields of picture $(-1,-1)$ we must insert $n-2$ ${\cal X}$ PCOs and $n-2$ $\widetilde{\cal X}$ PCOs to compute a correlation function of $n$ string fields. There are however, potential pitfalls with this approach if PCOs are allowed to collide \cite{Wendt:1987zh}, possible solutions for which have been proposed by various authors including \cite{Erler:2013xta,Erler:2014eba}. We shall not add anything substantive to this discussion, as will be primarily concerned with how the current technology of string field theory may be applied to the ambitwistor string.

\subsubsection{Picture changing operators}\label{Picture changing operators}

Since the picture changing operators come from the supergeometry, which in this case\footnote{As opposed to the action being given by integrating over a middle-dimensional cycle of $\Sigma_{2|1}\times\widetilde{\Sigma}_{2|1}$ in the conventional string \cite{Witten:2012bh}.} is $\Sigma_{2|2}$, we expect picture changing operators ${\cal X}$ \emph{and} $\widetilde{\cal X}$, one set coming from the gauge fixing of each of the two gravitini $\chi$ and $\widetilde{\chi}$. Thus, in contrast  to the conventional string, the ambitwistor string has two sets of \emph{holomorphic} picture changing operators
$$
{\cal X}(z)=\oint_z\rd\omega \;j(\omega)\xi(z),	\qquad	\widetilde{\cal X}(z)=\oint_z\rd\omega \;j(\omega)\tilde{\xi}(z),
$$
where $j(z)$ is the BRST current. Using the OPEs given in (\ref{OPE}) and $(\ref{OPE2})$, the expressions are straightforwardly evaluated
$$
{\cal X}(z)=c\partial\xi +e^{\phi}P_{\mu}\psi^{\mu}+ \frac{1}{2}\partial\eta\,e^{2\phi}\tilde{b}+\frac{1}{2}\partial\left(\eta\,e^{2\phi}\tilde{b}\right),
$$
and
$$
\widetilde{\cal X}(z)=c\partial\tilde{\xi} +e^{\tilde{\phi}}P_{\mu}\widetilde{\psi}^{\mu}+ \frac{1}{2}\partial\tilde{\eta}\,e^{2\tilde{\phi}}\tilde{b}+\frac{1}{2}\partial\left(\tilde{\eta}\,e^{2\tilde{\phi}}\tilde{b}\right).
$$
We take the picture changing operator to be integrated around the relevant punctures so that we have insertions
$$
{\cal X}_0=\int_{\cal C}\frac{\rd z}{z}{\cal X}(z),
$$
where ${\cal C}$ is a contour around the puncture where the picture-changed string field is inserted. The picture $(-1,0)$ string field is then given by
$$
\Psi^{(-1,0)}(z)=\widetilde{\cal X}_0\Psi^{(-1,-1)}(z)=\oint_z\frac{\rd \omega}{\omega-z}\widetilde{\cal X}(\omega)\Psi^{(-1,-1)}(z).
$$
Explicitly,
\begin{eqnarray}\label{-10}
\Psi^{(-1,0)}(z)&=&\int\rd k\Big(-e(k)\;\eta-\left(\tilde{e}(k)\;\partial\tilde{\xi}\partial^2c\,\tilde{c}  +i\tilde{f}_{\mu}(k)\;\widetilde{\Pi}^{\mu}\,\partial\xi\,\partial\tilde{c}\,\tilde{c}\right)\,e^{-2\phi}
\nonumber\\
&&+2\tilde{e}(k)\;(P\cdot\widetilde{\psi}+ik\cdot\partial \widetilde{\psi})\tilde{\eta}\,\partial\xi\,\tilde{c} \,e^{\tilde{\phi}-2\phi} +\left( E_{\mu\nu}(k)\;\widetilde{\Pi}^{\nu}\psi^{\mu}\tilde{c}+\frac{i}{2} f_{\mu}(k)\;\psi^{\mu} \partial \tilde{c} \right)e^{-\phi}\nonumber\\
&&+\frac{1}{2}E_{\mu\nu}(k)\;\tilde{\eta}\psi^{\mu}\widetilde{\psi}^{\nu}\,e^{-\phi-\tilde{\phi}}-\tilde{e}(k)\;\left(2\partial\tilde{\eta}\tilde{b}\,\tilde{c}+\frac{3}{2}\partial^2\tilde{\eta} \right)\tilde{\eta}\,\partial\xi\,e^{-2\phi+2\tilde{\phi}}\nonumber\\
&&+\frac{i}{2}\tilde{f}_{\mu}(k)\;\widetilde{\psi}^{\mu}(\tilde{\eta}\partial\tilde{c}-2\partial\tilde\,\tilde{\eta})\partial\xi\,e^{-2\phi+\tilde{\phi}}
\Big)c\;e^{ik\cdot X}
\end{eqnarray}
The picture $(0,-1)$ string field has a similar expression which may be found by inspection of the above result (\ref{-10}). Finally, the picture $(0,0)$ string field is given by
$$
\Psi^{(0,0)}(z)={\cal X}_0\widetilde{\cal X}_0\Psi^{(-1,-1)}(z):= \oint_z\frac{\rd \omega}{\omega-z} \oint_z\frac{\rd \omega'}{\omega'-z}{\cal X}(\omega){\cal X}(\omega')\Psi^{(-1,-1)}(z).
$$
Applying ${\cal X}_0$ to (\ref{-10}) above gives
\begin{eqnarray}\label{0}
\Psi^{(0,0)}(z)&=&\int\rd k \left( E_{\mu\nu}(k)\;\Pi^{\mu}\widetilde{\Pi}^{\nu}\tilde{c}+\frac{1}{2}e(k)\;\partial^2c+\frac{1}{2}\tilde{e}(k)\;\partial^2c+ \frac{i}{2}f_{\mu}(k)\;\Pi^{\mu}\partial\tilde{c}+  \frac{i}{2}\tilde{f}_{\mu}(k)\;\widetilde{\Pi}^{\mu}\partial\tilde{c}\right.
\nonumber\\
&&\left. -\left(e(k)\;\left(P\cdot\psi+ik\cdot\partial\psi\right)\eta-\frac{1}{2}E_{\mu\nu}(k)\;\eta\widetilde{\Pi}^{\nu}\psi^{\mu}+\frac{i}{2}f_{\mu}(k)\;\psi^{\mu}\partial\eta\right)\,e^{\phi}\right.
\nonumber\\
&& \left.- \left(\tilde{e}(k)\;\left(P\cdot\widetilde{\psi}+ik\cdot\partial\widetilde{\psi}\right)\eta-\frac{1}{2}E_{\mu\nu}(k)\;\Pi^{\mu}\tilde{\eta}\widetilde{\psi}^{\nu}+\frac{i}{2}\tilde{f}_{\mu}(k)\;\widetilde{\psi}^{\nu}\partial\tilde{\eta}\right)\,e^{\tilde{\phi}}\right.
\nonumber\\
&&\left.-e(k)\;\partial\eta\,\tilde{b}\,\eta\,e^{2\phi}-\tilde{e}(k)\;\partial\tilde{\eta}\,\tilde{b}\,\tilde{\eta}\,e^{2\tilde{\phi}}
\right)c\,e^{ik\cdot X},
\end{eqnarray}
where we have defined
$$
\Pi^{\mu}=P^{\mu}+(k\cdot\psi)\psi^{\mu},	\qquad	\widetilde{\Pi}^{\mu}=P^{\mu}+(k\cdot\widetilde{\psi})\widetilde{\psi}^{\mu}.
$$
It is reassuring to see that the leading term is what we would expect from the picture $(0,0)$ vertex operator $V(z)=c\tilde{c} \;\varepsilon^{\mu\nu}\;\Pi_{\mu}\widetilde{\Pi}_{\nu}e^{ik\cdot X}$ found in \cite{Mason:2013sva}.

\subsection{The Surface Superstate and Interactions}\label{The Surface Superstate and Cubic Interactions}

In this section we discuss the interaction terms. We shall focus on the cubic interactions to begin with as a number of technical complications enter beyond cubic order. Some of those issues will be discussed in section \ref{The Action Beyond Cubic Order}. Much of the discussion in the bosonic case carries over to the supersymmetric theory. The basic building block is the supersymmetric generalisation of the surface state constructed in section \ref{The Surface State}. The surface state for the supersymmetric theory was constructed in \cite{Reid-Edwards:2015stz} and is required to
give the correct scattering amplitude when contracted with asymptotic states with the appropriate ghost and PCO insertions.

\subsubsection{The surface superstate}\label{The surface superstate}

The surface state for the supersymmetric ambitwistor theory \cite{Reid-Edwards:2015stz} is simply an extension of that found for the bosonic theory to include the fermionic sector and the superghosts. We shall also refer to the supersymmetric surface state as $\langle \Sigma|$, hopefully without any confusion arising. It may be written as
$$
\langle \Sigma|=\int \prod_{i=1}^n\rd p_{(i)}\;\delta\left(\sum p_{(i)}\right)\;\langle q_1;p_{(1)}| ... \langle q_n; p_{(n)}|\;\exp(V_m+V_{\text{gh}}+\widetilde{V}_{\text{gh}}){\cal Z},
$$
where the matter contribution to $V$ is given by
$$
V_m=\sum_{m,n}\sum_{i,j}\left( {\cal S}^{mn}(z_i,z_j)\tilde{\alpha}_m^{(i)}\cdot\alpha^{(j)}_n+ {\cal K}^{rs}(z_i,z_j)\psi^{(i)}_r\cdot\psi^{(j)}_s+ {\cal K}^{rs}(z_i,z_j)\widetilde{\psi}^{(i)}_r\cdot\widetilde{\psi}^{(j)}_s\right),
$$
where ${\cal S}^{mn}(z_i,z_j)$ is identical to that given in the bosonic theory. The function ${\cal K}^{rs}(z_i,z_j)$ is given by
$$
{\cal K}^{rs}(z_i,z_j)=\oint_{t_i=0}\rd t_i\oint_{t_j=0}\rd t_j\, t_i^{-m-\frac{1}{2}}t_j^{-n-\frac{1}{2}}\;\sqrt{ h'_i h'_j}\;\frac{1}{h_i(t_i)-h_j(t_j)}.
$$
The ghosts contribute the $V_{\text{gh}}$ term, the explicit form of which is given in (\ref{V}) but now augmented by a similar expression involving the superghosts. The superghost term is identical to that in the conventional superstring \cite{AlvarezGaume:1988sj}. The role of the ${\cal Z}$, as in the bosonic theory, is to strip off the $c(z)$ and $\tilde{c}(z)$ ghosts respectively of three of the $n$ string fields that contract with $\langle\Sigma|$. The $q$ in the $\langle q_i;p_{(i)}|$ denote the picture number of the vaccuum being used. We shall usually take this to be $q=-1$, where picture changing operators are inserted to ensure that the overall picture number is $-2$ at tree level.

\subsubsection{The Action to cubic order}

A proposal for the 3-point interaction term is
\begin{equation}\label{s3}
\{\Psi^3\}= \langle\Sigma| |\Psi^{(-1,-1)}\rangle|\Psi^{(-1,-1)}\rangle |\Psi^{(0,0)}\rangle ,
\end{equation}
where the picture $(-1,-1)$ states are given by (\ref{state2}) and the picture $(0,0)$ state may in principle be derived by substituting (\ref{0}) into (\ref{limit}). Alternatively, the expression (\ref{s3}) could be written in terms of three $(-1,-1)$ picture string fields with a single pair of ${\cal X}$ and $\widetilde{\cal X}$ PCOs inserted. In principle, one could substitute these expressions into the (\ref{s3}) and derive a cubic correction to the linearised action (\ref{action2}). This would be a long process and would require evaluating the state corresponding to (\ref{0}) which, we expect, takes a rather complicated form.  A more useful approach is to use the string fields as written in (\ref{field}) and (\ref{0}) and to evaluate (\ref{s3}) as an off-shell correlation function in the worldsheet CFT. For a long time it was thought that there was no off-shell extension to on-shell amplitudes in conformal field theory \cite{Collins:1974bq}; however, a clear approach was later set out for the bosonic \cite{Samuel:1988ua} and supersymmetric \cite{Lechtenfeld:1988hx} string theories.

There are various subtleties that must  be addressed when computing CFT correlation functions off-shell. These issues have been explored in \cite{Samuel:1988ua} and, for the most part, are due to the fact that the formalism is no longer conformally invariant and so many of the tricks that are usefully employed in CFT no longer apply. A key feature is that conformal invariance is lost off-shell and so the mapping from the Riemann surface that describes the worldsheet to the complex plane $\C$ where we compute the Green's function is non-trivial (see Appendix \ref{Conformal maps and vertex functions} for related issues). Conformal transformation factors must be taken into account and, in general, the amplitude will not be independent of the location of the vertex insertion points.

The off-shell amplitudes computed by conformal field theory methods are the same as those computed by the string field theory and so provide an alternative method of computation. It is simplest to deal instead with the string field interactions as off-shell correlation functions in the conformal field theory, using the CFT description of the string field (\ref{field}) instead of (\ref{state2}). The cubic interaction term (\ref{s3}) may written as the off-shell correlation function
\begin{equation}\label{P3}
\{\Psi^3\}=\langle \Psi^{(-1,-1)}(z_1)\Psi^{(-1,-1)}(z_2)\Psi^{(0,0)}(z_3)\rangle.
\end{equation}
Though a lengthy calculation, some aspects have been checked in detail and are found to be consistent with the expected action of type II supergravity to cubic order
\begin{eqnarray}\label{S3}
S_3&=&\int\rd x\Big(-\frac{1}{8}E_{\mu\nu}\Big( -(\partial_{\lambda}E^{\lambda\nu})(\partial_{\rho}E^{\mu\rho}) -(\partial_{\lambda}E^{\lambda\rho})(\partial_{\rho}E^{\mu\nu}) -2(\partial^{\mu}E_{\lambda\rho})(\partial^{\nu}E^{\lambda\rho})
\nonumber\\
&& +2(\partial^{\mu}E_{\lambda\rho})(\partial^{\rho}E^{\lambda\nu}) +2(\partial^{\lambda}E^{\mu\lambda})(\partial^{\nu}E_{\lambda\rho}) \Big)
 +\frac{1}{2}E_{\mu\nu}f^{\mu}\tilde{f}^{\nu}-\frac{1}{2}f^{\mu}f_{\mu}\tilde{e}+\frac{1}{2}\tilde{f}^{\mu}\tilde{f}_{\mu}e
\nonumber\\
&&-\frac{1}{8}E_{\mu\nu}\Big( (\partial^{\mu}\partial^{\nu}e)\tilde{e}  - (\partial^{\mu}e)(\partial^{\nu}\tilde{e})     - (\partial^{\nu}e)(\partial^{\mu}\tilde{e})  +e\partial^{\mu}\partial^{\nu}\tilde{e} \Big)
\nonumber\\
&& -\frac{1}{4}f^{\mu}\Big( E_{\mu\nu}\partial^{\nu}\tilde{e}+\partial^{\nu}(E_{\mu\nu}\tilde{e})\Big) +\frac{1}{4}f^{\mu}\Big( (\partial_{\mu}e)\tilde{e}-e\partial_{\mu}\tilde{e})  \Big)
\nonumber\\
&& -\frac{1}{4}\tilde{f}^{\nu}\Big( E_{\mu\nu}\partial^{\mu}e+\partial^{\mu}(E_{\mu\nu}e)\Big) +\frac{1}{4}\tilde{f}^{\nu}\Big( (\partial_{\nu}e)\tilde{e}-e\partial_{\nu}\tilde{e})  \Big)
\Big).
\end{eqnarray}
 Adapting the steps given in \cite{Hull:2009mi}, one may show that, once the auxiliary fields $f_{\mu}$ and $\tilde{f}_{\mu}$ are eliminated, the correct cubic actions for the NS sector of the Type II string is recovered.

We expect (\ref{S3}) to be reproduced by the ambitwistor string field theory interaction (\ref{P3}). In detail the computation is lengthy and we have not checked it in full. The terms cubic in $E_{\mu\nu}$ in (\ref{S3}) follow from the fact that the operator formalism must reproduce the correct three-point on-shell scattering amplitude and so are very easy to check. The string field $\Psi$ was found to be suitable for the linearised theory; however, one outstanding question is whether or not additional contributions to $\Psi$ become necessary in the non-linear theory. A more detailed study of the cubic action contribution will shed some light on this issues and we hope to return to it in the future.

\subsubsection{The Action Beyond Cubic Order}\label{The Action Beyond Cubic Order}

In this section we briefly outline a proposal for the higher order interaction terms. At cubic order, picture changing must be considered. Beyond cubic order additional considerations enter. In particular, regions of moduli space must be integrated over. More specifically, because we are dealing with off-shell quantities, regions of a bundle over moduli space must be considered. Moreover, more picture changing operators must be inserted in order to have a meaningful result. In the context of the NS sectors of the Type II and Heterotic superstring field theories a proposal was given in \cite{Saroja:1992vw}, where $\{\Psi^n\}$ was given by
$$
\{\Psi^n\}=\int_{{\cal V}_n}\langle\Sigma|{\cal B}_{n-3}(\vec{\nu})|\vec{\Psi}\rangle.
$$
The object ${\cal B}_{n-3}(\vec{\nu})$ generalises the insertion of $B_{n-3}(\vec{\nu})$ (\ref{form}) in the bosonic theory to include information on the picture changing operators. The supersymmetric generalisation for the conventional string is given in \cite{Saroja:1992vw} and we propose the following generalisation for the ambitwistor string field theory
$$
{\cal B}_{n-3}=\sum_{r=0}^{n-3}B^{(r)}_{n-3}\wedge K_n^{n-3-r}\wedge \widetilde{K}^{n-3-r}_n,
$$
where $B_{n-3}^{(r)}$ is a form on $T^*{\cal M}_n$ given by $B_n^{(r)}=\prod_{a=1}^{r}\tilde{\mathbf{b}}(\vec{\nu}_a)\mathbf{b}(\vec{\nu}_a)\bar{\delta}\Big({\cal H}(\vec{\nu}_a)\Big)$, so that $B_{n-3}$ is the top form, and $K_n^{2n-6-r}$ is a $2n-6-r$ form encoding the location of the picture changing operators. The $K_n$ are defined by two conditions \cite{Saroja:1992vw}. Firstly, they must satisfy a descent equation $\rd K^{(r)}_n=\{Q,K_n^{r+1}]$, where the derivative is taken with respect to the moduli $\tau^a$. Secondly, at the boundary $\partial{\cal V}_n$ the $K_n^{(r)}$ must decompose as
$$
K_n^{(r)}\Big|_{\partial{\cal V}_n}=\sum_{s=0}^rK_{n_L}^{(r-s)}\wedge K_{n_R}^{(s)},
$$
where $n_L+n_R=n-2$. The lowest form $K^{(0)}_n$ is given by
$$
K^{(0)}_n=\sum_{\alpha}A^{(\alpha)}(\tau_1,...,\tau_{2n-6}) {\cal X}(w_1^{(\alpha)}(\tau))...{\cal X}(w_{n-2}^{(\alpha)}(\tau)),
$$
where ${\cal X}$ are picture changing operators and, for each $\alpha$, $w^{(\alpha)}(\tau)$ denote a set of $n-2$ coordinates. For a given $\alpha$ we can think of $A^{(\alpha)}(\tau_1,...,\tau_{2n-6})$ as an arbitrary function of the moduli and $w^{(\alpha)}$ an associated arbitrary location of a picture changing operator. There is a tremendous potential ambiguity in the choices of the locations $w^{(\alpha)}(\tau)$ and in the coefficients $A^{(\alpha)}$; however, as argued in \cite{Saroja:1992vw}, these choices do not lead to physically different results. It was advocated in \cite{Sen:2014pia} that a number of choices, each labelled by $\alpha$, be made which respect certain symmetry requirements and then the sum over $\alpha$ averages over them\footnote{It is required that $\sum_{\alpha}A^{\alpha}=1$.}. A simple choice is \cite{Sen:2014pia}
$$
K^{(r)}=\left[\prod_{i=1}^{n-3}\Big({\cal X}(z_i)-\partial\xi(z_i)\rd z_i\Big)\right]^r,
$$
where the $r$ superscript instructs us to pick the $r$-form from the expression. For this choice ${\cal B}_{n-3}=\sum_{r=0}^{n-3}B^{(r)}_{n-3}\wedge K^{(r)}\wedge \widetilde{K}^{(r)}$.

Following \cite{Sen:2014pia,Sen:2015hia}, we generalise the space $\widehat{\cal A}_n$ introduced in section \ref{Ambitwistor interactions as forms} to the space $\widetilde{\cal A}_n$ include the locations of the $n-2$ ${\cal X}$ PCOs and the $n-2$ $\widetilde{\cal X}$ PCOs in the fibre data. To integrate over this space one would need an analogue of the vertical integration described in \cite{Sen:2014pia,Sen:2015hia} to avoid  spurious singularities.

Note that, for $n=3$, we only have ${\cal B}_0=K^{(0)}\wedge\widetilde{K}^{(0)}$ and there are no ghost insertions as all punctures are fixed and so ${\cal B}_0={\cal X}(\omega_1)\widetilde{\cal X}(\omega_2)$, giving the insertion of a single PCO of each kind at an arbitrary point. It is worth mentioning that another approach that seeks to avoid potential spurious singularities associated with the location of the PCO was presented by \cite{Erler:2013xta,Erler:2014eba}, where the solution found was to smear the PCO's along closed paths around each vertex. This was done in a permutation invariant way to produce a generalised notion of the string vertex. The approach proposed there would work equally well for the string field theory considered in this article.

We stress that we have not worked out the details of dealing with PCOs in the ambitwistor string field theory. Our modest aim here is to suggest plausible ways in which the current machinery of conventional superstring field theory may be adapted to the ambitwistor case. It is possible that a more thorough analysis may yield subtleties that require further consideration. As a final comment, one should ideally treat the discussion in the section in terms of the cotangent bundle of the supermoduli space along the lines suggested in \cite{Ohmori:2015sha}.

\section{Discussion}\label{Discussion}

In this article, we have outlined how the ambitwistor string theory of \cite{Mason:2013sva} can be used to give a string field theory description of Type II supergravity. Some details remain to be worked out, particularly in the supersymmetric case and the details of the perturbation theory; however, the general structure is clear.

An important outstanding issue is providing a clear understanding of the propagator. This is an outstanding problem even in the first quantised ambitwistor string and, despite some promising avenues \cite{Ohmori:2015sha,Li:2017emw,Casali:2017zkz}, a satisfactory geometric understanding of the propagator still remains just out of reach. What hints there are suggest that this is one of a growing number of aspects in which our intuition for the conventional and the ambitwistor string differ in important ways. It is our hope that a consideration of the string field theory might shed some light on this important issue. Another important element we have not provided is a  proof of the main identity for the ambitwistor superstring field. This identity is key to understanding the algebraic structure of the string field theory and a proof this identity, or an analogous one, in the context of the ambitwistor theory would provide additional evidence of the self-consistency of the theory being proposed here.

In light of recent advances \cite{Geyer:2015bja,Geyer:2015jch} it would also be interesting to formally extend the theory to loop level. Although the supergravity which this superstring field theory describes does not exist as a quantum theory, the study of the loop integrands in such theories has provided striking proposals for simplifying loop calculations which may be applicable to other theories. It would be interesting if the operator formalism could shed light on some of these developments.

A clear omission has been any discussion of the Ramond sector of the theory. Recently, Sen demonstrated how a kinetic term for the Ramond sector may be introduced, giving a full BV master action for the type II and Heterotic string field theories \cite{Sen:2015uaa}. We anticipate that the construction may be extended fully to the ambitwistor theory considered here.  The proposed action is
$$
S=-\frac{1}{2}\langle\Phi|c_0Q\mathscr{G}|\Phi\rangle+\langle\Phi|c_0Q|\Psi\rangle+\sum_{n=3}^{\infty}\frac{1}{n!}\{\Psi^n\},
$$
where $|\Psi\rangle$ is a string field, now including the Ramond sector, of picture number $(q,\tilde{q})$ where $q$ and $\tilde{q}$ are $-1$ for the NS sector and $-1/2$ for the Ramond sector. For example, a contribution to $|\Psi\rangle$ from the RNS or RR sectors will be given by contributions in the $(-1/2,-1)$ or $(-1/2,-1/2)$ pictures respectively. By contrast $|\Phi\rangle$ is a string field of picture number $(q,\tilde{q})$ where $q$ and $\tilde{q}$ are again $-1$ for the NS sector, but $-3/2$ for the Ramond sector. This is a natural ambitwistor modification of the action presented in \cite{Sen:2015uaa}.

The key to writing down a kinetic term for the Ramond sector was the introduction of the operator $\mathscr{G}$. We adapt this to the ambitwistor string in the obvious way so that the action of $\mathscr{G}$ acts trivially on the Neveu-Schwarz sector and inserts a picture changing operator if the string field is in the Ramond sector. As described in \cite{Sen:2015uaa}, the equations of motion for the $\Phi$ gives rise to the condition $|\Psi\rangle=\mathscr{G}|\Phi\rangle$\footnote{Since $|\Phi\rangle$ does not appear in the interactions, this restriction may be imposed consistently on the full quantum theory.} The equation of motion for $\Psi(z)$ is
$$
Q|\Phi\rangle+\sum_{n=2}^{\infty}\frac{1}{n!}[\Psi^n]=0.
$$
Writing $|\Psi\rangle=\mathscr{G}|\Phi\rangle$, then gives a non-linear equation of motion for $|\Phi\rangle$, which in turn determines $|\Psi\rangle$.

It would be interesting to see whether a kinetic term for the self-dual RR five-form in type IIB supergravity can be recovered directly by incorporating Sen's construction into the ambitwistor string field theory presented here. A proposal for such a Lorentz-invariant construction was given in \cite{Sen:2015nph}, inspired by the conventional string field theory. The ambitwsitor theory described in this paper is only a theory of supergravity and as such one might imagine that the derivation of such a kinetic term would proceed in a much simpler way in this case.

Of particular interest are the similarities and differences with the conventional string field theory. An important difference is, as noted in \cite{Mason:2013sva}, that the $X(z)X(\omega)$ OPE is trivial in the ambitwistor theory and so we can sensibly discuss functions of $X$, including metrics on curved spacetimes. Remarkably, the generalisation of the ambitwistor worldsheet theory to general curved NS backgrounds is straightforward \cite{Adamo:2014wea}. The worldsheet theory is described by the action\footnote{The worldsheet fermions $\psi^{\mu}$ and $\bar{\psi}_{\mu}$ in (\ref{SG}) are linear combinations of the worldsheet fermions appearing in (\ref{sS}) and considered throughout this paper. The precise relationship between the two sets of fermions is given in \cite{Adamo:2014wea}.} \cite{Adamo:2014wea}
\begin{equation}\label{SG}
S=\int_{\Sigma}\Pi_{\mu}\bar{\partial}X^{\mu}+i\bar{\psi}_{\mu}\bar{\partial}\psi^{\mu}+\frac{1}{2}eP^2
\end{equation}
where $\Pi_{\mu}=P_{\mu}+i\Gamma_{\mu\nu}^{\lambda}\psi^{\nu}\bar{\psi}_{\lambda}$, and $P^2=g^{\mu\nu}(X)P_{\mu}P_{\nu}$. The conserved bosonic charges are the stress tensor $T(z)$ and the null condition $H(z)=\frac{1}{2}P^2(z)$. Treating the fields $\bar{\psi}_{\mu}$ and $\psi^{\mu}$ as fundamental, the constraint $T(z)$ does not depend on the background and so it seems possible to impose a constraint analogous to ${\cal L}_0|\Psi\rangle=0$ on the string fields. The null condition is then imposed perturbatively as before. A generalisation of the string field theory considered in this paper to this more general sigma model is expected to lead to a description of linearised supergravity in curved backgrounds, where the ambitwistor string field describes fluctuations $h_{\mu\nu}$ about a fixed background metric $\hat{g}_{\mu\nu}$, i.e. $g_{\mu\nu}=\hat{g}_{\mu\nu}+h_{\mu\nu}$. In particular, we expect the quadratic term to include the curved space version of the Fierz-Pauli theory
$$
\langle\Psi|c_0Q|\Psi\rangle=\int\rd x\Big(\frac{1}{4}\nabla_{\mu}h_{\nu\lambda}\nabla^{\mu}h^{\nu\lambda}-\frac{1}{2}\nabla_{\mu}h_{\nu\lambda}\nabla^{\nu}h^{\mu\lambda}+\frac{1}{2}\nabla_{\nu}h\nabla_{\mu}h^{\mu\nu}-\frac{1}{4}\nabla_{\mu}h\nabla^{\mu}h+...\Big),
$$
where $\nabla_{\mu}$ is constructed using the background metric $\hat{g}_{\mu\nu}$ and the ellipsis denote terms containing other massless fields. More ambitiously, one might hope to find a genuinely background independent description of the classical supergravity as a string field theory. It would be interesting to see how the standard ingredients of the Einstein-Hilbert action arise from the superstring field theory. 

Finally, we mention the possibility of constructing a string field theory explicitly in ambitwistor space itself. Throughout this paper we have worked in cotangent bundle variables $X^{\mu}$ and $P_{\mu}$ and as such have constructed a string field theory in terms of the zero modes of these variables, i.e. on spacetime. It would be interesting to see if, by working explicitly in terms of ambitwistor coordinates on $P\mathbb{A}$, we can recast supergravity in terms of the natural language of ambitwistors. 

\begin{center}
\textbf{Acknowledgements}
\end{center}

 RR would like to thank Eduardo Casali, Chris Hull, Lionel Mason, Ivo Sachs and David Skinner for helpful conversations.
RR would also like to thank the Isaac Newton Institute for
   Mathematical Sciences, Cambridge, for support and hospitality during
   the programme Gravity, Twistors and Amplitudes (supported by EPSRC grant no EP/K032208/1) where the initial stages of work on this paper was
   undertaken. DR would like to thank the E. A. Milne Centre for Astrophysics and the Department of Physics \& Mathematics, University of Hull for their support.

\appendix
 
\section{Conformal maps and surface states}\label{Conformal maps and vertex functions}

In this Appendix we give further details on the operator formalism. In particular, a method for calculating the functions of the punctures is given for the case of the conventional string. Further details may be found in \cite{Reid-Edwards:2015stz,AlvarezGaume:1988bg,LeClair:1988sp}. A standard mode expansion of a primary field of dimension $d$ is
$$
\phi(t)=\sum_n\phi_n\,t^{-n-d}
$$
Under a conformal transformation $t\rightarrow z=h(t)$, the primary field transforms as
\begin{equation}\label{1}
\phi(t)\rightarrow h[\phi(t)]=(h'(t))^d\;\phi(h(t)).
\end{equation}
where $h'$ is the derivative of $h$ with respect to $t$. Writing this new description of the field in terms of the `old' coordinates $t$, the mode expansion may be written as
\begin{equation}\label{2}
h[\phi(t)]=\sum_nh[\phi_n]\,t^{-n-d}.
\end{equation}
where the mode coefficients may be found in the standard way
$$
h[\phi_n]=\oint_{t=0}\frac{\rd t}{2\pi i}\,t^{n+d-1}\,h[\phi(t)]
$$
which may be written in terms of the transformed field $\phi(z)=\phi(h(t))$ using (\ref{2}) as
\begin{equation}\label{3}
h[\phi_n]=\oint_{t=0}\frac{\rd t}{2\pi i}\,t^{n+d-1}\,(h'(t))^d\,\phi(h(t))
\end{equation}
For example, the dimension one field $\partial X^{\mu}(z)=\sum_n \alpha^{\mu}_nz^{-n-1}$ gives
$$
h[\alpha^{\mu}_{-n}]=\oint_{t=0}\frac{\rd t}{2\pi i}\,t^{n+d-1}\,h'(t)\,\partial X^{\mu}(h(t)).
$$
One would like to write the $X^{\mu}$ contribution to the surface state in terms of the oscillator expansion $\langle\Sigma|=\langle 0|e^{V_X}$, where
$$
V_X=\sum_{i,j=1}^n\sum_{m,n>0}{\cal N}_{mn}(z_i,z_j)\alpha_n^{(i)}\cdot\alpha_m^{(j)}.
$$
It is a straightforward application of the commutation relations to show that, for $m,n>0$,
$$
{\cal N}_{mn}(z_i,z_j)=\frac{1}{2n}\langle 0|\exp\left(\sum_{k,l}\sum_{p,q>0}{\cal N}_{pq}(z_i,z_j)\,\alpha^{(k)}_p\cdot\alpha^{(l)}_q\right)\;\alpha^{(i)}_{-m}\cdot\alpha^{(j)}_{-n}|0\rangle
$$
Since only the contributions where $p$ and $q$ equal $-m$ or $-n$ and only the $i$'th and the $j$'th Fock spaces play a role, the above expression may be written compactly as $\langle V_2||\Phi_i\rangle|\Phi_j\rangle$ and is determined by the two-point function $\langle \partial X^{(i)}(z) \partial X^{(j)}(w)\rangle$ as described in \cite{LeClair:1988sp}. If we take $\langle \partial X(z) \partial X(w)\rangle=-\eta^{\mu\nu}(z-w)^{-2}$ then we find, for $m,n>0$,
$$
\langle h_i[\alpha_{-n}^{\mu(i)}]\,h_j[\alpha^{\nu(j)}_{-m}]\rangle=\frac{1}{n} \oint_{0}\frac{\rd t_i}{2\pi i}\,t^{-n}\,h'_i(t_i) \oint_{0}\frac{\rd t_j}{2\pi i}\,t^{-m}\,h'_j(t_j)\,\frac{-\eta^{\mu\nu}}{\left(h_i(t_i)-h_j(t_j)\right)^2}.
$$
Vertex functions for other contractions may be found in a similar way. Using the ghost contraction $\langle b(z)c(w)\rangle =(z-w)^{-1}$ and (\ref{3}) it is not hard to show that
$$
{\cal K}_{nm}(z_i,z_j)= -\oint\frac{\rd t_i}{2\pi i}\oint\frac{\rd t_j}{2\pi i}\; t_i^{-n+1}t_j^{-m-2} \;\left(h_i'(t_i)\right)^2 \left(h_j'(t_j)\right)^{-1}\;\frac{1}{h_i(t_i)-h_j(t_j)}
$$
and the contribution from the $c$ zero modes is found straightforwardly from
$$
\int_{\Sigma}\rd^2z\bar{\partial} b_{\text{cl}} (c_{-1}z^2+c_0z+c_1)=\sum_{i=1}^N\oint_{z_i}b_{\text{cl}}^{(i)} (c_{-1}z^2+c_0z+c_1)
$$
Using the standard expansion $b^{(i)}(z)=(h'_i(t_i))^{-2}\sum_nb_n^{(i)}t^{-n-2}_i$ and changing the integral to local $t_i$ coordinates gives
$$
\int_{\Sigma}\rd^2z\,\bar{\partial} b_{\text{cl}}(c_{-1}z^2+c_0z+c_1)=\sum_i\sum_n{\cal M}_{nm}(z_i)b_m^{(i)}{\cal C}^n
$$
where $n=-1,0,+1$, ${\cal C}=(c_{-1},c_0,c_1)$ and
$$
{\cal M}_{nm}(z_i)=\oint_{t_i=0} \frac{\rd t_i}{2\pi i}\;t_i^{-m-2}(h_i'(t))^{-1}(h_i(t))^{n+1}
$$
Similarly, for a fermionic system with fermions $\psi^{\mu}$ for which $\langle \psi^{\mu}(z)\psi^{\nu}(w)\rangle=(z-w)^{-1}$ and (\ref{3}) we have
$$
{\cal S}_{nm}(z_i,z_j)= -\oint\frac{\rd t_i}{2\pi i}\oint\frac{\rd t_j}{2\pi i}\; t_i^{-n-\frac{1}{2}}t_j^{-m-\frac{1}{2}} \;\sqrt{h_i'(t_i) h_j'(t_j)}\;\frac{1}{h_i(t_i)-h_j(t_j)}
$$

\section{Further details on the derivation of the scattering equations}\label{Further details on the derivation of the scattering equations}

In this appendix we give further details on the derivation of the scattering amplitudes given in section \ref{Scattering Amplitudes and the Scattering Equations}. We only consider the $(X,P)$-dependent parts and define
$$
\langle \Sigma_{X,P}|=\langle p_1|...\langle p_n|e^{V_{X,P}},
$$
where $V_{X,P}$ is given by (\ref{V2}). Consider
$$
\langle \Sigma_{X,P}|\alpha^{(i)}_{-p}=\langle p_1|...\langle p_n|\Big( \alpha^{(i)}_{-p}+[V_{X,P}, \alpha^{(i)}_{-p}]+\frac{1}{2!} [V_{X,P},[V_{X,P},  \alpha^{(i)}_{-p}]]+...\Big)e^{V_{X,P}}.
$$
The first commutator is
$$
[V_{X,P},\alpha^{(i)}_{-p}]= \sum_{j\neq i}\sum_{n\geq 0} {\cal S}_{pn}(z_i,z_j) \alpha^{(j)}_n,
$$
Since $ {\cal S}_{mn}(z_i,z_j)=0$ for $m\geq 1$ and $n\geq 0$, this requires that this commutator is only non-zero if $p>0$. We note also that $ [V_{X,P},[V_{X,P},  \alpha^{(i)}_{-p}]]$ and all higher commutators vanish, leaving
\begin{eqnarray}
\langle \Sigma_{X,P}|\alpha^{(i)}_{-p}=\left\{\begin{array}{cc} 
  \langle p_1|...\langle p_n|\sum_{j\neq i}\sum_{n\geq 0} {\cal S}_{pn}(z_i,z_j) \alpha^{(j)}_n\;e^{V_{X,P}}, & p>0,\\
 \langle p_1|...\langle p_n|\alpha^{(i)}_{-p}\;e^{V_{X,P}} & p\leq 0.
\end{array}\right.
\end{eqnarray}
Since $\alpha^{(i)}_{p}$ commutes with $V_{X,P}$ for $p>0$, we then have
$$
\langle \Sigma|\alpha^{(i)}_{-p}=\langle \Sigma|\sum_{j\neq i}\sum_{n\geq 0} {\cal S}_{pn}(z_i,z_j) \alpha^{(j)}_n,	\qquad	p>0.
$$
where we have replaced $\langle \Sigma_{X,P}|$ with the full surface state $\langle\Sigma|$. It is useful to note that, if we contract with a momentum eigenstate $|k_j\rangle$
$$
\langle k_j|e^{V_{X,P}}\alpha^{(i)}_{-p}|k_j\rangle=\sum_{j\neq i} {\cal S}_{p0}(z_i,z_j) k_j
$$ 
for $p>0$ and zero otherwise. Thus we find the result
$$
\langle k_j|e^{V_{X,P}}\alpha^{(i)}_{-1}|k_j\rangle=\sum_{j\neq i} \frac{k_j}{z_i-z_j},
$$
which is simply the classical momentum $P_{\text{cl}}(z)$. A computation similar to this is used to show that the $\alpha_{-1}$ insertions in the on-shell states of the scattering amplitude give rise to $P_{\text{cl}}(z)$ in the final expression for the amplitude (\ref{amplitude}).

A second, related identity may be proven along the same lines:
\begin{eqnarray}
\langle \Sigma_{X,P}|\alpha^{(i)}_{-p}\cdot\alpha_{-q}^{(i)}&=&\langle p_1|...\langle p_n|\left(\alpha^{(i)}_{-p}\cdot\alpha_{-q}^{(i)}+\alpha_{-q}^{(i)}\cdot S_p^{(i)}+\alpha_{-p}^{(i)}\cdot S_q^{(i)}+S_p^{(i)}\cdot  S_q^{(i)}\right)\;e^{V_{X,P}}
\end{eqnarray}
where
$$
S_p^{(i)}:=\sum_{j\neq i}\sum_{n\geq 0}{\cal S}_{pn}(z_i,z_j)\alpha_n^{(j)}
$$

\section{Alternative derivation of the superstring field}\label{Alternative derivation of the superstring field}

 The procedure used to find the NS string field is the same as that used in the bosonic case. We begin with the NS vertex operator for massless fields in the $(-1,-1)$ picture
$$
V=c\tilde{c}E_{\mu\nu}\psi^{\mu}\widetilde{\psi}^{\nu}e^{ip\cdot X}.
$$
This suggests a string field given by a weighted sum over all possible momenta
$$
|\Psi\rangle=\int\rd p\left(E_{\mu\nu}\psi^{\mu}_{-\frac{1}{2}}\widetilde{\psi}^{\nu}_{-\frac{1}{2}}+...\right)c_1\tilde{c}_1|\text{-1,-1},p\rangle,
$$
where $+...$ denote possible auxiliary fields and
$$
|\text{-1,-1},p\rangle\equiv e^{-\phi(0)-\tilde{\phi}(0)}|p\rangle.
$$
To determine these auxiliary fields we consider the gauge parameter
$$
|\Lambda\rangle=-\int\rd p \left( i\lambda_{\mu}\;\psi^{\mu}_{-\frac{1}{2}}\widetilde{\beta}_{-\frac{1}{2}}-  i\widetilde{\lambda}_{\mu}\;\widetilde{\psi}^{\mu}_{-\frac{1}{2}}\beta_{-\frac{1}{2}} +\Omega\;\tilde{c}_0\,\beta_{-\frac{1}{2}}\widetilde{\beta}_{-\frac{1}{2}} \right) c_1\tilde{c}_1 |\text{-1,-1},p\rangle.
$$
The gauge transformation, at linear order, is given by
$$
\delta|\Psi\rangle=Q|\Lambda\rangle,
$$
where $Q$ is the BRST operator. Substituting in the expression for the BRST operator in the NS sector
\begin{eqnarray}
Q|\Lambda\rangle&=&-\int\rd p\left( \frac{1}{2}\tilde{c}_0\alpha_0^2+ \gamma_{-\frac{1}{2}}\alpha_0\cdot\psi_{\frac{1}{2}}+ \gamma_{\frac{1}{2}}\alpha_0\cdot\psi_{-\frac{1}{2}} +\widetilde{\gamma}_{-\frac{1}{2}}\alpha_0\cdot\widetilde{\psi}_{\frac{1}{2}}+\widetilde{\gamma}_{\frac{1}{2}}\alpha_0\cdot\widetilde{\psi}_{-\frac{1}{2}}\right.
\nonumber\\
&&\left.-2\tilde{b}_0( \gamma_{-\frac{1}{2}}\gamma_{\frac{1}{2}}+ \tilde{\gamma}_{-\frac{1}{2}}\tilde{\gamma}_{\frac{1}{2}})+...
\right)\nonumber\\
&&\times \left( i\lambda_{\mu}(p)\psi^{\mu}_{-\frac{1}{2}}\widetilde{\beta}_{-\frac{1}{2}}-  i\widetilde{\lambda}_{\mu}(p)\widetilde{\psi}^{\mu}_{-\frac{1}{2}}\beta_{-\frac{1}{2}} +\Omega(p)\tilde{c}_0\beta_{-\frac{1}{2}}\widetilde{\beta}_{-\frac{1}{2}} \right) c_1\tilde{c}_1 |\text{-1,-1},p\rangle\nonumber
\end{eqnarray}
Using the standard commutation relations gives
\begin{eqnarray}
Q|\Lambda\rangle&=&\int\rd p\left(
\left(ip_{\mu}\widetilde{\lambda}_{\nu}+ip_{\nu}\lambda_{\mu}\right)\psi^{\mu}_{-\frac{1}{2}}\widetilde{\psi}_{-\frac{1}{2}}^{\nu}\right. \nonumber\\
&&\left.+2\left(-\frac{i}{2}p\cdot\lambda+\Omega\right)\gamma_{-\frac{1}{2}}\widetilde{\beta}_{-\frac{1}{2}} +2\left(\frac{i}{2}p\cdot\widetilde{\lambda}+\Omega\right)\widetilde{\gamma}_{-\frac{1}{2}}\beta_{-\frac{1}{2}}\right.\nonumber\\
&&\left. +\left(\frac{i}{2}p^2\lambda_{\mu}-p_{\mu}\Omega\right)\psi^{\mu}_{-\frac{1}{2}}\widetilde{\beta}_{-\frac{1}{2}}\tilde{c}_0 +\left(-\frac{i}{2}p^2\widetilde{\lambda}_{\mu}-p_{\mu}\Omega\right)\widetilde{\psi}^{\mu}_{-\frac{1}{2}}\beta_{-\frac{1}{2}}\tilde{c}_0
\right)c_1\tilde{c}_1|\text{-1,-1},p\rangle\nonumber
\end{eqnarray}
This suggests the string field must have the form
\begin{eqnarray}
|\Psi\rangle&=&\int\rd p \left(E_{\mu\nu}(p)\;\psi^{\mu}_{-\frac{1}{2}}\widetilde{\psi}^{\nu}_{-\frac{1}{2}} +2e(p)\;\gamma_{-\frac{1}{2}}\widetilde{\beta}_{-\frac{1}{2}}+2\tilde{e}(p)\;\widetilde{\gamma}_{-\frac{1}{2}}\beta_{-\frac{1}{2}}\right.\nonumber\\
&&\left.+if_{\mu}(p) \;\psi^{\mu}_{-\frac{1}{2}}\widetilde{\beta}_{-\frac{1}{2}}\tilde{c}_0 +i\tilde{f}_{\mu}(p) \;\widetilde{\psi}^{\mu}_{-\frac{1}{2}}\beta_{-\frac{1}{2}}\tilde{c}_0 \right)c_1\tilde{c}_1|\text{-1,-1},p\rangle.\nonumber
\end{eqnarray}


\bibliography{newbib}
\bibliographystyle{utphys}


\end{document}